\documentstyle[psfig]{aa}

\begin{document}


\title{A LONG-WAVELENGTH VIEW ON GALAXY EVOLUTION FROM DEEP SURVEYS
BY THE INFRARED SPACE OBSERVATORY
}

\offprints{A. Franceschini}

\author{A. Franceschini\inst{1}, H.~Aussel\inst{2}, C.J. Cesarsky\inst{3}, 
D.~Elbaz\inst{4}, D.~Fadda\inst{5}
}

\institute{$^1$ Dipartimento di Astronomia, Vicolo Osservatorio 5, I-35122 Padova,
Italy; E-mail: franceschini@pd.astro.it \\
$^2$ Osservatorio Astronomico, Vicolo Osservatorio 5, I-35122 Padova\\
$^3$ European Southern Observatory, Germany\\ 
$^4$ Service d'Astrophysique, CEA/DSM/DAPNIA Saclay \\
$^5$ Instituto de Astrofisica de Canarias, La Laguna, Tenerife, Spain\\
}

\date{Received February 15, 2001/ Accepted August 22, 2001}

\titlerunning{Galaxy Evolution at Long Wavelengths}
\authorrunning{Franceschini A. et al. }

\abstract{
We discuss the constraints set on galaxy evolution by a variety of data from
deep extragalactic surveys performed in the mid-IR and far-IR with the {\it Infrared 
Space Observatory} and with millimetric telescopes at longer wavelengths.
These observations indicate extremely high rates of evolution for IR galaxies, 
exceeding those measured for galaxies at other wavelengths and
comparable or larger than the rates observed for quasars.
We also match the modelled integrated emission by IR galaxies at any redshifts
with the observed spectral intensity of the extragalactic IR background (CIRB), 
as a further constraint. 
The multi-wavelength statistics on IR galaxies can be reconciled with each other by 
assuming for the bulk of the population spectral energy distributions (SED) as 
typical for starbursts, which we take as an indication that stellar 
(rather than AGN, see also Fadda et al. 2001) activity powers
IR emission by faint galaxies. According to our model and following the analysis of
Elbaz et al. (2001), the deep ISO surveys at 15 $\mu$m 
may have already resolved more than 50\% of the bolometric CIRB intensity: 
the faint ISO 15 $\mu$m source samples, relatively easy to identify in deep
optical images (Aussel et al. 1999), can then allow to investigate
the origin of the CIRB background. From our fits to the observed
optical-IR SEDs, these objects appear to mostly involve massive galaxies hosting
luminous starbursts ($SFR\sim 100$ M$_\odot$/yr).
The evolutionary scheme we infer from these data considers a bimodal star formation 
(SF),
including a phase of long-lived quiescent SF, and enhanced SF taking place during
transient events recurrently triggered by interactions and merging.
We interpret the strong observed evolution as an increase with $z$ of the rate
of interactions between galaxies ({\sl density evolution}) and an increase of their
IR luminosity due to the more abundant fuel available in the past ({\sl luminosity
evolution}): both factors enhance the probability to detect a galaxy during the "active"
phase at higher $z$. Very schematically, we associate the origin of the bulk of 
the optical/NIR 
background to the quiescent evolution, while the CIRB is interpreted as mostly due
the dusty starburst phase. The latter possibly leads to the formation
of galaxy spheroids, when the dynamical events triggering the starburst 
re-distribute already present stellar populations.
The large energy contents in the CIRB and optical backgrounds are not
easily explained, considering the moderate efficiency of energy generation by stars:
a top-heavy stellar IMF associated with the starburst phase (and compared with a more 
standard IMF during the quiescent SF) would alleviate the problem.
The evolution of the IR emissivity of galaxies from the present time to $z\sim 1$ is 
so strong that the combined set of constraints by the observed z-distributions and 
the CIRB spectrum impose it to turn-over at $z>1$: 
scenarios in which a dominant fraction of stellar formation occurs at very high-z 
are not supported by our analysis.
}

\maketitle

\section{Introduction}

High-redshift galaxies and the generation of stars during the past cosmic history are
most usually investigated by observing the stellar integrated emissions in the
UV/optical/near-IR.
 These studies, based on a variety of selection techniques and exploiting 
very large telescopes on ground and efficient photon detectors, achieved extraordinary successes
during the last ten years in discovering large numbers of high-z galaxy candidates,
most of which were later confirmed by high-sensitivity optical spectroscopy
(Madau et al. 1996; Steidel et al. 1999). The outcomes of these {\sl unbiased}
surveys are galaxies characterized by typically moderate luminosities and a
modest activity of star-formation (few M$_\odot$/yr on average).

However, that this view could be to some extent incomplete is illustrated by the 
fact that {\sl biased} optical surveys emphasize a quite more {\sl active} universe,
as revealed by
the existence of heavily metal-enriched environments around quasars and active
galaxies at any redshifts (Omont et al. 1996; Padovani \& Matteucci 1993; Franceschini \& 
Gratton 1997); by the presence of populations of massive elliptical galaxies 
up to $z>1$, with dynamically relaxed profiles and complete exhaustion of the ISM 
(e.g. Moriondo, Cimatti, Daddi 2000; Rodighiero, Franceschini, Fasano 2001), 
expected to originate from violent starbursts; 
and by the intense activity of massive stars required to explain 
the metal-polluted hot plasmas present in galaxy clusters and groups 
(Mushotzky, Loewenstein, 1997).
What is essentially missing from optical observations is the evidence of cosmic
sites where active transformations of baryons are taking place at rates high enough
to explain the above findings, among others.

Hints on such a possible missing link between the {\sl active} and {\sl quiescent} 
universe come from inspection of the local universe.
The IRAS long-wavelength surveys, in particular, have revealed 
that in a small fraction of local massive galaxies 
(the so-called luminous [LIRG] and very luminous 
[ULIRG] infrared galaxies) star-formation is taking place at very high rates
($\geq 100$ M$_\odot$/yr) (Sanders et al. 1988; Kormendy \& Sanders 1992). 
Interesting to note, the reddened optical spectra of these
objects do not contain manifest signatures of the dramatic phenomena revealed
by the far-IR observations (Poggianti and Wu 1998). All this emphasizes the role of 
extinction by
dust, which is present wherever stars are formed, but is increasingly important in 
the most
luminous objects. This also illustrates the power of extending the selection waveband
for cosmological surveys from the optical (tracing the stellar-dominated emission)
to the IR where ISM-dominated emission is observable.

It is only during the last five years that new powerful instrumentation has
allowed to start a systematic exploration of the distant universe at long
wavelengths. Three major developments have allowed this.
Firstly, the discovery in the COBE all-sky maps of a bright isotropic 
background in the far-IR/sub-mm, of likely extragalactic origin (CIRB) and
interpreted as the integrated emission by dust present
in distant and primeval galaxies (Puget et al. 1996; Hauser et al. 1998).

The second important fact was the start of operation of the bolometer
array SCUBA on the 15m sub-millimetric telescope JCMT, able to resolve
a substantial fraction of the CIRB background at long wavelengths
into a population of very luminous IR galaxies at $z\sim 1$ or larger
(Smail et al. 1997; Hughes et al. 1998; Barger et al. 1998; Blain et al.
1999).
A new powerful bolometric imaging camera (MAMBO) has also recently become 
operative on the
IRAM 30m telescope, and started to provide clean deep images of the distant
universe at 1300 $\mu$m (Bertoldi et al. 2000, 2001).

Finally, the Infrared Space Observatory (ISO) allowed for the first time to
perform sensitive surveys of distant IR sources  in the mid- and far-IR
(Elbaz et al. 1999; Puget et al. 1999) and to characterize in detail the evolution 
of the IR emissivity of galaxies up to redshift $z\sim 1$ and above.

The present paper is devoted to the analysis of a large dataset
including number counts, redshift distributions, and luminosity functions
for faint IR sources selected between $\lambda_{\rm eff}\simeq 10$ and 
$\simeq 1000\ \mu$m. We emphasize in our analysis deep survey data from
the ISO mission, particularly in the mid-IR where the number count statistics are
robust (Elbaz et al. 1999; Altieri, Metcalfe and Kneib 1999; Aussel et al. 1999;
Oliver et al. 1997). 

Our approach is different from those of previous models fitting
the IR galaxy counts. In most cases (e.g. Devriendt et al. 1999, Roche and Eales 
1999; Pearson \& Rowan-Robinson 1996; Rowan-Robinson 2001; Xu et al. 2001) 
attempts have been made to provide 
combined descriptions of the IR and optical-UV data on faint galaxies.
This approach could produce even misleading results
whenever the optical data would constrain the global solution to
give very poor fits of the IR data.
Only an extremely detailed description of the complex relationship between
optical and IR emissions could provide meaningful results at some
stages. In our view, published models (e.g. Guiderdoni et al. 1998, including 
sophisticated modelling of the formation of structures) illustrate more 
the inconsistencies emerging when comparing optical and IR statistics
on faint sources than the benefits of a combined analysis.

Another way to see the problem is to consider in some details the optical spectral 
properties of luminous and ultra-luminous IR galaxies. Poggianti \& Wu (2000),
Poggianti, Bressan \& Franceschini (2001) and Rigopoulou et al. (2000) 
have studied rest-frame optical spectra for both
local and high-redshift objects, and consistently found that $\sim 70\%-80\%$
of the energy emitted by young stars and reprocessed in the far-IR leaves no 
traces in the optical spectrum (even after correction for dust extinction), 
hence can only be accounted for by 
long-wavelength observations. Altogether, the ratio of IR to optical emissions is
very broadly distributed and no clear empirical, nor physical, relationships 
have yet been established between the two.

Consequently,
we have chosen to confine our analysis to data at long-wavelengths (10 to 1000 $\mu$m) and to search for simple parametrizations of the evolution of galaxies 
in the IR, as an attempt to provide guidelines for future physical models of galaxy 
activity and its evolution. 
Only at the end we will relate these results with optical data on faint galaxies,
by comparing the observed integrated emissions in the forms of the optical
and CIRB backgrounds.
In spite of the schematicity of such an approach,
our results already contain new critical information on galaxy formation and evolution.

The paper is organized as follows. We set first the general framework by discussing 
in Sect. 2 the information contained in the CIRB spectral intensity.
Indeed, the availability of the CIRB measurements, providing a solid constraint
on the integrated IR emission of galaxies at any epochs, is a rather unique
feature of the IR domain, making it of extreme interest for studies of galaxy
evolution. We mention here some recent measurements of the cosmic
opacity at very high ({\rm TeV}) energies, allowing to set relevant constraints
at wavelengths where the CIRB is not directly measurable.
Sect. 3 is devoted to summarize results from the most relevant survey projects 
with ISO, while in Sect. 4 some relevant data obtained with millimetric telescopes
are summarized.
Sect. 5 illustrates our attempt to reproduce the multi-wavelength data with
simple prescriptions. Our present understanding of the physical nature of the
IR source populations is discussed in Sect. 6, together with a simple physical
interpretation of their previously described evolution.
Sect. 7 contains a discussion of the global properties of high-redshift IR
galaxies, like the evolutionary SFR density, and of the energy constraints set by the
observed IR and optical backgrounds. Our conclusions are summarized in Sect. 8.

For consistency with previous analyses, we adopt for H$_0$ the value of 50 Km/sec/Mpc
(note that this choice has no impact
on our inferred evolution properties, since the dependences on H$_0$ of the 
luminosity functions and number counts cancel out). In the following we indicate 
with the symbol L$_{\rm 12}$ the luminosity $\nu L(\nu)$ calculated at $\lambda=12 \mu$m
and expressed in solar units. The same terminology is used for L at other wavelengths.
The symbol S$_{\rm 12}$ indicates the monochromatic flux (in Jy) at 12 $\mu$m
(and similarly for other wavelengths).

\section{SETTING THE FRAMEWORK: THE COSMIC INFRARED BACKGROUND}

Cosmic background radiations provide a fundamental channel of information
on high-redshift sources, particularly when, for technological limitations,
observations at faint flux levels in a given waveband are not possible
(as it is largely the case in the IR/sub-mm domain). 
We briefly review in this Section the observational status about
the recently discovered cosmological background at IR and sub-millimetric 
wavelengths (CIRB), providing important constraints on galaxy formation and evolution.
 
The discovery of the CIRB -- anticipated by a modellistic prediction by 
Franceschini et al. (1994), and made possible by the NASA's COBE mission -- 
was viewed as {\sl the first chance to determine, or at least constrain,
the integrated emission of distant galaxies} (Puget et al. 1996; Guiderdoni et al. 
1997; Hauser et al. 1998; Fixsen et al. 1998).
For comparison, extragalactic backgrounds at other wavelengths appear
to contain only moderate contributions by distant galaxies:
the Radio, X-ray and $\gamma$-ray backgrounds, apparently dominated by distant 
quasars and AGNs (Giacconi et al. 2001; Tozzi et al. 2001),
and the Cosmic Microwave Background including photons generated at $z\sim 1500$.
Also, direct measurements of the optical-UV backgrounds are hampered by the intense 
starlight reflected by high latitude "cirrus" dust and Zodiacal-reflected Sun-light.

\subsection{Observational status about the CIRB}

In spite of the presence of bright foregrounds (Zodiacal and Interplanetary
dust emission, Galactic Starlight, high-latitude "cirrus" emission), there are two
relatively clean spectral windows in the IR
where these summed emissions produce two minima:
the near-IR (3-4 $\mu$m) and the sub-mm (100-500 $\mu$m) cosmological windows.
Redshifted photons by the two most prominent galaxy emission features, 
the stellar photospheric peak at $\lambda \sim 1\ \mu$m and the one at
$\lambda \sim 100\ \mu$m due to dust re-radiation, are here observable in principle.

Particularly favourable for the detection of an extragalactic signal turned out to be 
the longer-wavelength channel. By exploiting the different 
spatial dependencies of the various dust components and the observed 
correlations with appropriate dust tracers like the neutral and ionized hydrogen 
(throug the HI 21 cm and H$_\alpha$ lines), 
Puget {\it et al. } (1996) have identified in the
all-sky FIRAS/COBE maps an isotropic signal with an intensity following the law
$\nu$B$_{\rm \nu}\simeq 3.4\times 10^{-9} (400\ \mu$m/$\lambda)^{3} \ $
W m$^{-2}$ sr$^{-1}$ in the 400--1000 $\mu$m interval.

\begin{figure*}[!ht]
\vspace{1cm}
\psfig{figure=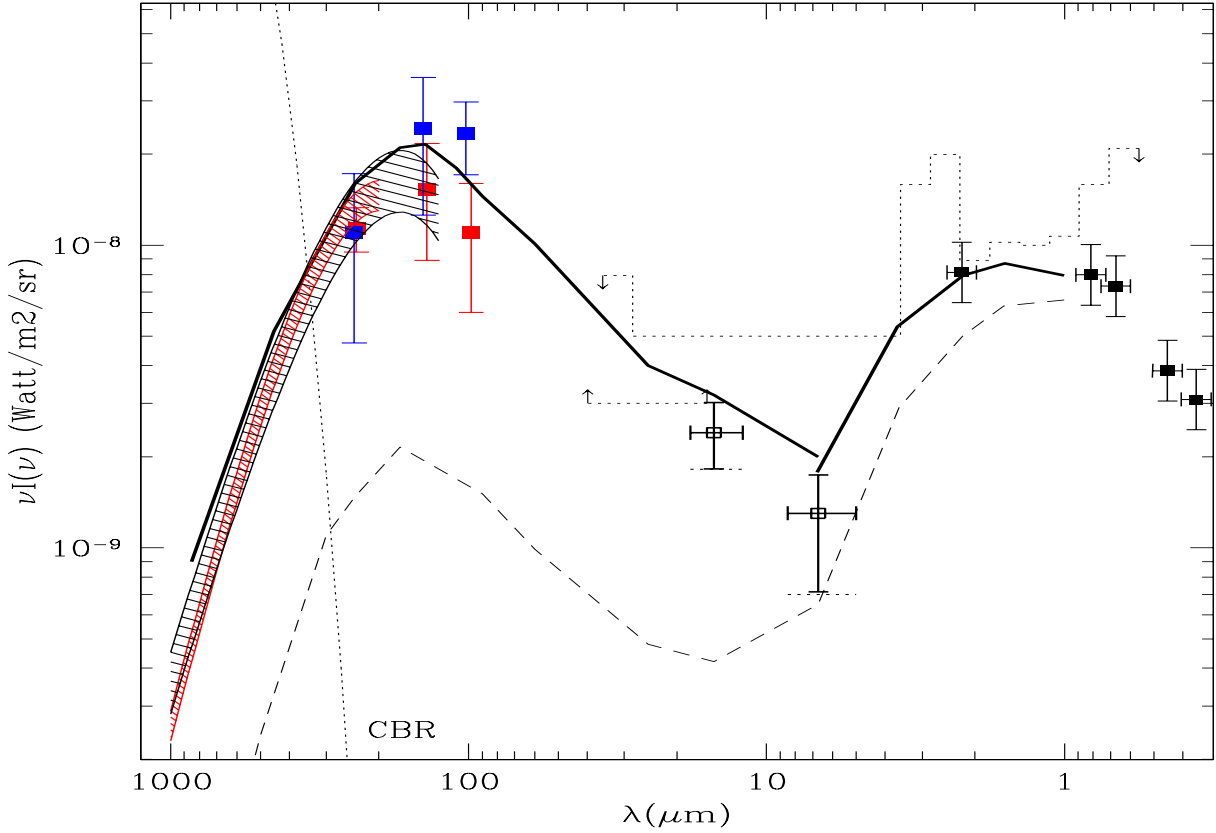,height=120mm,width=170mm}       
\caption{The Cosmic Infrared Background (CIRB) spectrum as measured by independent
groups in the all-sky COBE maps (e.g. Hauser et al. 1998), compared 
with estimates of the optical extragalactic background based on ultradeep
optical integrations by the HST in the HDF (Madau \& Pozzetti 2000). 
The three lower datapoints in the far-IR are from a re-analysis
of the DIRBE data by Lagache et al. (1999), the shaded areas from Fixsen 
et al. (1998) and Lagache et al. The two mid-IR points are the resolved fraction of the
CIRB by the deep ISO surveys IGTES (Elbaz et al. 2001), while the dashed histograms are 
limits set by {\rm TeV} cosmic opacity measurements (Sect. 2.2). 
The lower dashed line is the expected intensity based on the assumption that the 
IR emissivity of galaxies does not change with cosmic time. 
The thick line is the predicted CIRB spectrum of the presently discussed reference 
model for IR galaxy evolution.
 The dotted line marked CBR corresponds to the Cosmic Microwave Background spectrum.
}
\label{bkg}
\end{figure*}

This tentative detection has been confirmed by various other groups 
with independent analyses of data from FIRAS on COBE (e.g. Fixsen et al. 1998), 
as well as from the DIRBE experiment in two broad-band
channels at $\lambda = 140$ and $240 \mu m$ (Hauser et al. 1998).
The most recent results from DIRBE (Hauser et al. 1998; Lagache et al. 1999; 
Finkbeiner, et al. 2000) and FIRAS (Fixsen et al. 1998) are reported in Fig.
\ref{bkg}.

Finkbeiner, Davies \& Schlegel (2000), 
after a very delicate subtraction of the far dominant Galactic and IPD foregrounds,  
found an isotropic signal at 60 and 100 $\mu$m with intensities at the
level of $\sim$ 30\ $10^{-9}$ W m$^{-2}$ sr$^{-1}$. This controversial result 
(see Puget \& Lagache 2001 for a critical assessment)
appears to conflict, in any case, with independent estimates based on observations 
of the cosmic high-energy opacity (see below).

Recent analyses by Dwek \& Arendt (1998) and Gorjian, Wright \& Chary (2000) 
have claimed tentative detections in the near-IR cosmological window at 
3.5 $\mu$m and in the J, H and K DIRBE bands, however with 
large uncertainties because of the very problematic evaluation of the Zodiacal
scattered light. Because of this, CIRB estimates particularly in J, H and K
are to be taken more reliably as upper limits.

\subsection{Constraints from observations of the cosmic high-energy opacity}

No significant isotropic signals were detected at 
$4 \mu m<\lambda < 60 \mu m$, any cosmological flux being far dominated here by the 
Interplanetary dust (IPD) emission 
(to reduce it, missions to the outer Solar System would be needed).
In this wavelength interval the CIRB energy density can be presently constrained with
high-energy observations of Blazars, by measuring the optical depth at
{\rm TeV} energies due to the $\gamma \rightarrow \gamma$ interaction
with the background CIRB photons (Stecker, de Jager \& Salomon 1992).

The absorption cross-section of $\gamma$--rays of energy E$_\gamma$ 
has a maximum for IR photons with energies obeing the condition: 
$\epsilon_{\rm max} = 2 (m_e c^2)^2/E_\gamma$, or equivalently:
$\lambda_{\rm peak} \simeq 1.24\pm 0.6(E_\gamma [{\rm TeV}])\; \mu m.$
The optical depth for a high-energy photon $E_0$ travelling through a cosmic medium
filled of low-energy photons with density $\rho(z)$ from $z_{\rm e}$ to the present time is
\begin{equation}
  \tau(E_0,z_{\rm e})  =  
\label{tau}
\end{equation}
\[ 
c\int_0^{z_{\rm e}} dz {dt \over dz } \int_0^2 dx {x \over 2} \int_0^\infty d\nu
(1+z)^3 {\rho_\nu (z) \over h\nu } \sigma_{\rm \gamma\gamma} (E_0,\nu[1+z])   \]
where $\sigma_{\rm \gamma\gamma}$ is the cross-section for photon-photon interaction.
Coppi \& Aharonian (1999) report the following analytical approximation, good
to better than 40\%, to eq.(\ref{tau}): 
\[   \tau(E_0,z_{\rm e}) \simeq 0.24 \left( E_\gamma \over {\rm TeV} \right) \left(\rho(z=0) \over 
10^{-3}{\rm eV/cm^3} \right) \left( z_{\rm e} \over 0.1 \right) h_{\rm 60}^{-1} \]
\begin{equation}
\simeq 0.063 \left( E_\gamma \over {\rm TeV} \right) \left(\nu I_\nu \over {\rm nW/m^2/sr}\right) 
\left( z_{\rm e} \over 0.1 \right) h_{\rm 60}^{-1}
\label{inu}
\end{equation}

Applications of this concept have been possible when data from
the Gamma Ray Observatory and X-ray space telescopes 
have been combined with observations at {\rm TeV} energies by the Whipple, HEGRA and other 
Cherenkov observatories on Earth. Stanev \& Franceschini (1998) have discussed upper 
limits on the CIRB with minimal a-priori guess on the CIRB spectrum, using HEGRA data
for the Blazar MKN 501 (z=0.034) during an outburst in 1997, on the assumption that 
the high-energy source spectrum is the flattest allowed by the data.
These limits (dotted histogram in Fig. \ref{bkg}) get quite close to the 
background flux already resolved by the ISO mid-IR deep surveys (see Sect. 5.4 below).

More recently, Krawczynski et al. (1999) have combined the observations by Aharonian 
et al. (1999) of the MKN501 1997 outburst with X-ray data from RossiXTE and 
BeppoSAX, providing a simultaneous high-quality description of the whole high-energy 
spectrum. These data are well fit by a Synchrotron Self Compton (SSC) model 
in which the {\rm TeV} spectrum ($\nu\sim 10^{27}$ Hz) is produced
by Inverse Compton of the hard X-ray spectrum ($\nu\sim 10^{18}$ Hz): the combination
of the two constrains the shape of the "primary" (i.e. before
cosmic attenuation) spectrum at {\rm TeV} energies. This is used to derive 
$\tau_{\rm \gamma\gamma}$ as a function of energy and, after eqs. (\ref{tau}) and 
(\ref{inu}), a constraint on the spectral intensity of the CIRB.
The result is compatible with the limits by Stanev \& Franceschini (1998, see also
Renault et al. 2001)
and allows a tentative, model dependent, estimate of the CIRB intensity in the 
interval from $\lambda=20$ to 40 $\mu$m (see Fig.[\ref{bkg}]).
 
The observations of purely power-law Blazar spectra around $E_\gamma \simeq 1$ {\rm TeV}
translate into a fairly robust upper limit
of about $10^{-8}$ Watt/m$^2$ sr at $\lambda \sim 1 \mu$m shown in Fig. \ref{bkg}.
Substantially exceeding it, as sometimes suggested (Bernstein et al. 1998, 
Gorjian et al. 2000), would imply either very "ad hoc" $\gamma-$ray source spectra or new 
physics (Harwit, Proteroe \& Bierman 1999).

\subsection{Constraints by CIRB and optical backgrounds on galaxy evolution}

Altogether, after years of active debate among various teams working 
on the COBE data, first about the existence and later on the intensity and spectral
shape of CIRB, there is now ample consensus, at least from 140 to 500 $\mu$m where 
the CIRB spectrum is most reliably measured and where two completely independent datasets
(FIRAS and DIRBE, with independent absolute calibrations) are available.
The CIRB flux has in particular stabilized at values $\nu I_\nu \simeq 20 \pm 5$
and $\nu$ I$_\nu \simeq 15 \pm 5\ 10^{-9}$ Watt/m$^2$/sr at $\lambda=140$ and
240 $\mu$m.  Modest differences in the calibration of 
FIRAS and DIRBE around 100 $\mu$m have been reported (Hauser et al. 1998),
but these do not affect the overall result.

The measurement of the CIRB provides the
global energy density radiated in the IR by cosmic sources at any redshifts.
Two concomitant facts -- the very strong K-correction for galaxies in the far-IR/sub-mm
due to the very steep and featureless dust spectra, and their 
robustness due to the modest dependence of dust equilibrium temperature $T$ on
the radiation field intensity -- have suggested to use the CIRB spectrum to infer
the evolution of the long-wavelength galaxy emissivity as a function of redshift
(Gisper, Lagache \& Puget 2000). 
Indeed, while the peak intensity at $\lambda =100$ to 200 $\mu$m constrains it
at $z\leq 1$, the low foreground contamination at $\lambda>$ 
200 $\mu$m allows to set important constraints on the universal emissivity at $z>1$. 

Between 100 and 1000 $\mu$m the observed integrated CIRB intensity turns out to be 
$\sim (30\pm 5)\ 10^{-9}$ Watt/m$^2$/sr. In addition to this measured part of the CIRB, 
one has to consider the
presently un-measurable fraction resident between 100 and 10 $\mu$m.
Adopting modellistic extrapolations as in Fig. \ref{bkg}, consistent with the 
constraints set by the cosmic opacity observations, the total energy density 
between 7 and 1000 $\mu$m rises to 
\begin{equation}
\nu I(\nu)|_{\rm FIR} \simeq 40 \ 10^{-9}\ {\rm Watt/m}^2{\rm /sr} .
\label{CIRB}
\end{equation} 

This flux is to be compared with the integrated bolometric emission by distant galaxies 
between 0.1 and 7 $\mu$m (the "optical background"), for which we adopt the value
given by Madau \& Pozzetti (2000):
\begin{equation} 
\nu I(\nu)|_{\rm opt} \simeq (17 \pm 3) \ 10^{-9}\ {\rm Watt/m}^2{\rm /sr} .
\label{optical}
\end{equation} 
This latter has been obtained from HST source counts between 0.3 and 3 $\mu$m 
down to the faintest detectable limits, by exploiting the number count convergence 
at magnitudes $m_{\rm AB}\geq  22$.
A significant upwards revision of this optical background suggested by Bernstein et
al. [1998] to account for low surface brigtness emission by galaxies is not
confirmed and would tend to conflict with measurements of the cosmic high-energy
opacity.

Already the directly measured part of the CIRB sets a relevant constraint on
the evolution of cosmic sources, if we consider that for local galaxies only 30\% on 
average of the bolometric flux is absorbed by dust and re-emitted in the far-IR. 
{\sl The CIRB's intensity exceeding the optical
background tells that galaxies in the past should have been much more
"active" in the far-IR than in the optical, and very luminous in an absolute sense.
A substantial fraction of the whole energy emitted by high-redshift galaxies should 
have been reprocessed by dust at long wavelengths.}

\section{RESOLVING THE CIRB INTO SOURCES: 
DEEP SKY SURVEYS WITH THE INFRARED SPACE OBSERVATORY (ISO)}

The ISO Observatory (a 60cm cryogenic telescope operated by ESA between 1995 and
1998) included two instruments of cosmological interest: a mid-IR 32$\times$32 array 
camera (ISOCAM), and a far-IR imaging photometer (ISOPHOT) with small
3$\times$3 and 2$\times$2 detector arrays from 60 to 200 $\mu$m.
The main extragalactic results from the 30-month ISO mission have been summarized by 
Genzel \& Cesarsky (2000).

The improvement in sensitivity
offered by ISO with respect to the previous IRAS surveys motivated to spend a 
relevant fraction of the observing time to perform a set of deep surveys
at mid- and far-IR wavelengths, with the aim to parallel optical searches 
of the deep sky with observations at wavelengths where dust is not only far less 
effective in extinguishing optical light (relevant for estimating 
the very uncertain extinction corrections for high redshift 
galaxies, e.g. Meurer et al. 1997), but is also an intense source of emission.
ISO observations then provided an important complementary tool 
to evaluate the global energy output by stellar populations and active nuclei.

\subsection{Overview of the main ISO surveys}

Deep surveys with ISO have been performed in two wide mid-IR
(LW2: 5-8.5\,$\mu$m and LW3: 12-18\,$\mu$m)
and two far-IR ($\lambda=90$ and 170 $\mu$m) wavebands. All surveys are
performed through repeated raster pointings to achieve the best spatial 
resolution and sensitivity.
The diffraction-limited spatial resolutions were $\sim$4.6 arcsec FWHM at 15 
$\mu$m and $\sim$50 arcsec at 100 $\mu$m. Mostly because of the better imaging,
ISO sensitivity limits in the mid-IR are three orders of magnitude
deeper in flux density than at long wavelengths (0.1 mJy versus 100 mJy).
To some extent, these different performances are counter-balanced by the typical 
FIR spectra of galaxies and AGNs, which are almost two orders of magnitude 
brighter at 100 $\mu$m than at 10 $\mu$m.
We summarize in the following the most relevant programs of ISO surveys.


Five extragalactic surveys with the LW2 and LW3 filters have been performed 
in the ISOCAM Guaranteed Time (IGTES), including large-area 
shallow surveys (S$_{\rm 15}[lim]\simeq 0.5-0.7$ mJy) and small-area deep integrations 
(S$_{\rm 15}[lim]\simeq 0.1$ mJy). A total area of 1.5 square degrees 
have been surveyed in the Lockman Hole and the "Marano" southern field, where more
than one thousand sources have been detected (Elbaz et al. 1999). These two areas 
were selected for their low zodiacal and cirrus emissions and because of
the existence of data at other wavelengths (optical, radio, X).
Since the LW2 band at 7\,$\mu$m does not sample dust emission in high-z sources 
and includes a large fraction of Galactic stars,
we will confine our analysis in the following to data in the LW3 15 $\mu$m.


The European Large Area ISO Survey (ELAIS) was the largest program in the ISO Open Time 
(Oliver et al. 2000a).
A total of 12 square degrees have been surveyed at 15 $\mu$m with ISOCAM
and at 90 $\mu$m with ISOPHOT (6 and 1 sq. degrees have been
covered at 6.7 and 170 $\mu$m respectively).
To reduce the effects of cosmic variance, ELAIS was split into
3 fields of comparable size, 2 in the north (N1, N2), one in the south (S1), 
plus six smaller areas.
While data analysis is still in progress, a source list of over 1000 
(mostly 15 $\mu$m)
sources is being published, including starburst galaxies and AGNs (type-1 and type-2), 
typically at z$<$0.5, with several quasars (including various BAL QSOs) 
found up to the highest z.

\begin{figure*}[!ht]
\psfig{figure=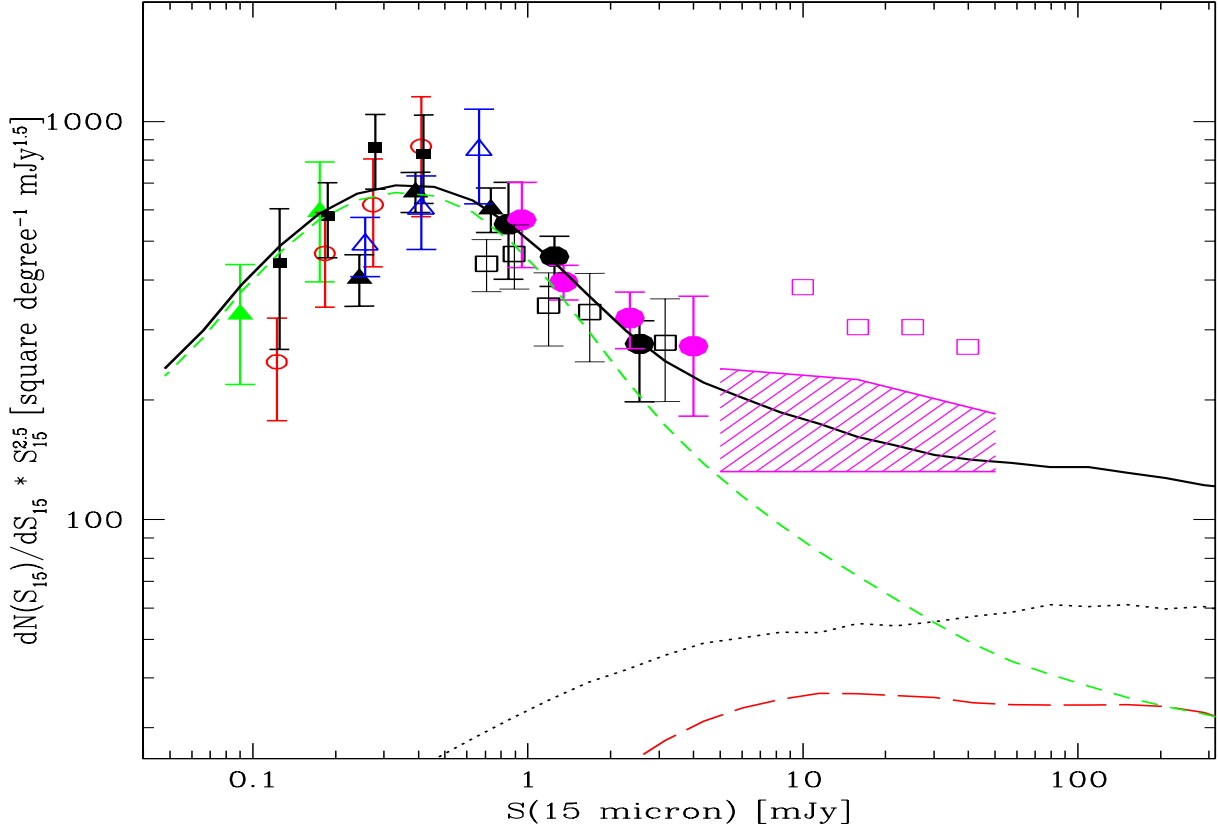,height=120mm,width=170mm}
\caption{Differential counts at $\lambda_{\rm eff}=15\ \mu$m normalized to the 
Euclidean law ($N[S]\propto S^{-2.5}$; the differential form is preferred here because
all data points are statistically independent). 
The data come from an analysis of the IGTES surveys by Elbaz et al. (1999).
The dotted line corresponds to the expected counts for a population of
non-evolving spirals. The short dashed line comes from our modelled population
of strongly evolving starburst galaxies, the long-dashed one are type-I AGNs. 
The shaded region at S$_{\rm 15}>10$ mJy
comes from an extrapolation of the faint 60 $\mu$m IRAS counts (Mazzei et al. 2001).
}
\label{cdif15}
\end{figure*}


The two ultradeep blank-field exposures by the Hubble Space Telescope 
(one in the North and the other in the South, the Hubble Deep Fields, HDF) have 
promoted a substantial effort of multi-wavelength studies aimed at
characterizing the SEDs of distant and high-z galaxies. These areas,
including the Flanking Fields for a total of $\sim 50$ sq. arcmin, 
have been observed by ISOCAM at 6.7 and 15 $\mu$m down to
a completeness limit of 100 $\mu$Jy at 15 $\mu$m.
These sensitive ISO surveys have allowed to detect dust emission from
luminous starburst galaxies up to a redshift $z=1.3$ (Rowan-Robinson et al. 1997; 
Aussel et al, 1999).     In the inner 11 sq. arcmin, the HST provides a detailed
morphological information for ISO galaxies at any redshifts.
Thanks to the variety of photometric data and an almost complete redshift
information available (Aussel et al. 1999; Cohen et al. 2000), these surveys are 
allowing the most detailed characterization of the faint IR source population.
The redshift distributions at the LW3 survey limits show an excess number of
sources between z=0.5 and z=1.2, 
partly an effect of the K-correction as explained below.

\begin{figure*}[!ht]
\psfig{figure=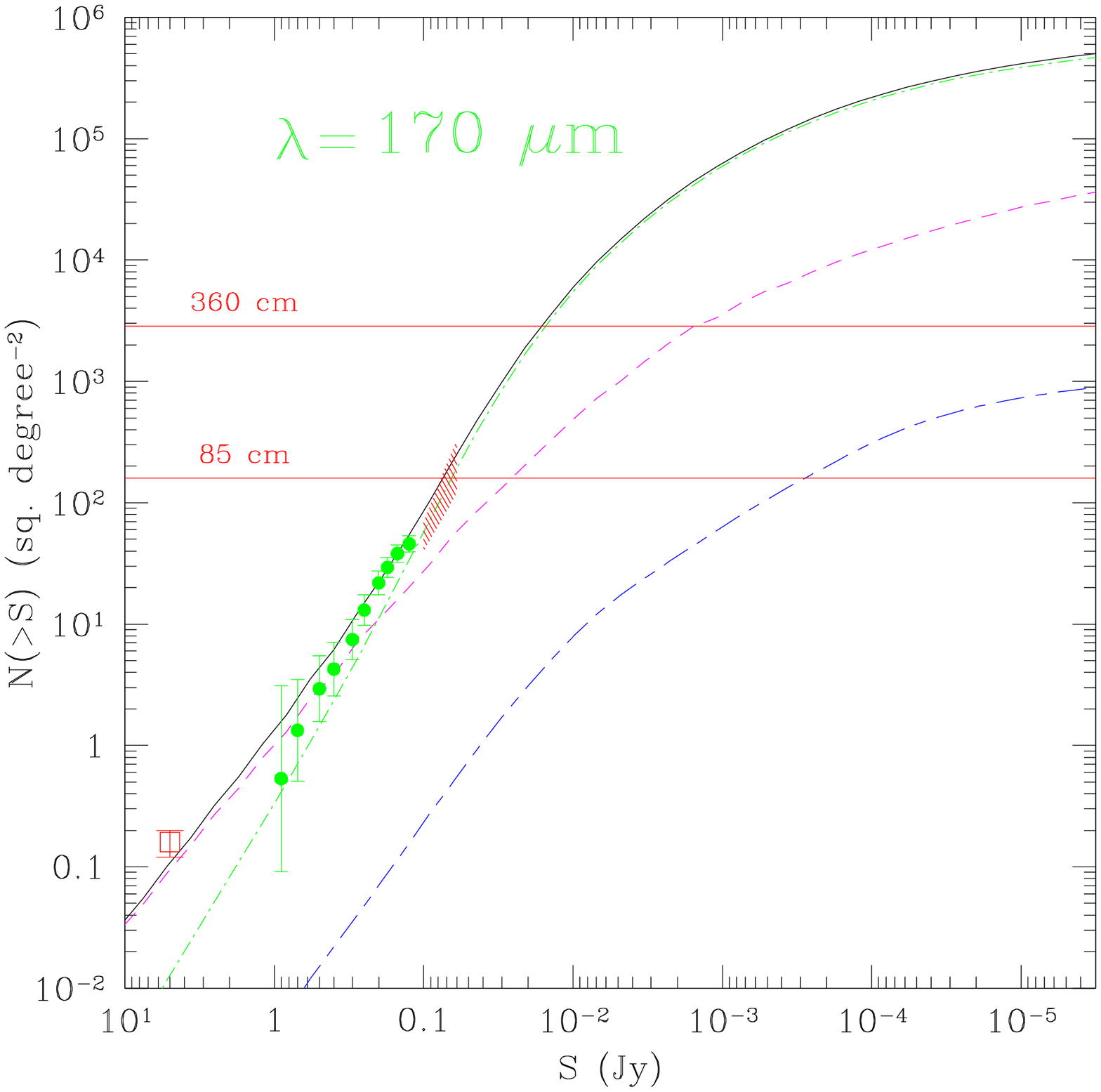,height=100mm,width=150mm}
\caption{Integral counts based on the ISOPHOT FIRBACK survey (Sect. 3.1) at 
$\lambda_{\rm eff}=170\ \mu$m (filled circles, from Dole et al. 2000) and on the ISOPHOT 
Serendipitous survey (open squares, 
Stickel et al. 1998).  The shaded region is a constraint coming from a fluctuation
analysis by Matsuhara et al. (2000). The dashed and dot-dashed lines correspond to
the non-evolving and the strongly evolving populations as in Fig.\ref{cdif15}.
The lowest curve is the expected contribution of type-I AGNs.
The horizontal lines mark the confusion limits (assumed to correspond to 27
beams/source) for various telescope sizes: the lines marked "85cm" and "360cm" 
correspond to the SIRTF and Herschel/FIRST limits for faint source detection, assumed 
diffraction-limited spatial sampling. The confusion limit scales up on the y-axis
proportionally to the square of the diameter of the primary reflector.
}
\label{c175}
\end{figure*}


Two fields from the Canada-France Redshift Survey (CFRS) have been
observed with ISOCAM to intermediate depths: the '1415+52' field (observed at 6.7 and 
15 $\mu$m) and the '0302+00' field (with only 15 $\mu$m data, but twice as deep). 
Studies of ISOCAM sources detected in both fields have provided the first
tentative interpretation of the nature of distant IR galaxies (Flores et al. 1999). 
The LW3 survey displays a redshift distribution similar to those of the HDF surveys
(see Fig. \ref{dz15} below).


FIRBACK is a set of deep cosmological surveys carried out with ISOPHOT,
specifically aimed at detecting at 170 $\mu$m the sources of the far-IR background
(Puget et al. 1999). 
Part of this survey was done in the Marano area and in ELAIS N1, and
part in collaboration with the ELAIS team in ELAIS N2.
This survey is limited by extragalactic source confusion in the large ISOPHOT beam 
(90 arcsec) to S$_{\rm 170}\geq$ 135 mJy (see for more details Puget et al. 1999 and
Dole et al. 2001). Constraints
on the counts below the confusion limit obtained from a fluctuation analysis of
one Marano/FIRBACK field are discussed by Lagache \& Puget (1999). 
The roughly 200 sources detected are presently targets of follow-up observations, 
especially using deep radio exposures to help reducing the ISO errorbox and 
identifying the optical counterparts.  Also an effort is being made
to follow-up these sources with sub-mm telescopes (JCMT, IRAM) to derive constraints
on their redshifts.


Three lensing galaxy clusters, Abell 2390, Abell 370 and Abell 2218,  
have received very long integrations by ISOCAM (Altieri et al 1999; Lemonon
et al. 1999; Biviano et al. 2001).  The lensing has been exploited to achieve even 
better sensitivities with respect to ultradeep blank-field surveys (e.g. the HDFs),
and allowed detection of sources between 30 and 100 $\mu$Jy at 15 $\mu$m
over a total area of 56 square arcmin (obviously at the expense of an additional
uncertainty introduced by flux amplification and area distortion).
The lensing-corrected 
number counts at 15 $\mu$m were used by Biviano et al. (2001) to estimate a
lower limit to the CIRB of $3.3\pm 1.3$ nW/m$^2$/sr, close to the upper limit 
by Stanev \& Franceschini (1998).


Ultra-deep surveys in the Lockman Hole and SSA13 with the LW2 7\,$\mu$m 
ISOCAM band were performed by Taniguchi {\it et~al.} (1997).
The Lockman region was also surveyed with ISOPHOT
by the same team: constraints on the source counts at 90 and 170 $\mu$
are derived by Matsuhara et al. (2000) based on a fluctuation analysis.

\subsection{ISO data reduction and analyses}

ISOCAM data needed particular care to remove -- in addition to the usual photon, readout, 
flat-field and dark current noises -- the effects of glitches induced
by the frequent impacts of cosmic rays on the 
detectors (the 960 pixels registered on average 4.5 events/sec during the mission).
This badly conspired with the need to keep them cryogenically cooled to reduce
the instrumental noise, which implied a slow electron reaction time
and long-term memory effects. 
For the deep surveys this implied a problem to
disentangle faint sources from trace signals by cosmic ray impacts.

To correct for that, tools have been developed by various groups for the
two main instruments (CAM and PHOT),
essentially based on identifying patterns in the time history
of the response of single pixels, which are specific to either astrophysical
sources (a jump above the average background flux when a source
falls on the pixel) or cosmic ray glitches (transient spikes followed by
a slow recovery to the nominal background). 
The most common "normal" glitches, induced by cosmic electron impacts and 
lasting only one or two readouts, where the easier to identify and remove.
Other less frequent impacts by protons and alpha particles 
leave longer-lasting spurious signals, from typically several to occasionally 
one hundreds or more readouts. The long integrations adopted for deep ISO
surveys were needed not only to reduce the instrumental noise, but even more to
achieve enough redundancy (number of elementary integrations per sky position)
to separate spurious from astrophyical signals in the pixel time history.

A rather performant non-parametric algorithm for ISOCAM data reduction is PRETI 
(Stark et al. 1999), exploiting multi-resolution 
wavelet transforms in the 2-D observable plane of the position on the detector 
vs. time sequence). A competely independent parametric method, based on a physical 
model for the detector transients, has been devised by Lari et al. (2001), 
and especially taylored for the shallow ELAIS integrations.
Independent tools have been developed by D\'esert et al. (1999) and by Oliver et al. 
(2000a).

The PRETI and LARI detection algorithms have been tested by means of 
Monte Carlo simulations including all artifacts introduced by the analyses. 
Such simulations have been performed
on real datasets, including both a long staring observation of more than 500
readouts (Elbaz et al. 1998) and the deep survey frame itself
(Lari et al. 2000). Test sources with known fluxes were introduced with their PSF
and model transients, against which the detection algorithm has been tested.
With these simulations it is has been possible to control as a function 
of the flux threshold: the detection reliability, the completeness, the Eddington 
bias and photometric accuracy ($\sim$10-20\% when enough redundancy was available, 
as in the CAM HDFs and IGTES ultradeep surveys). 
The PRETI and LARI methods have been applied in particular to the HDF North
dataset (Aussel et al. 1999, C. Lari, private communication), and they showed
excellent agreement down to the faintest fluxes.

The astrometric accuracy is of order of 2 arcsec for deep highly-redundant
and properly registered images, allowing relatively easy identification of the 
sources (Aussel et al. 1999, 2001; Elbaz et al. 2001; Fadda et al. 2001). 
For example, among the complete ISOCAM sample of 41 galaxy identifications in the 
HDF North studied by Aussel et al. (1999, 2001), all have an optical counterpart 
brighter than I=23 within 3", and only one source appears to be confused by the
presence of more than one optical galaxy in the ISO errorbox.
The quality and reproducibility of the results for the CAM surveys is also
proven by the good consistency of the counts from independent surveys 
(see Fig.[\ref{cdif15}] below).
 
A reliable reduction of the longer wavelength ISOPHOT observations proved to be
more difficult.
The $170\mu m$ counts from PHOT C200 surveys are 90\% reliable above the 5-$\sigma$
confusion
limit S$_{\rm 170}\sim$ 130 mJy, and required relatively standard procedures for
baseline corrections and "de-glitching".
Quite more severe are the noise problems for the C100 channel (60 to $90\mu m$, 
which would 
otherwise benefit by a better spatial resolution), preventing so far to achieve
significantly better sensitivities than IRAS. The C100 PHOT survey dataset
is still presently under analysis (C. Lari and G. Rodighiero, work in progress).

\begin{figure}[!ht]
\psfig{figure=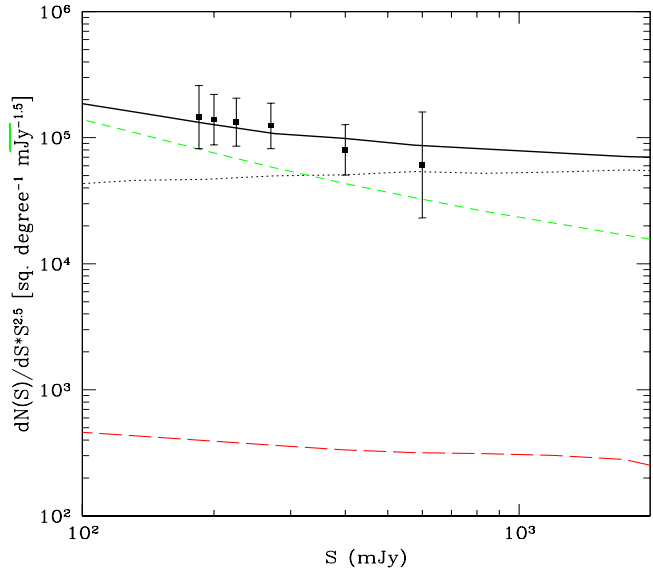,height=80mm,width=90mm}
\caption{Differential counts from the ISOPHOT FIRBACK survey at 
$\lambda_{\rm eff}=170\ \mu$m, from Dole et al. (2001), compared with the prediction 
of the reference model. As in Fig. \ref{cdif15}, the dotted line corresponds to the
non-evolving spiral population, the short-dashed to evolving starbursts,
long-dashes are type-I AGNs.
}
\label{c170dif}
\end{figure}

\subsection{Mid-IR and far-IR source counts from ISO surveys}

IR-selected galaxies have typically red colors, partly because of extinction by dust 
responsible for the excess IR flux. When found at substantial redshifts,
these sources are also quite faint in the optical. 
For this reason the redshift information is presently available 
only for limited subsamples (e.g. the HDF North and CFRS samples).
In this situation, the source number counts provide crucial
constraints on the evolution properties.


Particularly relevant information comes from the mid-IR surveys based on the 
ISOCAM 15 $\mu$m LW3 filter, because they include the faintest, most distant 
and most numerous ISO-detected sources with reliable identifications.
To cover with LW3 a wide dynamic range in flux with good source statistics, 
Elbaz et al. (1999) performed a variety of surveys with sky coverages decreasing 
as a function of the flux limits.
Including ELAIS and the IRAS data, the range in fluxes reaches four 
orders of magnitude. 

The differential counts (normalized to the Euclidean law $dN \propto S^{-2.5} dS$)
based on data from seven independent sky areas, shown in Fig. \ref{cdif15}, 
reveal a remarkable agreement. 
Of the various samples considered by Elbaz et al., only sources in flux bins for 
which 
the survey was better than 80\% complete were used (for a total of 614 sources).

At fluxes fainter than 1 mJy the contamination by stars is of the order of only few \%,
while at brighter fluxes in the Lockman Shallow Survey it reaches $\simeq 10\%$, and 
further increases at increasing flux.

In addition to the data reported by Elbaz et al., the shaded region at S$_{\rm 15}>5$ mJy
in Fig. \ref{cdif15} corresponds to an estimate of the extragalactic counts by 
Mazzei et al. (2001) overriding the problem to account for the large fraction 
of galactic stars at these bright fluxes. This estimate is based on the
60 $\mu$m IRAS galaxy counts translated to the LW3 band by using ISOCAM photometry of
a complete sample of faint IRAS sources in the North Ecliptic Pole (Aussel et al.
2000). This evaluation of the bright 15 $\mu$m counts helps to constrain the
level of non-evolving galaxies and the normalization of the local luminosity
function (see Sect. 5.1), untill a systematic identification of the ELAIS catalogues 
will be available.

The combined 15 $\mu$m  differential counts display various remarkable features
(Elbaz et al. 1999):
a roughly euclidean slope from the brightest fluxes sampled by IRAS down to 
S$_{\rm 15}\sim 10$ mJy; a fast upturn at S$_{\rm 15}< 3$ mJy, where the counts increase 
as $dN\propto S^{-3.1}dS$ to S$_{\rm 15}\sim 0.4$ mJy; and finally the evidence for a 
flattening below S$_{\rm 15}\sim 0.3$ mJy (where the slope becomes quickly sub-Euclidean,
$dN\propto S^{-2} dS$). 
Note that the sudden change in slope and the faint flux convergence is supported by 3 
independent surveys.

The areal density of ISOCAM 15\,$\mu$m sources at the limit of $\sim$50$\mu$Jy is 
$\sim$5 arcmin$^{-2}$.  If we consider that the
diffraction-limited diameter of a point-source is $\sim 50$ arcsec$^2$
and for a slope of the counts $\beta\sim -2$, 
this density is close to the ISO confusion limit at 15 $\mu$m of $\sim 0.1$ 
sources/areal resolution element, or $7/arcmin^2$ in our case (Franceschini 2000).
Confusion will likely remain a limitation for the NASA's SIRTF mission,
in spite of the moderately larger primary collector.

Obviously, far-IR selected samples are even more seriously affected by confusion.
The datapoints on the 170$\mu$m integral counts reported in Fig. \ref{c175}
come from the FIRBACK survey (Dole et al. 2001), 
while Fig. \ref{c170dif} shows the same counts in
differential units.  Similarly deep observations at 90, 150 and 180 $\mu$m
are reported by Juvela, Mattila \& Lemke (2000). These surveys have essentially 
attained at the ISOPHOT confusion limit.
Additional constraints at slightly fainter fluxes have been attempted using  
background fluctuation analyses (Lagache \& Puget 1999; Matsuhara et al. 2000,
see shaded region in Fig. \ref{c175}). 

\begin{figure}
\psfig{figure=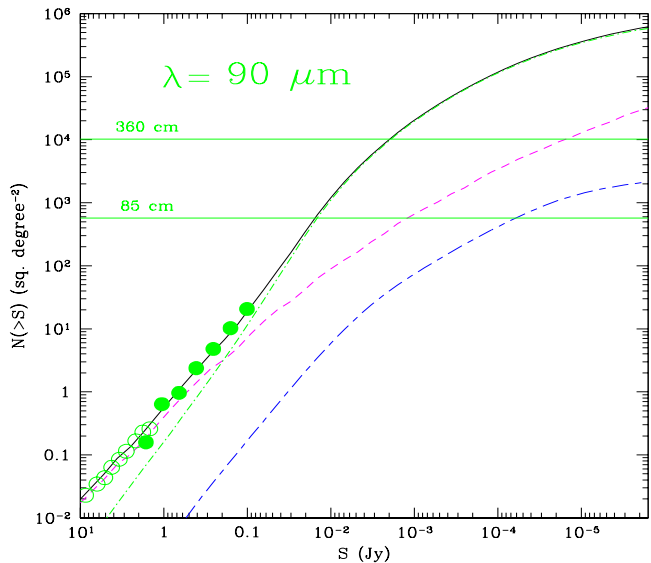,height=80mm,width=90mm}
\caption{Integral counts at $\lambda_{\rm \rm eff}=90\ \mu$m based on the IRAS (open circles) 
and the ELAIS 90 $\mu$m (filled circles) surveys, as reported by Efstathiou 
et al. (2000). Line symbols as in Fig. \ref{c175}. The lines marked "60cm" and "360cm" 
correspond to the ISO and Herschel/FIRST limits for faint source detection.
}
\label{c90}
\end{figure}

Shorter wavelength ISOPHOT C100 observations could in principle benefit by a less
severe confusion limitation.
Preliminary results of ISOPHOT ELAIS surveys at $\lambda_{\rm eff}=90\ \mu$m 
(Efstathiou et al. 2000), as well as counts derived from the IRAS 100 $\mu$m survey, 
are reported in Fig. \ref{c90}, showing nice agreement in the overlap flux range.

\section{GALAXY SOURCE COUNTS FROM SURVEYS AT MILLIMETRIC WAVELENGTHS}

Surveys in the sub-millimeter offer a unique advantage 
to naturally generate volume-limited samples from flux-limited observations.
This property is due to the peculiar shape of galaxy spectra, 
with an extremely steep slope from 1 mm to 200 $\mu$m 
[roughly $L(\nu)\propto \nu ^{3.5}$, Andreani \& Franceschini (1996)]. 
Then, as we observe in the sub-mm galaxies
at larger and larger redshifts, the selection waveband in the source rest-frame 
moves to higher frequencies along a steeply increasing
spectrum, and the corresponding K-correction almost completely counter-balances 
the cosmic dimming of the flux, for sources at $z \geq 1$ and up to $z\sim 10$
(Blain \& Longair 1993).
A further related advantage of sub-mm surveys is that local galaxies 
emit very modestly at these wavelengths. Altogether, a sensitive sub-mm survey
will avoid local objects (stars and nearby galaxies) and will select preferentially 
sources at high and very high redshifts: a kind of direct picture of the 
high-redshift universe, impossible to obtain at other frequencies.

Important discoveries have come from the operation of a powerful array of bolometers 
(SCUBA) on JCMT, due to a combined effect of sensitivity, large multiplexing 
capability, an efficient 15m primary mirror, and an operation site allowing to 
observe at short mm wavelengths ($\lambda=850\ \mu$m).    In such way,
SCUBA on JCMT has allowed to resolve more than $20\%$ of the long-$\lambda$ 
CIRB background into a population of faint distant, mostly high-z, sources.

\begin{figure*}[!ht]
\psfig{figure=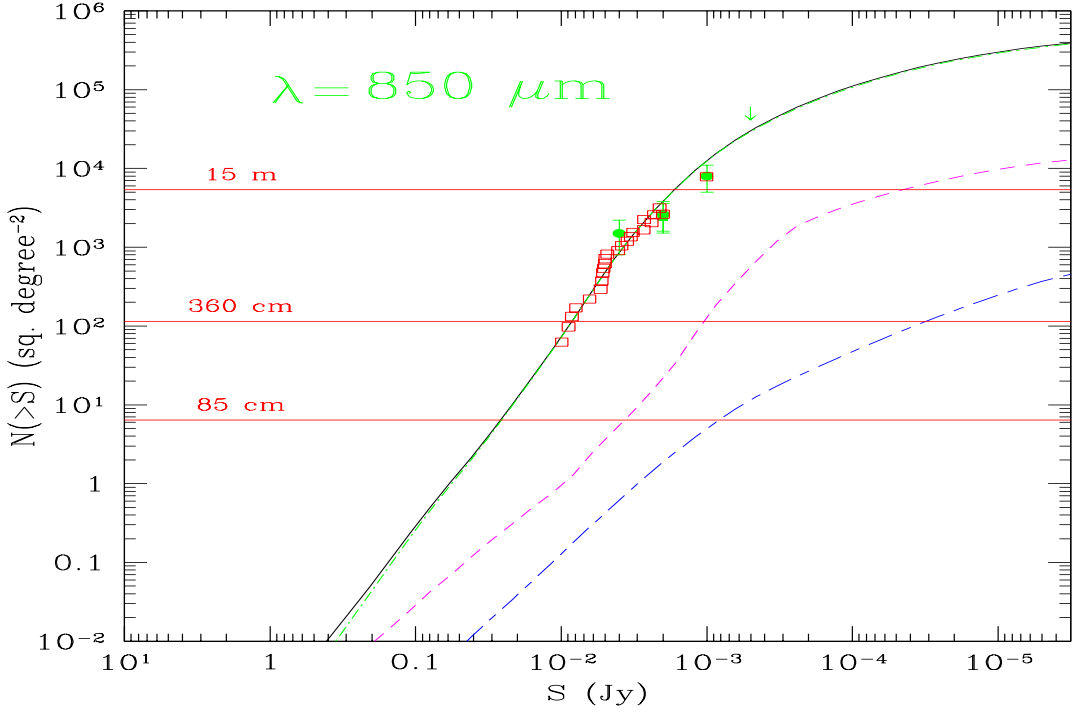,height=100mm,width=150mm}
\caption{Integral counts at $\lambda_{\rm eff}=850\ \mu$m. 
The filled squares are from Blain et al. (1999), the open squares from Barger,
Cowie and Sanders (1999).
See caption to Fig. \ref{c175} for meaning of the lines.
}
\label{c850}
\end{figure*}

Various groups have used SCUBA for deep cosmological surveys. 
Smail et al. (1997, 1999) have exploited 7 distant galaxy clusters as cosmic 
lenses, obtaining a sample of 17 (3$\sigma$) sources with S$_{\rm 850}>6$ mJy.
Hughes et al. (1998) published a single very deep image of the HDF North
containing 5 ($4\sigma$) sources at S$_{\rm 850} \geq 2$ mJy.
Lilly et al. (1999) and Eales et al. (2000) have published samples including
$\sim$30 ($3\sigma$) sources to 3.5 mJy. Scott, Dunlop et al. (2001, in prep) have recently completed a survey 
in the Lockman and ELAIS N2 regions
and detected approximately 20 ($4\sigma$) sources. 
In addition to deep sub-mm surveys (Barger et al. 1998, 1999), Barger, Cowie and 
co-workers have carried out an extensive program of follow-up of SCUBA 
sources with optical telescopes. Other systematic identification efforts 
have been attempted by Ivison et al. (2000, see also Sect. 6.3).

\begin{figure}[!ht]
\psfig{figure=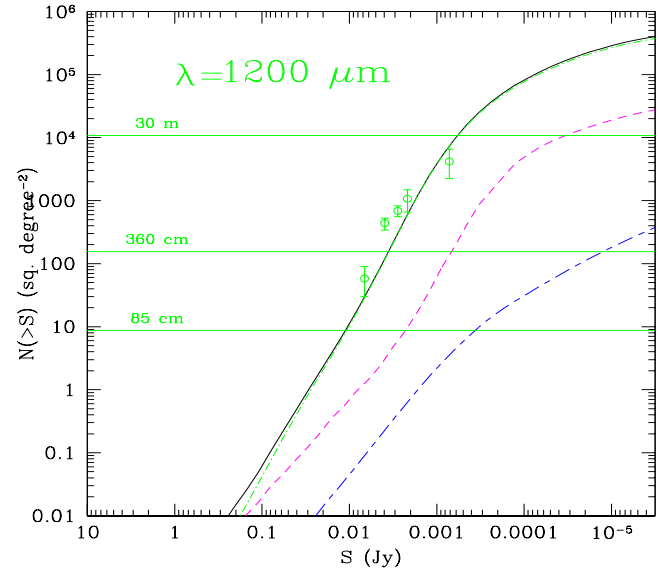,height=80mm,width=90mm}
\caption{Integral counts at $\lambda_{\rm eff}=1200\ \mu$m by Bertoldi et al. (2001).
See also caption to Fig. \ref{c175}.
}
\label{c1200}
\end{figure}

Each of these deep integrations required many tens of hours each of
especially good weather, which meant approximately 20\%
of the JCMT observatory time since 1997.
In spite of this effort, the surveyed areas (few tens of $arcmin^2$) and number of 
detected sources (from 100 to 200 sources) are modest, which illustrates the 
difficulty to work from ground at these wavelengths.
Much fewer sources have been detected in the 450 $\mu$m channel, 
for which the atmospheric transmission at JCMT is usually poor.

The extragalactic source counts at 850 $\mu$m, reported in Fig. \ref{c850}, 
show a dramatic departure from the Euclidean law [$dN \propto S^{-3}\ dS$
in the crucial flux-density interval from 2 to 10 mJy], a clear signature of the 
strong evolution and high redshift of SCUBA-selected sources.

A new bolometer array (MAMBO) has become recently operative on the largest existing
mm telescope, the IRAM 30m. 
A large survey at $\lambda_{\rm eff}=1.2$ mm of 3 fields with a total area of over 
200 arcmin$^2$ to a flux limit of few mJy has been performed with MAMBO
(Bertoldi et al. 2000): preliminary galaxy counts and evaluations of the redshift 
distributions are reported by Bertoldi et al. (2001), both confirming the SCUBA 
results at 850 $\mu$m (see Fig. \ref{c1200}).

\section{INTERPRETATIONS OF FAINT IR/MM GALAXY COUNTS AND REDSHIFT-DISTRIBUTIONS}

\subsection{Predictions for non-evolving source populations in the mid-IR}

A zero-th order interpretation of data on deep counts is obtained by
comparing them with the expectations of models assuming no-evolution for cosmic sources. 
When referred, in particular, to the mid-IR galaxy counts of Fig. \ref{cdif15},
this calculation has to account in detail for the effects of the complex mid-IR 
spectrum of galaxies (including strong PAH emission and silicate absorption features, 
see Fig. \ref{spectra} and \ref{source8} below) in the K-correction factor relating
observed flux $S$ and rest-frame luminosity $L$:
\begin{equation}
S_{\rm \nu} = {L_{\rm \nu} K(L,z) \over 4\pi d_{\rm L}^2}, \label{S}
\end{equation}
where $d_{\rm L}$ is the luminosity distance 
\footnote{
The luminosity distance can be evaluated for $\Omega_\lambda=0$ and any 
value of $\Omega_{\rm m}$ from the usual Mattig relation
$d_{\rm L} = z[1+z(1-\Omega_{\rm m}/2)/(\sqrt{1+\Omega_{\rm m} z}+1+\Omega_{\rm m} z/2)]$.
 For $\Omega_\lambda + \Omega_{\rm m}=1$,
then $d_{\rm L}=(1+z) \int_0^z [(1+z^\prime)^2(1+\Omega_{\rm m} z^\prime) + 
z^\prime(2+z^\prime)\Omega_\lambda]^{-1/2} dz$.
}
and $K(L,z)={(1+z) L[\nu (1+z)]\over L(\nu)}$ the K-correction. 
The differential number counts (sources/unit flux interval/unit solid angle) at a given flux 
$S$ write as:
\begin{equation}
{dN \over dS} = 
\int_{\rm 0}^{z_{\rm max}}\,dz\,{dV\over dz}\,{d \log L(S,z)\over dS}\, 
\rho[L(S,z),z] \label{dNdS}
\end{equation}
where $\rho[L(S,z),z]$ is the epoch-dependent luminosity function and 
$dV/dz={4\pi\over 3} {d_{\rm L}^2\over 1+z} {d (d_{\rm L})\over dz}$ is the differential volume 
element. The moments of various order of the distribution $dN/dS$ of eq.(\ref{dNdS})
provide respectively the integral counts $N(>S)=\int {dN \over dS} dS$,
the contribution to the background intensity $I(\nu)=\int {dN \over dS} S_\nu\ dS_\nu$, and 
to the background fluctuations $<[\delta I(\nu)]^2>=\int {dN \over dS} S^2_\nu\ dS_\nu$.

Taking into account the system transmission function $T(\lambda)$, the
K-correction is more appropriately written as:
\begin{equation}
 K(L,z)={(1+z)\  \int_{\rm \lambda_1}^{\lambda_2} d\lambda\ \left( \lambda_0 \over \lambda \right)\ 
T(\lambda)\ L[\nu(1+z)]            \over
\int_{\rm \lambda_1}^{\lambda_2} d\lambda\ \left( \lambda_0 \over \lambda \right)\ 
T(\lambda)\ L[\nu,z=0]} .  
\label{Kco}
\end{equation}

\begin{figure}
\psfig{figure=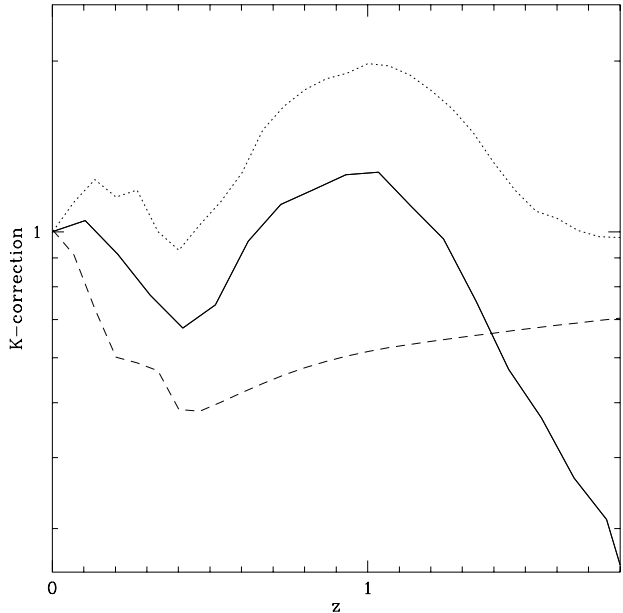,height=90mm,width=90mm}
\caption{K-correction as a function of the source redshift (eq.\ref{Kco}) 
for different filters
and source spectra. Dotted line: inactive spiral spectrum observed with the LW3 filter; 
continuous line: M82-like spectrum with LW3; dashed line: M82 spectrum observed with LW2.
Note that the spiral spectrum (dotted line) implies the strongest K-correction at $z\sim 1$
because of the lack of hot-dust emission depressing the rest-frame LW3 flux compared
with the redshifted PAH-dominated emission.
}
\label{Kcorr}
\end{figure}

Fig. \ref{Kcorr} illustrates the effects of the mid-IR emission/absorption
features on K-correction,
for different mid-IR spectra (an inactive spiral [dotted line], a M82-like starburst
[continuous line], both observed with ISOCAM LW3) and different filters (the starburst M82 
observed with LW2 [dashed line]). The detailed spectral response functions of the
various filters, as given in the ISOCAM Observer Manual (ESA Publications),
have been used in eq. (\ref{Kco}).
The effect of the prominent emission features on the counts is particularly 
important in the wide LW3 (12-18 $\mu$m) filter, as they imply an enhanced sensitivity
to sources at $0.5<z<1.3$, and a diminished one outside this redshift interval.
Correspondingly, the LW3 selection provides improved capability to study in detail
the evolution of galaxies between $z=0.5$ and 1, a redshift-interval that current
cosmogonic models (e.g. Kauffmann \& Charlot 1998) predict 
to be critical for the formation of structures.

\begin{figure*}[!ht]
\psfig{figure=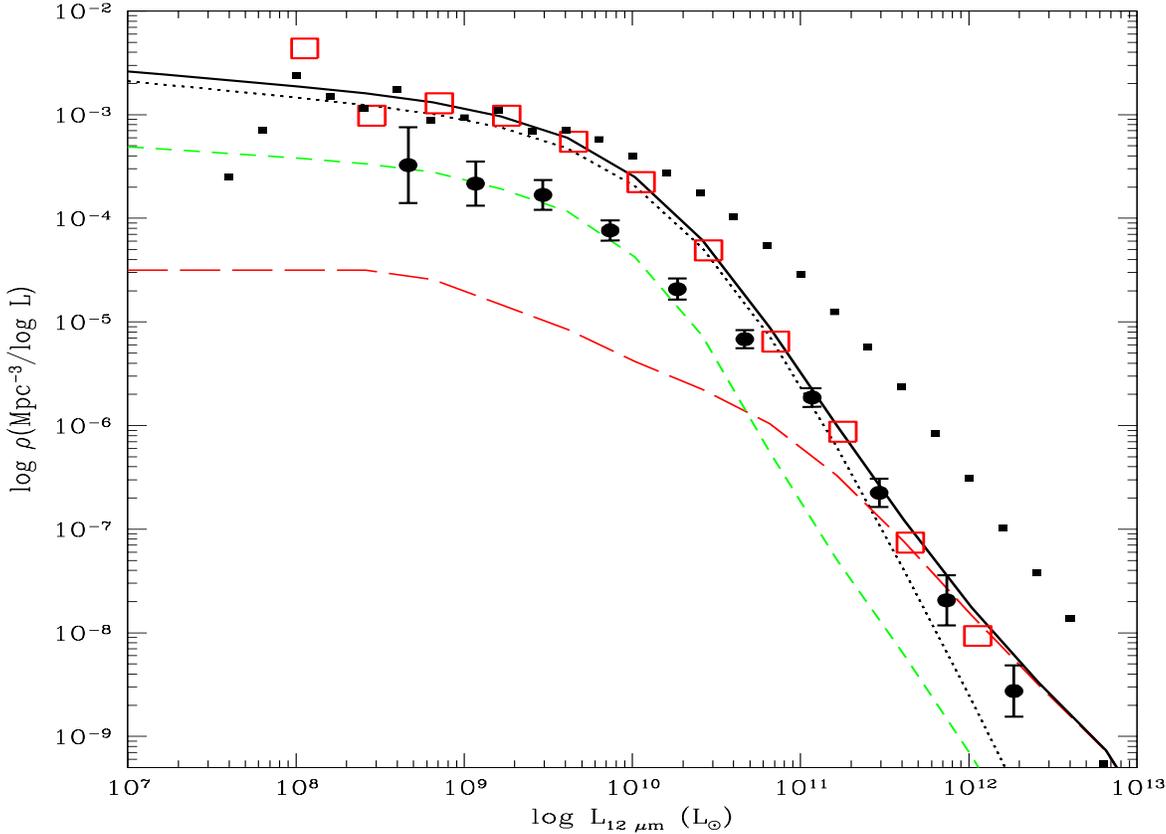,height=120mm,width=160mm}
\caption{Galaxy LLF's at 12 $\mu$m from Fang et al. (1998, red open squares)
and adapted from Xu et al. (1998) in the low-luminosity regime. A comparison
is made with the IRAS 60 $\mu$m LLF by Saunders et al. (1990, small filled squares).
Black circles are an estimate, based on the Rush et al. (1993) catalogue, 
of the contribution to the 12 $\mu$m LLF by 
active galaxies (including type-I AGNs [red long-dashed line] and type-II AGNs plus 
starbursts [green short-dash line]). Type-II AGNs and starbursts are included in the
same population on the assumption that in both classes the IR spectrum is dominated 
by starburst emission.
The dotted line is the separate contribution of normal spirals, while the continous line
is the total LLF. In our reference model (Sect. 5.2), only type-I AGNs 
and active galaxies dominated by starburst emission (short-dashes) are allowed
to evolve.  
}
\label{llf12}
\end{figure*}

The second ingredient for the no-evolution prediction is the local galaxy luminosity
function (LLF). In the mid-IR, LLFs have been published by Rush et al.
(1993), Xu et al. (1998) and Fang et al. (1998), all based on the 12 $\mu$m 
all-sky IRAS survey. Unfortunately, in spite of the proximity of the 
CAM LW3 and IRAS 12 $\mu$m bands, our knowledge of the 15 $\mu$m LLF is still 
somehow uncertain, because of: a) uncertainties in the IRAS 12~$\mu$m photometry;
b) the effect of local inhomogeneities, particularly the local Virgo super--cluster, 
in the shallow IRAS survey; and c) the uncertain flux conversion between the IRAS and 
CAM-LW3 bands (Alexander \& Aussel  2000).
Because of (a) and (b), the Rush et al. LLF determination was affected by an improper
flux normalization and a too steep faint-end slope (e.g. inconsistent with the IRAS 
60 $\mu$m LLF). 
These various effects have been discussed by Fang et al. (1998) and Xu et al. (1998): 
a re-calibrated 12 $\mu$m luminosity function based on these analyses
is reported in Fig. \ref{llf12}. Open squares
at L$_{\rm 12}>10^{10}$ L$_\odot$ come from Fang et al. (1998). For lower luminosity
values the Fang et al. LLF persisted to show excess density, and for this reason
we used here the determination by Xu et al. (1998, we  neglect the small correction
from 15 to 12 $\mu$m, given the flat shape of the LLF here).
The shallower low-luminosity slope of this new LLF determination is in particular 
consistent with the 60 $\mu$m one (small filled squares in Fig.\ref{llf12},
see also Sect. 5.3 below).
We fit these data (dotted line in Fig. \ref{llf12}) with an analytic form
$$ \rho(L) = \rho_\star \times L^{\rm (1-a)} \times ( 1 + L/L_\star/b)^{\rm -b} $$
similar to the one suggested by Rush et al., but with different slopes (a=1.15 vs. 1.7,
b=3.1 vs. 3.6) and different normalizations.

The dotted line in Fig. \ref{cdif15} corresponds to the predicted counts assuming
a non-evolving population using this best-estimate LLF. The correction to the CAM 
LW3 band is made by adopting a 12 $\mu$m to LW3 flux ratio which is a function 
of the 12 $\mu$m luminosity: for the less luminous objects the ratio
is based on the observed mid-IR spectrum of quiescent spirals, while for
the highest luminosity galaxies the ratio is the one expected for ultraluminous
IR galaxies, and for intermediate objects is close to a typical starburst spectrum
like the one of M82 (continuous line in Fig. \ref{source8}).
The 15 to 12 $\mu$m flux ratio is then assumed to increase 
continuously with luminosity, the
15 $\mu$m flux being increasingly dominated by the starburst emission at increasing L
(see Sect. 5.3 for details).
Note that in the absence of evolution, different values for the cosmological
parameters have negligible influence on the no-evolution expectation.

It is clear that the no-evolution prediction, even taking into account the
effects of the PAH features on the K-corrections, falls quite short of the
observed counts at fluxes fainter than a few mJy. Also the observed slope
in the 0.4 to 4 mJy flux range ($dN[S]\propto S^{-3\pm0.1} dS$ in differential units)
is very significantly different from the no-evolution prediction ($dN(S)\propto S^{-2}dS)$. 
The extrapolation to the bright fluxes is instead consistent, within the uncertainties,
with the IRAS 12 $\mu$m counts with a slope close to Eclidean.

Another clear sign of a serious mis-match is provided by the comparison of the no-evolution
prediction with the observed redshift distributions in Fig. \ref{dz15}, where
the model keeps a large factor below the observed peak at $z\sim 0.85$.

\subsection{Evidence for a strongly evolving population of mid-IR galaxies:
the reference model}

A first robust conclusion of the previous Section was the need for an
evolving source population dominating the counts.
The shape of the differential counts shown in Fig. \ref{cdif15} contains 
some indications about the properties of the evolving populations. 
In particular the almost flat (Euclidean) normalized
counts extending from the bright IRAS fluxes down to a few mJy, followed by the sudden 
upturn below, suggests that is not likely the whole population of IR galaxies that evolve:
in this case and for the adopted IR galaxy LLF, the super-Euclidean 
increase in the counts would appear at brighter fluxes and not be as abrupt as observed. 
The observed behaviour is more consistent with a locally
small fraction of IR galaxies to evolve with high rates back in cosmic time.

\begin{figure}[!ht]
\psfig{figure=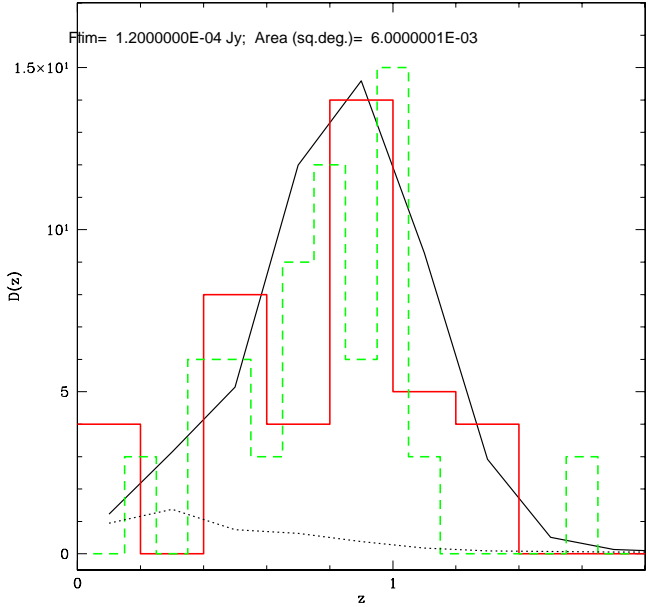,height=90mm,width=90mm}
\caption{Redshift distribution of 15 $\mu$m sources with S$_{\rm 15}>120\ \mu$Jy in the
HDF North (continuous red histogram; Cohen et al. 2000; Aussel et al. 2001), 
compared with our best-fit evolutionary model (continuous line). 
The dashed green histogram is the z-distribution of sources in the CFRS field reported
by Flores et al. (1999). The dotted line at the bottom corresponds to the no-evolution
prediction.
}
\label{dz15}
\end{figure}

We have reproduced the IR counts in Fig. \ref{cdif15} with the contribution of three
population components characterized by different physics and evolutionary properties. 
The main contributions come from non-evolving normal spirals
(with a 12 $\mu$m LLF as the dotted line in Fig. \ref{llf12}), 
and from a fast evolving population (LLF as the green dashed line in Fig. \ref{llf12}).  
The evolving population includes type-II AGNs and starburst galaxies, with the idea
that for both classes the IR spectrum may be dominated by starburst emission.
A third component considered are type-I
AGNs, whose LLF is the long-dashed line in Fig. \ref{llf12}. Based on optical and X-ray
surveys (e.g. Franceschini et al. 1994b), the latter are assumed to evolve
in luminosity as $L(z)=L(0)\times (1+z)^3$ up to $z=1.5$ and $L(z)=L(0)\times 2.5^3$ 
above.
This different treatment of type-I and II AGNs is not necessarily in contradiction
with the Unified AGN Model: for type-II objects the inclined dusty torus implies a
self-absorbed emission in the mid-IR by the AGN, and an overall IR spectrum
dominated by a circum-nuclear starburst.

A similar multi-population modelling was proposed time ago by Danese et al. (1987)
to explain the IRAS and faint radio counts, and more recently by Roche \& Eales (1999)
and Xu et al. (2001).

As shown in Fig. \ref{llf12}, the fraction of the evolving starburst population 
is assumed 
to be $\sim 10$ percent of the total, roughly consistent with the local observed fraction
of interacting galaxies. The quick upturn in the counts then
requires a strong increase with redshift of the average emissivity of the evolving
population to match the peak in the normalized counts around S$_{\rm 15}\simeq 0.5$ mJy.

In the presence of strong evolution, the fit to the observed counts has a sensible
dependence on the assumed values for the geometrical parameters of the universe. 
For a zero-$\Lambda$ open universe ($\Omega_{\rm m}=0.3, \ \Omega_\Lambda=0$),
a physically plausible solution would require a redshift increase of the comoving
density of the starburst sub-population as
$  \rho(L[z],z) = \rho_0(L_0)\times (1+z)^6$ and of their luminosity as
$L(z)=L_0\times (1+z)^3$   for $z<z_{\rm break}$, with $\rho$ and $L$ constant above,
and $z_{\rm break}=0.9$.
%
%
%
%
These are quite extreme evolution rates, if compared for example with those observed in 
optically-selected samples of merging and interacting galaxies (e.g. 
Le Fevre et al. 2000).

The inclusion of a non-zero cosmological constant, with the consequent increase 
of the cosmic timescale and volumes at $z\sim 1$, tends to make the best-fitting 
evolution rates less extreme.
For $\Omega_{\rm m}=0.3, \ \Omega_\Lambda=0.7,$ a best-fit to the counts requires:
\[   \rho(L_{\rm 12}[z],z) = \rho_0(L_{\rm 12})\times (1+z)^{4} \ \  \ \ \ z<z_{\rm break}  \]
\[  \rho(L_{\rm 12}[z],z) = \rho_0(L_{\rm 12})\times (1+z_{\rm break})^{4} \ \  \ \ \ z_{\rm break}<z<z_{\rm max}  \]
\[ L_{\rm 12}(z) = L_{\rm 12}\times (1+z)^{3.8}   \ \  \ \ \ z<z_{\rm break}  \]
 \begin{equation}
 L_{\rm 12}(z) = L_{\rm 12}\times (1+z_{\rm break})^{3.8}   \ \  \ \ \ z_{\rm break}<z<z_{\rm max}
\label{solu}
\end{equation}
with $z_{\rm break}=0.8$ and $z_{\rm max}=3.7$. 

Note that, although there are margins for variations of the relative weight for the 
number density and luminosity evolution, only a combination of the two provides
credible solutions. Assuming for example evolution only in source number density 
or only in L would require unplausibly high evolution rates for the evolving
population ($\rho\propto [1+z]^{11}$, or alternatively $L[z]\propto [1+z]^{6}$). 
Such extreme solutions would also encounter problems when fitting the radio or 
far-IR counts as counterpart to the 15 $\mu$m population counts.

The vast majority ($>90\%$) of the sources in the HDFN and CFRS 1415 ISO surveys have 
spectroscopic redshifts, and for the remaining objects photometric redshifts are 
easy to estimate. The redshift distributions $D(z)$ for the HDF 
North (Cohen et al. 2000; Elbaz et al. 2001; Aussel et al. 2001) and 
CFRS 1415 surveys (Flores et al. 1999) 
are compared in Fig. \ref{dz15} with our best-fitting model, which 
appears to properly account for these data. They set stringent limits on the rate of 
cosmological evolution for IR galaxies above $z\sim 1$, and force it to level off to 
avoid exceeding the observed $D(z)$ on the high-$z$ tail.
Also, a consequence of the fast evolution observed at $z<1$ is that the observed 
CIRB intensity is quickly saturated at moderate z, and requires again the evolution 
rate to turn over at $z\geq 1$.

Unfortunately, apart from these constraints on the high-z IR emissivity,
ISO surveys do not offer an accurate sampling of the hidden SF in the interval 
$1\leq z \leq 2$, which will only be possible with the longer-wavelength surveys
by SIRTF at $\lambda_{\rm eff}=24\ \mu$m and by the Herschel Space Observatory 
(formerly FIRST) in the far-IR, probing dust and PAH emission at $z>1$. 

In our present evolutionary scheme, any single galaxy would be expected to spend most 
of its life in the quiescent (non-evolving) phase, while being occasionally put by 
interactions in a short-lived (few to sevral $10^7$ yrs) starbursting mode. 
The cosmological evolution characterizing this phase may simply be due
to an increased probability in the past to find a galaxy during such an excited mode.
The density evolution in eq. (\ref{solu}) scales with redshift approximately
as the rate of interactions due to a simple geometric effect following the 
increased source volume density. 
The luminosity evolution may be interpreted as an effect of the larger
gas mass available to form stars at higher z.

%
%
%
%
%

\subsection{A panchromatic view of IR galaxy evolution}

Deep surveys at various IR/sub-mm wavelengths can be exploited to simultaneously 
constrain the evolution properties and broad-band spectra of faint IR sources.
We report in this Section a comparison of 
the 15 $\mu$m survey data with those coming from longer-wavelength
surveys, in particular the IRAS 60 $\mu$m, the FIRBACK 90 and 170 $\mu$m, 
and the SCUBA 
850 $\mu$m data, which are the deepest, most reliable available at the moment.
Information on both number counts and the source redshift distributions,
whenever available, were used for these comparisons. 

\begin{figure}[!ht]
\psfig{figure=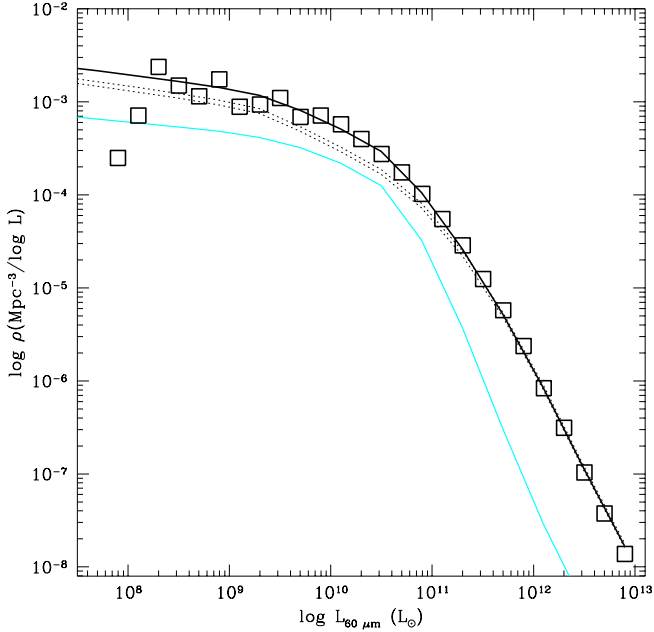,height=90mm,width=90mm}
\caption{Local luminosity function of galaxies at $\lambda_{\rm eff}=60\ \mu$m
(filled squares) by Saunders et al. (1990). The lines are the best-fit model predictions
(upper continuous line: total; dotted: quiescent population; lower continuous: starbursts).
}
\label{llf60}
\end{figure}

\begin{figure}[!ht]
\psfig{figure=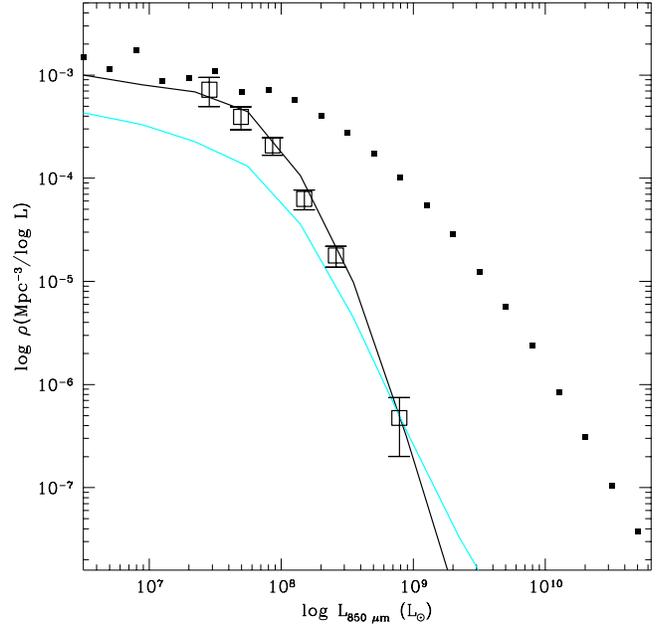,height=90mm,width=90mm}
\caption{Local luminosity function of galaxies at $\lambda_{\rm eff}=850 \$mu$m
(open squares) by Dunne et al. (2000), compared with the IRAS 60 $\mu$m LLF
by Saunders et al. (1990). Note the completely different slopes of the two.
}
\label{llf850}
\end{figure}

Further essential constraints, providing the local boundary conditions on the 
evolutionary histories, are given by the multi-wavelength local luminosity functions. 
In addition to the 12 and 15 $\mu$m LLF's discussed in Section 5.1, 
the galaxy LLF is particularly well known at 60 $\mu$m after the IRAS all-sky 
survey and the extensive spectroscopic follow-up (Saunders et al. 1990),
and is illustrated in Fig. \ref{llf60}. 
Dunne et al. (2000) also attempted to constrain the galaxy LLF in the millimeter
based on mm observations of complete samples of IRAS 60 $\mu$m galaxies. 
The results are shown in Fig. \ref{llf850}.

As previously mentioned, the properties of LLF's observed at various IR/sub-mm 
wavelengths can be explained by assuming that the galaxy IR SED's depend on bolometric 
luminosity. 
The comparisons of LLF's made in Figs. \ref{llf12} and \ref{llf850} show that the 
60$\mu$m LLF has a much flatter power-law shape at high-luminosities compared with 
both the mid-IR and millimetric LLF's. This is clearly an effect of the spectra for 
luminous 
active galaxies showing excess 60 $\mu$m emission compared with inactive galaxies,
as also illustrated by the luminosity-dependence of the IRAS far-IR colours.
We defer to Franceschini (2000, Sect. 6.6) for further discussion on this effect.

Consequently, we have modelled the redshift-dependent multi-wavelength LLF's of 
galaxies by assuming spectral energy distributions dependent
on luminosity, with spectra ranging from those typical of low-luminosity inactive 
objects, to those peaked at 80 $\mu$m of luminous and ultra-luminous IR galaxies
as previously described. 

\begin{figure}[!ht]
\psfig{figure=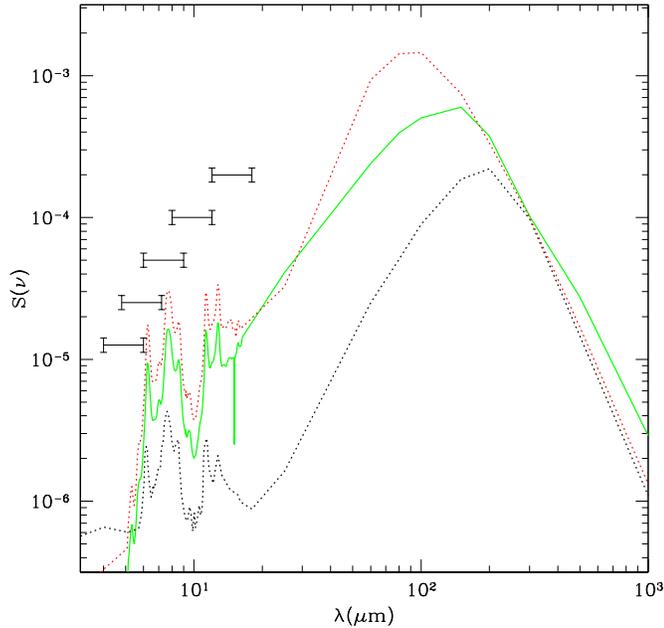,height=90mm,width=90mm}
\caption{Our adopted range of mid- to far-IR spectra of galaxies. The lower dotted line
corresponds to a low-luminosity inactive spiral (S$_{\rm low}[\nu]$ in eq. \ref{S}), 
the top dotted line is typical of ULIRGs (S$_{\rm high}[\nu]$).
The intermediate one (cyan continuous line) corresponds to our adopted spectrum for
the starburst population: this spectrum is similar to the one of the prototype 
star-forming galaxy M82 (in the range from 5 to 18 $\mu$m it is precisely the ISOCAM 
CVF spectrum of M82).
}
\label{spectra}
\end{figure}

We have taken as reference for our multi-wavelength LF
the one by IRAS at 12 $\mu$m ($\rho(L_{\rm 12})$) discussed in Sect. 5.1. 
This is transformed to longer wavelengths
according to spectral energy distributions which vary according to the value of L$_{\rm 12}$.
The assumption was that for L$_{\rm 12}<L_{\rm min}$ the spectrum is that of an inactive
spiral (S$_{\rm low}[\nu]$, the lower dotted line in Fig. \ref{spectra}), 
while for L$_{\rm 12}>L_{\rm max}$ it is a typical
ULIRG spectrum (S$_{\rm high}[\nu]$, top line in Fig. \ref{spectra}).
For intermediate luminosity objects, the assumed SED S$(\nu)$ is a linear 
interpolations between the two:
\begin{equation}
S(\nu) |_{\rm L_{\rm 12}} = 
\label{S}
\end{equation}
\[ {S_{\rm low}(\nu) (\log L_{\rm 12}-\log L_{\rm min}) + 
S_{\rm high}(\nu) (\log L_{\rm 12}-\log L_{\rm min}) \over \log L_{\rm max}-\log L_{\rm min}} .
\] 
From $S(\nu)|_{\rm L_{\rm 12}}$ it is immediate to compute the luminosity at
any frequencies and the K-correction from eq. (\ref{Kco}), taking into account the
detailed filter response functions.
The multi-wavelength LLFs at any wavelengths $\lambda$ are easily computed from the 
relation
\begin{equation}
\rho(L_{\rm \lambda}|L_{\rm 12},z) = \rho(L_{\rm 12},z) \left( d \log L_{\rm \lambda} \over 
d \log L_{\rm 12}\right)^{-1}.
\end{equation}
Good fits to the multi-wavelength LF's
are found by setting L$_{\rm min}=2\ 10^9$ L$_\odot$ and L$_{\rm max}=10^{12}$ L$_\odot$.

For the evolving active starburst galaxies we adopted both a single average spectral energy 
distribution (independent on luminosity) and a luminosity-dependent
spectral shape as discussed for the non-evolving population.
For simplicity and for a better controlled parametrization, 
our best-fit model for the active starburst population assumes
the single spectrum solution. 

If we adopt as representative for the starburst spectrum
the IR SED of the ultra-luminous galaxy Arp 220, the consequence would be
that all far-IR counts (and the CIRB intensity, see Elbaz et al. 2001) 
would be exceeded by substantial factors.
On the contrary, if we assume for the IR evolving sources a more typical starburst
spectrum (see continuous line in Fig. \ref{spectra}, which is similar to those of M82
and other luminous starbursts observed by ISO), then most of the observed 
properties of far-IR galaxy samples (number counts, redshift distributions, 
luminosity functions) are appropriately reproduced. 

\begin{figure}[!ht]
\psfig{figure=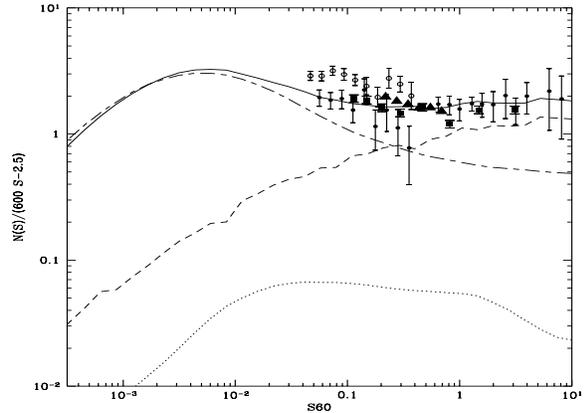,height=60mm,width=80mm}
\caption{Differential counts at 60 $\mu$m from IRAS surveys, normalized 
to the Euclidean law $600\ S^{-2.5}\ (sr^{-1}J^{-1})$, versus the 60 $\mu$m
flux in Jy. Meaning of the lines as in previous figures.
}
\label{c60}
\end{figure}

\begin{figure}[!ht]
\psfig{figure=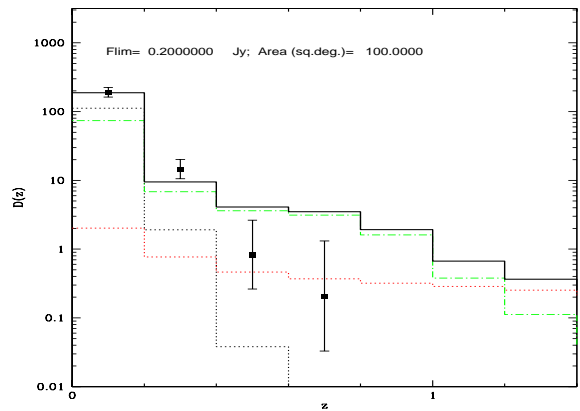,height=60mm,width=80mm}
\caption{Predicted redshift distribution for the 60 $\mu$m IRAS Faint Source Survey 
identified by Oliver et al.(1996). The survey depth and area are indicated. 
Meaning of the lines as in the previous figures.
}
\label{dz60}
\end{figure}

Best-fits to the counts based on our reference
model and adopting a spectral template as in Fig. \ref{spectra},
are given in Figs. \ref{cdif15}, \ref{c175}, \ref{c90}, \ref{c850} and \ref{c1200}.
Fig. \ref{c60} reports the fit to the 60 $\mu$m differential
counts derived from various IRAS 
surveys, while Fig. \ref{dz60} compares our model prediction with the z-distribution
for faint galaxies selected from the IRAS Faint Source Survey at 60 $\mu$m by Oliver 
et al. (1996, see also Saunders et al. 2000), with a flux limit of S$_{\rm 60}=200 $ mJy.
In addition to the 60 $\mu$m counts, 
the model clearly reproduces the observed z-distribution up to $z=0.4$,
while it predicts a somewhat excess fraction of higher-z sources.
We do not interprete this as to necessarily be a problem for the model, since due 
to the large IRAS errorbox,  
the most distant and optically faint sources could have been systematically 
mis-identified with
brighter more local galaxies falling by chance in the errorbox: our model would 
predict that of order of 4\% of the FSS could be starbursts at $z>0.4$.

It is evident that, with various degrees of significance which depend on the
survey depths, all the observed long-wavelength counts require a substantial 
increase of the IR volume emissivity of galaxies with redshift.

\begin{figure}[!ht]
\psfig{figure=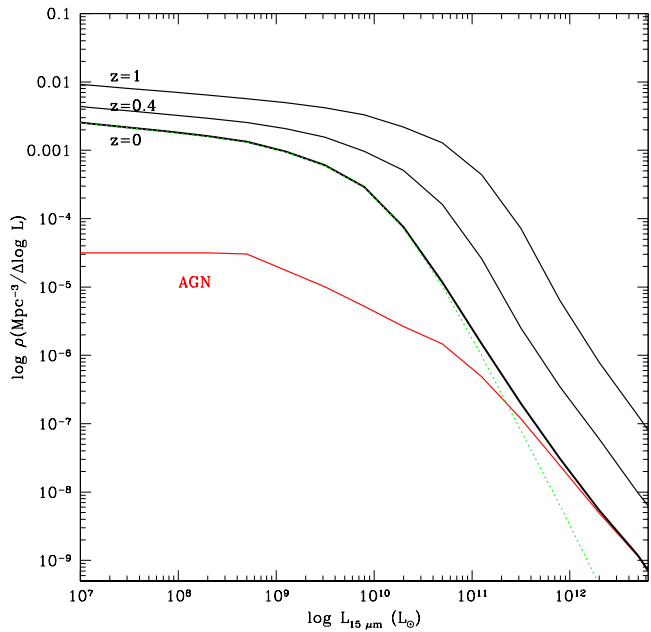,height=90mm,width=90mm}
\caption{Evolution of the 15 $\mu$m luminosity function of IR sources, 
following our best-fitting reference model. The lower continuous and dotted
lines are the contribution of AGN and of normal galaxies to the LLF. The three continuous
lines labelled z=0, 0.4 and 1 are the global (AGN + galaxy) LFs at the corresponding 
redshifts.
}
\label{evol}
\end{figure}

We report in Fig. \ref{evol} the evolutionary 15 $\mu$m luminosity function 
of IR sources in the critical redshift interval from z=0 to 1. The 
evolution of the high-luminosity end is partly driven by the type-I AGN population, 
which dominates the LF above L$_{\rm 15}=10^{12}\ L_{\rm \sun}$.
Note that $\rho(L,z)$ is required to exceed at $z\sim 1$ the local galaxy density (as 
evident in particular in the luminosity interval L$_{\rm 15}\simeq 10^7$ to 
L$_{\rm 15}\simeq 10^{10}$ L$_\odot$), which we interprete as an effect of merging,
implying an increase of the comoving number of objects in the past.

The evolution pattern of the global LF is clearly luminosity dependent:
the largest increase happens at L$_{\rm 15}\simeq 10^{11}\ L_{\rm \sun}$, whereas at very
low and very large luminosities the evolution is lower.
The model predicts that the evolution of galaxies 
with L$_{\rm 15}>10^{11.5}\ L_{\rm \sun}$ (the Ultra-Luminous Galaxies, ULIRGs) is lower than 
that of the Luminous IR galaxies (LIRGs) around $10^{11}\ L_{\rm \sun}$.
The assumption of similar evolution for LIRGs and ULIRGs would tend to imply exceeding 
the observed fraction of $z>0.4$ galaxies in the 60 $\mu$m IRAS Faint Source Surveys 
and the North Ecliptic Pole survey (Ashby et al.1996, Aussel et al. 2000).

\subsection{Alternative evolution patterns}

The available data at long$-\lambda$ on faint galaxies, although 
constraining the evolution pattern, do not allow to determine it univocally.
We have attempted, in particular, evolutionary schemes considering a single evolving 
population with combined luminosity/density evolution, as an alternative to the 
previously discussed two-population model.

\subsubsection{Luminosity-independent evolution for a single population}

This model allows the LF for the whole population of IR galaxies to evolve in both
luminosity and volume density following a pattern as in eq. (\ref{solu}), with reduced 
evolution
rates: $\rho(L_{\rm 12},z)\propto (1+z)^{1.6}$, L$_{\rm 12}\propto (1+z)^{3}$, 
$z_{\rm break}=0.8$, $z_{\rm max}=3.5$,
and L$_{\rm min}=3\ 10^9$, L$_{\rm max}=10^{12}$ in eq. (\ref{S}), S$_{\rm low}(\nu)$ and
S$_{\rm high}(\nu)$ similar to those plotted in Fig. \ref{spectra}.

Forcing it to best-fit to the 15 $\mu$m counts and D(z),
the model appears to be too rigid when fitting data at longer wavelengths.
In particular, the predicted 60 $\mu$m counts show a fast convergence already below
S$_{\rm 60}\sim 0.1\ Jy$ and a very modest contribution to the CIRB, while at the same 
time the 170 $\mu$m counts would be very steep and exceeding the observations. 
The predicted 850 $\mu$m counts are also steeper than observed.

\subsubsection{Luminosity-dependent evolution rates}

Under this scheme, the evolution rates are assumed to depend linearly on luminosity:
$\rho(L_{\rm 12},z)\propto (1+z)^{j(L_{\rm 12})}$, with 
$j(L_{\rm 12})=J\times log (L_{\rm 12}/L_1)/log(L_2/L_1)$, and
L$_{\rm 12}\propto (1+z)^{k(L_{\rm 12})}$, with 
$k(L_{\rm 12})=K\times log (L_{\rm 12}/L_1)/log(L_2/L_1)$ [with
$J$, $K$, $L_1$ and $L_2$ as adjustable parameters]. This law
is intended to add freedom to the single-population model, by considering that
galaxies with higher L$_{\rm 12}$ (hence more "active") may evolve with faster rates
with respect to lower-L$_{\rm 12}$ (less "active") objects.

As a matter of fact, the larger freedom for this model is of no help to
improve the quality of the fits (the parameters $L_1$ and $L_2$ tend to assume
low values, i.e. to go back to the luminosity-independent solution).

Altogether, evolutionary schemes alternative to our reference two-population model
appear to provide worse fits to the multi-wavelength combined IR data.
Our previously discussed evolution model provides the best statistical
fit of the data with a plausible physical interpretation (but certainly not the only 
acceptable solution).

\subsubsection{Variations in the model population combination}

Finer variations of the relative fractions of sources belonging to the
different physical components (AGNs, normal spirals and evolving active galaxies),
with respect to the values adopted in our best-fit model, 
are clearly allowed by the present data. However, the 15 $\mu$m number counts, local 
LF and redshift distributions, in particular, together with looser constraints from 
longer-wavelength observations, allow rather narrow margins to such variations.
For example, if the normalization of the evolving population is incresed and that
of the non-evolving one decreased to still keep fitting the 15 $\mu$m local LF,
we may obtain steeper counts in Fig. \ref{cdif15} in the flux interval from 10 to
0.5 mJy (with a suitable change of the evolution rates). This, however, would 
start forcing the fit to the observed counts in this flux range, 
and would also tend to spoil those to the longer-$\lambda$ data (Figs. 3, 11, 14). 

Note also that our modelling of the AGN component cannot be but rough at this
stage. Our adopted procedure treats type-I and type-II AGNs separately. The latter
are simply included in a single evolving population together with the active
starbursts. This reflects our view that dust-extinguished AGN and starburst emissions 
happen concomitantly in the same sources during the "active" phase
and reflects our present inability to quantitatively disentangle the two. 
This concomitant "activity" is now revealed by a variety of facts, including the
hard X-ray observations of ISO sources reported by Fadda et al. (2001, see
also Elbaz et al. 2001), and some evidence that the brightest sub-mm objects
are associated with AGNs (Ivison et al. 2000a).

On the contrary, our modelling of the type-I AGNs is simple and robust, and exploits
the 12 $\mu$m LLF by Rush et al. (1993) and the evolution rate found from 
optical and X-ray surveys. Based on this, type-I AGN are expected to contribute
$\sim 20-30\%$ of the bright 15 $\mu$m counts (S$_{\rm 15}>10$ mJy, see Fig. \ref{cdif15}), 
and negligible fractions at fainter fluxes or longer wavelengths.

\subsection{Contributions of IR galaxies to the CIRB}

The 15$\mu$m counts in Fig. \ref{cdif15} display a remarkable convergence
below S$_{\rm 15}\sim 0.2$ mJy, proven by at least three independent surveys. The observed
asymptotic slope flatter than $-2$ in differential count units implies a modest 
contribution to the integrated CIRB flux by sources fainter than 0.1 mJy, unless 
a sharp upturn of the counts would happen at much fainter fluxes
with a very steep number count distribution, which seems rather unplausible.
A meaningful estimate of the CIRB flux can then be obtained from direct integration 
of the observed mid-IR counts:
this computation has been done by Elbaz et al. (2001), who find a value at 15 $\mu$m
of $2.6\pm 0.5$ nW/m$^2$/sr contributed by LW3 sources brighter than S$_{\rm 15}=40\ \mu$Jy
(corresponding to the datapoint at 15 $\mu$m in Fig. \ref{bkg}, and consistent with
the results by Biviano et al. 2001; the other closeby point in the figure
comes from a similar integration of ISO counts at 7 $\mu$m). 
From our reference evolutionary model we expect that the 
contribution of fainter sources would bring the total background to 3.3 nJy/m$^2$/sr.

Comparing these values with the upper limits set by the observed {\rm TeV} cosmic opacity 
(dotted histogram in Fig. \ref{bkg}) confirms that the ISOCAM surveys have resolved 
a significant fraction (50-70\%) of the CIRB intensity in the mid-IR.

Unfortunately, we do not have a way to test directly how much of the bolometric CIRB these
faint 15 $\mu$m surveys contribute. In particular, the depths of the ISO far-IR surveys 
(FIRBACK and ELAIS, see Dole et al. 2001) are not enough to resolve more than few 
percent of the CIRB at its peak wavelength.

Using the locally established good correlation between the mid-IR and the
far-IR fluxes for local IR galaxies and after a careful analysis of the
incidence of AGNs among the faint ISOCAM sources (this is essential because
starbursts and AGNs have different IR SEDs, with peak emissions at 
$\sim 60-100\ \mu$m and $\sim 20-30\ \mu$m, respectively), 
Elbaz et al. (2001) estimate that the 
sources resolved by CAM LW3 contribute 60\% at least of the COBE/DIRBE background 
at 140 and 240 $\mu$m.

The good match to the IR multi-wavelength counts and related statistics that we found 
in our previous analysis by 
assuming a typical starburst spectrum for the evolving population indicates that
these IR sources are likely dominated by star-formation processes (see also Sect. 6.4).
Our best-fit model of the multi-wavelength statistics implies that the ISOCAM sources
with S$_{\rm 15}>40\ \mu$Jy contribute a CIRB intensity at 170 $\mu$m of $\nu I(\nu)\simeq
10^{-8}$ Watt/m$^2$/sr, or $\sim 50\%$ of the observed flux (Fig. \ref{bkg}).
At 90 $\mu$m the fraction rises to $\sim 59\%$.    All this supports the conclusion 
that {\sl the population detected by ISO in the mid-IR not only contributes a major
fraction of CIRB at 15$\mu$, but is also likely responsible for a  
majority contribution to the photon energy density contained in the CIRB.} 

On the contrary, sources at much higher redshifts, as the SCUBA ones are observed to
be (see Sect. 6.3 below), likely produce a much less significant contribution to the 
bolometric CIRB whatever their comoving volume energy production rate might be.
The high-z energy production can only last for a short cosmic time interval
($\Delta t \propto (1+z)^{-5/2}\Delta z$), and the generated photons are degraded
in energy by $(1+z)$, for a total penalty factor of $(1+z)^{3.5}$ (Harwit 1999).
Nevertheless, because of the K-correction, high-redshift objects can dominate the 
CIRB at the
longer wavelengths, as SCUBA sources are observed to do. All in all, the background
radiation is not a sensitive tracer of the very ancient cosmic phases.

\begin{figure}
\psfig{figure=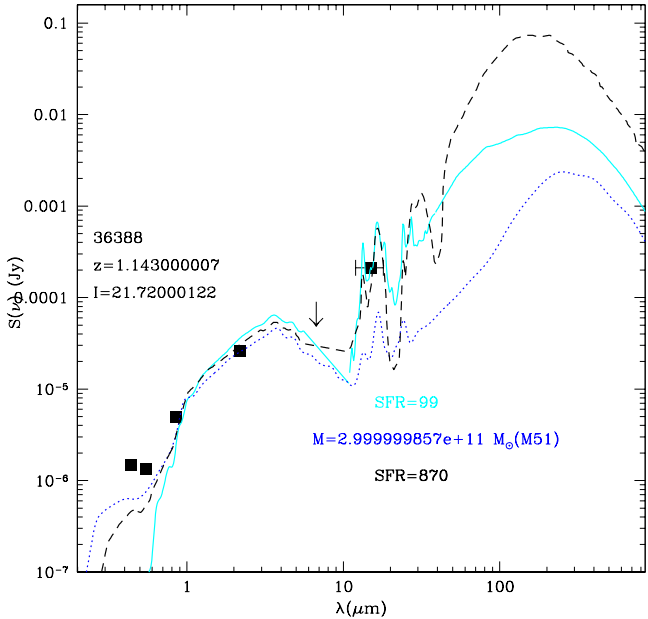,height=90mm,width=90mm}
\caption{Broad-band spectrum of a mid-IR source selected by ISOCAM LW3 
in the Hubble Deep Field North (Aussel et al. 1999), compared with
the SED's of M82 (thick continuous line), Arp 220 (dashed line), and M51
(dotted line). Estimates of the SF rate [based on the M82 and Arp 220
templates] and of the stellar mass [based on the M51 template] are indicated.
}
\label{source8}
\end{figure}

\section{NATURE OF THE FAST EVOLVING IR SOURCE POPULATION}

\subsection{The mid-IR selected sources}

Given the variety of multi-wavelength imaging data, the ISO surveys in the Hubble 
Deep Fields and the CFRS areas (Sect. 3.1) provide ideal tools for tests of
the evolving population responsible for the upturn of the ISO mid-IR counts and for
a substantial fraction of the CIRB.

Aussel et al. (1999 and 2001) report reliably tested complete samples of 
49 and 63 sources to S$_{\rm 15}\geq 100 \mu$Jy in the HDF North and South respectively.
Flores et al. (1999) analyse a sample of 41 sources brighter than S$_{\rm 15}\sim 
300 \mu$Jy in the CFRS 1415+52 area.
These surveys have been performed with highly redundant spatial and temporal sampling
with ISOCAM LW3, allowing to achieve a precise astrometric registering.
These sources are identified with optical galaxies having typical magnitudes
from $V\simeq 21$ to $V\simeq 24.5$ (Aussel et al. 2001; see also Sect. 3.2).

The optical-IR SED of a typical faint LW3 source at $z=1.14$ is reported in Figure
\ref{source8}, where the LW3 flux and the LW2 upper limit are plotted together with
optical-NIR fluxes.

The dotted line fitting the optical-NIR spectrum and corresponding to the SED of
a quiescent spiral (M51) falls short by a factor $\sim 10$ of explaining the mid-IR
emission, whereas SEDs of IR starbursts (Arp220, dashed line; M82 continuous line)
provide more consistent fits to the observed mid-IR flux after normalizing
to the optical/NIR spectral intensity. 
The vast majority of faint ISO sources show similar mid-IR flux excesses.

A clue to the nature of ISO sources can be obtained from HST imaging data, providing 
detailed morphological information, and spectroscopic follow-up available 
in these fields. 
Flores et al. (1999) find that at least 30 to 50\% of them show 
evidence of peculiarities and multiple structures, in keeping with the local evidence
that galaxy interactions are the primary trigger of luminous IR starbursts.
The Caltech redshift survey in the HDF North by
Cohen et al. (2000) showed that over 90\% of the faint LW3 ISO sources are members 
of galaxy concentrations and groups, which they identify as peaks in their redshift
distributions.  It is in such dense galaxy environments with low velocity dispersion 
that interactions produce resonant perturbation effects on galaxy dynamics
and the most efficient trigger of SF.
However, both morphological and clustering properties of ISO sources need further
investigation, which is currently in progress.


Flores et al. (1999) report a preliminary analysis of optical spectra for IR sources
in CFRS 1415+52, noting that a majority of these display both weak emission
(OII 3727) and absorption ($H_\delta$) lines, as typical of the {\sl e(a)} galaxy spectral
class.         Rigopoulou et al. (2000) have observed with the ISAAC spectrograph
on VLT a sample of 13 high-z (0.2$<z<$1.4) galaxies selected in the HDF South with
S$_{\rm 15}>100\ \mu$Jy:
{\sl a prominent (EW$>50\ \AA$) $H\alpha$ line is detected in almost all of the 
sources, indicating substantial rates of SF 
after de-reddening corrections, and demonstrating that these optically faint but IR luminous 
sources are indeed powered by an ongoing massive dusty starburst}.

The {\sl e(a)} spectral appearence found by Flores et al. (1999) is interpreted by 
Poggianti \& Wu (1999) and
Poggianti, Bressan, Franceschini (2001) as due to selective dust attenuation,
extinguishing more the newly-formed stars than the older ones which have already
disrupted their parent molecular cloud.
These papers independently found that $\sim 70 - 80\%$ of the energy emitted by young
stars leaves no traces in the optical spectrum, hence can only be accounted for with
long-wavelength observations.

\begin{figure}
\psfig{figure=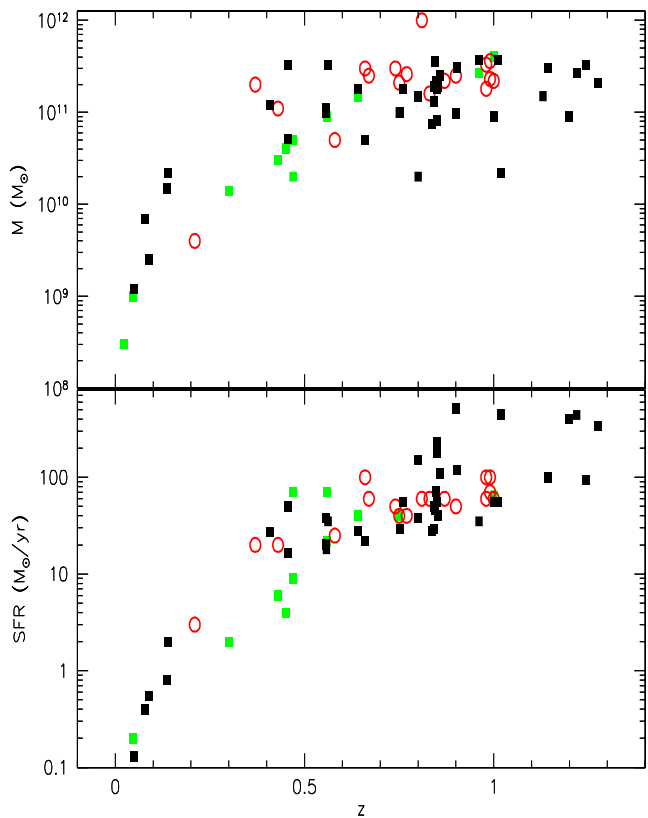,height=120mm,width=90mm}
\caption{Evaluations of the star formation rates [from an estimator based on the 
mid-IR flux] and baryonic masses [from fits of the NIR SED] as a function of redshift
for galaxies selected by ISOCAM LW3 at 15$\mu$m in the HDFN (filled squares)
and CFRS 1415+52 (open circles).
}
\label{sfr_z}
\end{figure}

Further efforts of optical-NIR spectroscopic follow-up of faint
ISO sources are presently ongoing, including attempts to address the 
source kinematics and dynamics based on line studies with IR spectrographs on large
telescopes (Rigopoulou et al. 2001, in preparation).
At the moment, for an evaluation of the main properties of the IR population
we have to rely on indirect estimates exploiting the near-IR and far-IR fluxes.
We have estimated the baryonic mass in stars from fits of template SEDs of
local galaxies to the observed near-IR broad-band spectrum.
Our adopted templates come from the modellistic analysis of Silva et al. (1998)
of a sample of both inactive spirals and starbursts of various masses and 
luminosities.
The Initial Mass Function is assumed to be a Salpeter with standard low- and 
high-mass cutoffs (0.15 to 100 $M_\odot$).
Our estimated values of the baryonic masses ($\sim$ 10$^{11}$ M$_\odot$ at $z>0.4$, 
with 1 dex typical spread, see Fig. \ref{sfr_z}) 
indicate that already evolved and massive galaxies preferentially host the 
powerful starbursts.

For estimating the other fundamental indicator of the physical and 
evolutionary status of the sources -- the ongoing rate of star-formation (SFR) -- 
we have exploited the mid-IR flux as an alternative to 
the (heavily extinguished) optical-UV emissions.
The capability of the mid-IR flux (from both LW3 and LW2 ISOCAM observations) 
as a tracer of the SFR is discussed by
Vigroux et al. (1999), Elbaz et al. (2001) and Aussel et al. (2001):
the fluxes in these IR bands appear tightly correlated with the bolometric (mostly far-IR) 
emission, which is the most robust measure of the number of massive reddened 
newly-formed stars, and is also correlated with the radio emission of stellar origin.  
Only very extinguished peculiar sources (e.g. Arp 220), for which the mid-IR 
spectrum is self-absorbed, escape this correlation.
Note that the ISO mid-IR flux at these very faint limits provides advantages over 
the radio to be a more reliable  
(having a tighter correlation with the bolometric
flux, see e.g. Cohen et al. 2000) and more sensitive indicator of star formation.
Of the 49 ISO sources in the HDFN only 7 are detected in an ultra-deep radio map 
at 1.4 GHz by Richards et al. (1998): the ISO 15 $\mu$m completeness limit of 0.1 mJy
would correspond to a flux of few $\mu$Jy at 1.4 GHz (i.e. quite below the
Richards et al. limit of 40 $\mu$Jy),
taking into account the radio/far-IR luminosity correlation for typical starbursts.

The rates of SF indicated by the fits to the mid-IR flux for sources at $z>0.5$
range from several tens to few hundreds solar masses/yr, i.e. a
substantial factor larger than found for optically-selected galaxies at similar
redshifts (e.g. Ellis 1997).

Altogether, the galaxy population dominating the faint mid-IR counts and substantially
contributing to the bolometric CIRB intensity (assumed typical SB SEDs)
appears to be composed of luminous (L$_{\rm bol}\sim 10^{11}-10^{12.5}$ L$_\odot$) 
starbursts in massive ($M\sim 10^{11}$ M$_\odot$) galaxies at $z\sim 0.5-1.3$,
observed during a phase of intense stellar formation ($SFR\sim 100$ M$_\odot$/yr).
The typically red colors of these systems suggest that they are mostly unrelated to
the faint blue galaxy population dominating the optical counts (Ellis 1997),
and should be considered as an independent manifestation of (optically hidden)
star formation (Elbaz 1999, preprint; Aussel 1998).

%
%
%
%
%
%

\subsection{Far-IR selected galaxy samples }

Surveys at longer wavelengths suffer quite severe problems in the identification 
of the optical counterparts of the IR sources.

The FIRBACK/ELAIS surveys have resolved a modest fraction ($\sim 5 \%$, see Dole
et al. 2001) of the CIRB at its peak wavelength of 170 $\mu$m, the limit being 
imposed by source confusion.
Because of the missing information on the far-IR LLF, the interpretation of the counts 
in Figs. \ref{c175}, \ref{c170dif} and \ref{c90} is subject to some uncertainties. 
Our best-fit multi-wavelength model implies that the observed counts at the faint
flux limit are a moderate factor ($\sim 2-3$) above the source areal density
corresponding to no-evolution (see Fig. \ref{c170dif}). The model predicts that the 
detected sources lye at moderate-redshifts (the majority at $z\leq 0.5$, see Fig. 
\ref{A3}).   The multi-wavelength follow-up performed at 1.4 GHz, 1.3 mm, 850 and
450 $\mu$m, as well as optical/NIR identifications and spectroscopy based on
cross-correlations with deep radio surveys (e.g. Sanders 2001), seem to show that
the majority of the sources are local ($z<0.5$), with 10\% or so being found at
$z\sim 1$ or higher (Dole et al. 2001).

Scott et al. (2000), in particular, 
have obtained data at 450 and 850 $\mu$m for 10 FIRBACK sources
with accurate radio astrometry: although the combined FIR/radio selection and the
sub-mm follow-up may somehow bias the result, the FIR-mm SEDs compared with 
plausible far-IR template spectra tentatively indicate mostly low redshifts for these 
sources, with a minority being at $z\sim 1$.

%
In the future, deeper far-IR observations will be possible with SIRTF, 
while a proper characterization
of the faint far-IR population will require the Herschel's better spatial resolution.

\subsection{Faint millimetric sources }

Thanks to the unique $K-$correction for dusty spectra, deep millimetric surveys
are capable to detect starburst galaxies over an extremely wide redshift interval.
However, the sensitivities achievable by present-day instruments and the modest
surveyed areas imply that only sources at $z>1$ are selected; 
consequently, only the very high luminosity tail
of the population of IR sources (essentially ULIRGs with L$_{\rm bol}> 10^{12}$ L$_\odot$)
is detectable in this way.

Therefore, ISOCAM and millimetric telescopes provide extremely complementary
sampling capabilities in terms of redshift coverage (typically $z<1$
for sources selected by ISO and $z>1$ by SCUBA)
and source luminosities (mostly L$_{\rm bol}< 10^{12}$ L$_\odot$ by ISO, and larger by SCUBA).

Unfortunately, the extreme properties of the mm-selected sources entail a
dramatic difficulty to identify the sources. Several factors contribute: the
very high redshifts imply that the optical counterparts are extremely
faint and red, hence largely unaccessible by optical spectroscopy
(Smail et al. 1999 identify a fraction of SCUBA sources with Extremely Red Objects,
ERO's, see also discussion in Dey et al. 1999). 
Furthermore, the dominance of dust emission implies very extinguished
optical-NIR spectra. Although the diffraction-limited beams of both SCUBA (15 arcsec)
and IRAM (11 arcsec) are much sharper than the ISO far-IR beam ($\sim$60-90 arcsec),
the faintness of optical counterparts implies a very high chance
of mis-identification.   It is remarkable that, in spite of the important effort
for SCUBA source identification, of the 100-200 sources only three have reliable
identifications and redshifts at present. All these three have been observed to
contain massive reservoirs of gas (Frayer et al. 1999; Ivison et al. 2000a).

Given the extreme difficulty to get the redshift from optical spectroscopy, 
some millimetric estimators have been devised.
The most promising of such techniques, exploiting the S$_{\rm 850\mu}/S_{\rm 20cm}$ flux ratio
as a monotonic function of redshift (Carilli \& Yun 1999), confirms in a statistical 
sense that faint SCUBA sources are ultra-luminous galaxies at typical 
$z\sim 1$ to $\sim 3-4$ (Barger et al. 1999;   Smail et al. 2000).  
The predicted z-distribution by
our reference model (Fig. \ref{A3}) is in rough agreement with these estimates.

As suggested by several authors (Franceschini et al. 1994; Lilly et al. 1999;
Granato et al. 2001), the similarity in properties (bolometric luminosities, SEDs) between 
this high-z population and local ultra-luminous IR galaxies argues in favour
of the idea that these represent the long-sought "primeval galaxies", those in particular
originating the local massive elliptical and S0 galaxies.
This is also supported by estimates of the volume density of these
objects in the field $\sim 2-4\times 10^{-4}$ Mpc$^{-3}$, high enough to allow most
of the E/S0 to be formed in this way (Lilly et al. 1999).
As for the E/S0 galaxies in clusters, the recent discovery by SCUBA of a 
significant excess of very luminous ($L\sim 10^{13}$ L$_\odot$) sources at 850  $\mu$m
around the z=3.8 radiogalaxy 4C41.17 (Ivison et al. 2000b) may
indicate the presence of a forming cluster surrounding the radiogalaxy, where the 
SCUBA sources would represent the very luminous ongoing starbursts.

{\sl By continuity, the less extreme of the starbursts 
(those with $L\sim 10^{11}-10^{12}$ L$_\odot$) 
discovered by ISOCAM at lower redshifts may be related to the origin of lower
mass spheroids and spheroidal components in later morphological type galaxies.}

\subsection{AGN contributions to the source energetics}

A natural question arises as of how much of the bolometric flux in these IR/mm 
sources is contributed by gravitational accretion rather than stellar emission. 
Almaini, Lawrence and Boyle (1999) have suggested that a minimum of $10-20\%$ of the
CIRB at 850 $\mu$m (and a similar fraction of the bolometric one) may be
due to obscured AGNs, and that this fraction could even be quite larger.
Unfortunately, probing the nature of the faint IR-mm sources at high-redshifts
turned out to be exceedingly difficult, since the optical--UV--soft-X-ray primary 
emission is almost completely re-processed by dust into 
an IR spectrum barely sensitive to the properties of the primary incident one. 
Ivison et al. (2000a) find indications for the presence of (type-II) AGN components 
in three of the seven SCUBA sources in cluster fields, a fraction not inconsistent
with that observed in local ULIRGs of similar extreme luminosity.

Preliminary inspection of the $H\alpha$ line profiles for the faint 
ISO mid-IR sources (Rigopoulou et al. 2000), together with constraints set by the 
15 to 7 $\mu$m flux ratios (for sources at $z\sim 0.5-1$ this ratio measures the
presence of very hot dust heated by the AGN, which is absent in starbursts), indicate 
that the large majority of sources are also mostly powered by a starburst rather 
than an AGN. 
Tran et al. (2001) consistently find that the contribution of gravitational
accretion to the IR emission by local ULIRGs becomes important, or even dominant, only 
in the very high luminosity regime (L$_{\rm bol}>10^{12.4}$ L$_\odot$).

An important diagnostic is being offered by observations of the hard X-ray flux,
since starbursts are weaker X-ray emitters than any kinds of AGNs. 
The {\sl Chandra} X-ray observatory has performed several deep investigations 
of the high-z SCUBA population (Fabian et al.  2000;  Hornschemeier et al.
2000; and Barger et al. 2001). Only a very small percentage of the objects turn 
out to be in common, the two classes of sources being largely orthogonal.  
The two high-z sub-mm sources detected with {\sl {\sl Chandra}} by Bautz et al. (2000)
were both previously classified as AGN based on optical spectra.
Unless the large majority of sub-mm sources are Compton-thick and any hard X-ray 
scattered photons are also 
photo-electrically absorbed, the conclusion is that the bulk of the emission by 
high-luminosity SCUBA sources is due to star formation (in agreement with a 
dominant stellar emission in local ULIRGs inferred by Genzel et al. 1998).
The fraction of the CIRB at 850 $\mu$m due to AGNs was estimated by 
Barger et al. (2001) to be not likely larger than 10\%.

Fadda et al. (2001) and Elbaz et al. (2001) discuss cross-correlations of faint 
ISO samples with deep hard X-ray maps from XMM and {\sl Chandra}. They find that
typically 10\% of the faint ISO sources show hard X-ray evidence for the
presence of an AGN. 

{\sl
Although the detailed interplay between starbursts and AGNs is still an open issue, 
it is quite likely that minor AGN contributions are present in a substantial
fraction of the active IR population: 
Risaliti et al. (2000) and Bassani et al. (2001) find trace
AGN emission in $60-70\%$ of local ULIRGs, based on BeppoSAX hard X-ray data.
Similarly, a significant fraction of the high-z IR starbursts discovered by
ISO and SCUBA may contain (energetically-negligible) low-luminosity AGNs, detectable 
in hard X-rays, and possibly responsible for the bulk of the X-ray background. 
Still their IR emission is likely to be mostly of stellar origin.}

\subsection{Discussion}

Among various samples of faint sources selected at long wavelengths,
those detected by ISOCAM in deep and ultra-deep surveys allow
the most precise quantification of the cosmic history of the IR population.
The ISOCAM LW3 extragalactic counts, extending over 4 orders of magnitude in flux 
(Fig. \ref{cdif15}) when combined with the IRAS 12 $\mu$m local surveys, provide
rather detailed constraints on the evolution pattern. The LW3 sources not only 
contribute a dominant fraction of the CIRB in the mid-IR,  
but they are also likely important contributors to the CIRB at longer wavelengths.
The extremely high luminosities and redshifts and modest volume densities of SCUBA 
sources indicate that they probably produce only a small fraction of the bolometric
CIRB energy. 
It should not be forgotten, however, that these inferences will remain model dependent
untill we will be able to resolve into sources a significant fraction of the CIRB
at its peak wavelengths, and this will have to wait for the operation of Herschel/FIRST.

Because of their non-extreme properties, ISOCAM LW3
sources can be fairly unambiguously identified and investigated in the optical.
The outcome of our spectroscopic observations is that the faint population 
making up the CIRB in the mid-IR appears dominated by actively star-forming galaxies
with substantial $H\alpha$ emission. 

The LW3 ISOCAM counts and redshift distributions require extremely
high rates of evolution of the 15$\mu$m luminosity function up to $z\sim 1$,
with preference for evolution (in both source luminosity and spatial density)
of a population of IR starbursts contributing little to the local LF.   
A natural way to account for this would be to assume that this population consists of 
otherwise normal galaxies, but observed during a dust-extinguished luminous starburst 
phase, and that its extreme evolution is due to {\sl an increased probability 
with z to observe a galaxy during such a transient luminous starburst event.}

The widespread evidence that starbursts are triggered by interactions and merging 
suggests that the number density evolution could be interpreted as an 
increased probability of interaction back in time. 
Assuming that the phenomenon is dominated by interactions in the field and a 
velocity field constant with $z$, than this probability would scale roughly as
$\propto \rho(z)^2 \propto (1+z)^6$, $\rho$ being the number density in the proper 
(physical) 
volume. A more complex situation is likely to occur, since also the velocity field 
evolves 
with $z$ in realistic cosmological scenarios and if we consider that the most 
favourable
environment for interactions are galaxy groups, which are observed to host the
majority of ISOCAM distant sources (Cohen et al. 2000).
In turn, the increased luminosity with $z$ of the typical starburst is due, 
qualitatively, to the larger amount of gas available in the past to make stars.

How this picture of a 2-phase evolution of faint IR sources compares with results
of optical and near-IR deep galaxy surveys has not been investigated by us.
Since, because of dust, most of the bolometric emission during a starburst comes out in 
the far-IR, we would not expect the optical surveys to see much of this violent 
starbursting phase revealed by IR observations. 
Indeed, B-band counts of galaxies and spectroscopic surveys are interpreted in
terms of number-density evolution, consequence of merging, and essentially no evolution 
in luminosity.
The Faint Blue Object population found in optical surveys can be interpreted in our
scheme as 
the post-starburst population, objects either observed after the major event of SF,
or more likely ones in which the moderately extinguished intermediate age 
($\geq 10^7\ yrs$)
stars in a prolonged starburst (several $10^7\ yrs$) dominate the optical spectrum.
In this sense optical and far-IR selections trace different phases of the evolution of 
galaxies, and provide independent sampling of the cosmic star formation.

A lively debate is currently taking place about the capabilities of UV-optical
observations to map by themselves the past and present
star-formation, based on suitable corrections for dust extinction in distant galaxies.
Adelberger et al. (2000) suggest that the observed 850 $\mu$m galaxy counts and the 
background could be explained with the optical Lyman drop-out high-z population 
by applying a proportionality correction to the optical flux and by taking into 
account the locally observed distribution of mm-to-optical flux ratios.

On the other hand, a variety of facts indicate that optically-selected and 
IR/mm-selected faint high-redshift sources form almost completely disjoint samples.
Chapman et al. (2000) observed with SCUBA a subset of $z\simeq 3$ Lyman-break 
galaxies having the highest predicted rates of SF as inferred from the optical 
spectrum, but detected only one object out of ten. 
van der Werf et al. (2000) found that the procedures adopted to correct 
the optical-UV spectrum and to infer their sub-mm fluxes failed in the case of two 
strongly lensed Lyman-break galaxies observed with SCUBA.
A similar dichotomy is observed in the local universe, where the bolometric flux by
luminous IR galaxies is mostly unrelated with the optical emission spectrum 
(Sanders \& Mirabel 1996).

Rigopoulou et al. (2000), Poggianti \& Wu (2000) and Poggianti 
et al. (2001) report {\sl independent evidence on both local and high-z luminous 
starbursts that typically 70\% to 80\% of the bolometric flux from young stars
leaves no traces in the UV-optical spectrum, because it is completely obscured
by dust}. As there seems to be no robust "a priory" way to correct for this missing 
energy, we conclude that only long-wavelength observations, with the appropriate 
instrumentation, can eventually {\sl measure} SF in galaxies at any redshifts.

\section{GLOBAL PROPERTIES: THE COMOVING SFR DENSITY AND CONSTRAINTS FROM BACKGROUND 
OBSERVATIONS}

\subsection{Evolution of the comoving luminosity density and SFR}

We are not presently in the position to derive an independent assessment 
of the evolutionary SFR density based on the available complete samples of faint 
IR sources, since the process of optical identifications is far from complete.

\begin{figure}
\psfig{figure=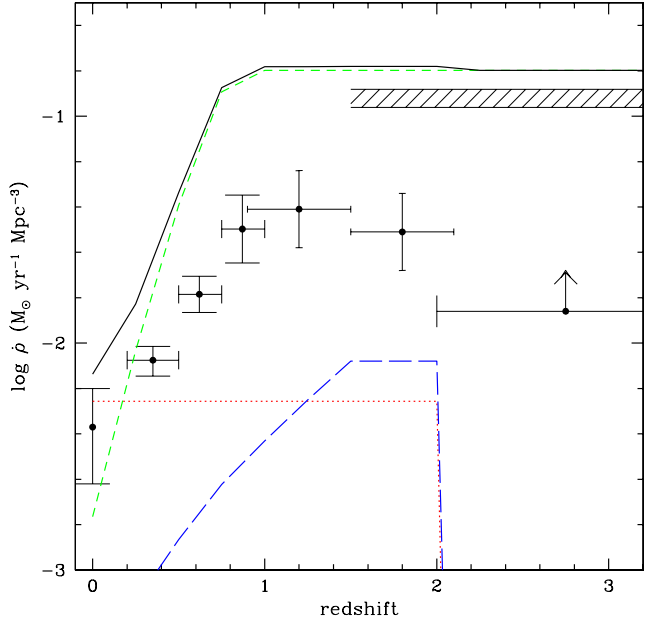,height=90mm,width=90mm}   
\caption{Evolution of the comoving luminosity density for the IR-selected population
based on the model of IR evolution discussed in Sect 5. The luminosity density
is expressed here in terms of the star formation rate density
(computed from the far-IR
luminosity assuming a Salpeter IMF, according to the recipes reported in
Rowan-Robinson et al. (1997, their eq. [7]).
The IR evolution is compared with data coming from optical observations by
Lilly et al. (1996), Connolly et al. (1997) and Madau et al. (1996), transformed 
to our adopted $\Omega_{\rm m}=0.3, \ \Omega_\Lambda=0.7$ cosmology. 
Dotted line: quiescent population. Short-dash line: evolving starbursts. 
Long dashes: type-I AGNs (with arbitrary normalization).
The continuous line is the total.
The shaded horizontal region is an evaluation of the average SFR in spheroidal 
galaxies by Mushotzky \& Loewenstein (1997).
}
\label{sfr}
\end{figure}

Again, only model-dependent estimates of the SFR density as a function of redshift 
are possible at the moment. The prediction based  
on our previously described evolution scheme is reported in Fig. \ref{sfr}.
There is a clear indication here that the contribution of IR-selected sources 
to the luminosity density significantly exceeds those based on optically 
selected sources, and that the excess may be progressive with redshift up to $z\sim 1$.
{\sl Consider that, in any case, the optical and IR
estimates of the SFR in Fig. \ref{sfr} refer to largely distinct source populations
(extinction-corrected optical fluxes account on average for only 20-30\%
of the bolometric emission by young stars in IR galaxies, see Sect. 6).}

The fast evolution inferred from the IR observation should however level off at $z>1$, 
to allow consistency with the observed z-distributions for faint ISOCAM sources 
(Fig. \ref{dz15}, se also Chary \& Elbaz 2001).
Another important constraint in this sense comes from the observed spectral 
shape of the CIRB, with its apparent sharp peak around $\lambda\simeq 100-200\ 
\mu m$ and the fast turnover longwards: fitting it with standard dust-emission
spectra implies a maximum in the comoving galaxy IR emissivity close to
$z=1$. This is also confirmed by attempts to derive the average time-dependent 
IR volume emissivity from deconvolution of the CIRB spectrum, see Gisper et al. (2000)
and Takeuchi et al. (2001).

Surveys of H$_\alpha$ line emission (the best optical indicator of SF) from high-z
galaxies indicate similar evolution of the SFR density, with a similarly sharp
change in slope occurring at $z\sim 1$ (see van der Werf, Moorwood and Yan 2001 
for a review). Our present results offer the advantage, however, to be unaffected 
by the uncertain extinction corrections of the optical SF indicators, and to exploit
the very robust constraint set by the CIRB.

Altogether these results indicate that the history of galaxy long-wavelength emission
does probably follow a general path not much dissimilar from that revealed 
by optical-UV observations, 
by showing a similar peak activity around $z\sim 1-1.5$, rather than being
confined to the very high-redshifts as sometimes was suggested based on SCUBA
results. This confirms that the bulk of the galaxy activity is to 
be placed around $z=1$, which is evident from Fig. \ref{sfr} if the dependence of the 
cosmological timescale on redshift is considered (see also Harwit 1999 and
Haarsma \& Partridge 1998).

As a final note, the estimated rate of evolution of the IR volume 
emissivity of galaxies appears in Fig. \ref{sfr} to be even higher than the evolution 
rate for type-I AGNs.

\subsection{Energy constraints from background observations}

Further constraints on the high-redshift far-IR/sub-mm population can be inferred
from observations of the global energetics residing in the CIRB and optical backgrounds. 
These imply a very substantal demand on contributing sources,
as detailed below in schematic terms.

Let us assume that a fraction $f_\ast$ of the universal mass density in baryons
$ \rho_{\rm b} = {3H_0^2 (1+z)^3 \over 8 \pi G } \Omega_{\rm b}$ 
$ \simeq  7\ 10^{10}(1+z)^3 h_{\rm 50}^2 \Omega_{\rm b}\ {\rm [M_\odot/Mpc^3]} $
undergoes at redshift $z_\ast$ a transformation (either processed in stars or by 
gravitational fields) with radiative efficiency $\epsilon$. The locally observed 
energy density of the remnant photons is 
\[  \rho_\gamma = \rho_{\rm b} {c^2\epsilon f_\ast \over (1+z_\ast)^4} \simeq  \]
\begin{equation} 
5\ 10^{-30} h_{\rm 50}^2 {\Omega_{\rm b} \over 0.05} {f_\ast \over 0.1} 
{2.5 \over (1+z_\ast)} {\epsilon \over 0.001} [{\rm gr/cm^3}] .
\label{energy}
\end{equation} 
For stellar processes, $\epsilon$ is essentially determined
by the IMF: $\epsilon = 0.001$ for a Salpeter IMF and a low-mass cutoff 
$M_{\rm min}=0.1$ M$_\odot$, $\epsilon = 0.002$ and $\epsilon = 0.003$
for $M_{\rm min}=2$ and $M_{\rm min}=3$, while $\epsilon$ gets the usually quoted value of
$\epsilon = 0.007$ only for $M_{\rm min}=10$ M$_\odot$.


\subsubsection{Constraints from the integrated optical background}

Let us adopt for the optical/near-IR bolometric emission by distant galaxies 
between 0.1 and 7 $\mu$m the value given in eq.(\ref{optical}).
We discussed evidence that in luminous starbursts the optical spectra are only 
moderately contributed by the starburst emission itself, the latter being largely 
hidden in the far-IR.   Then let us assume that the optical/NIR background 
mostly originates by moderately active SF in spiral disks and by
intermediate and low-mass stars. As observed in the Solar Neighborhood, a good
approximation to the IMF in such relatively quiescent environments is the
Salpeter law with standard low-mass cutoff, corresponding to a mass--energy
conversion efficiency  $\epsilon \sim 0.001$. With these parameter values, we
can reproduce the whole optical BKG intensity of eq.(\ref{optical}) by
transforming a fraction $f_\ast \simeq 10 \%$ of all nucleosynthetic baryons into
(mostly low-mass) stars, assumed the bulk of this process happened at 
$z_\ast \sim 1.5$ and 5\% of the closure value in baryons (for our adopted 
$H_0=50\ Km/s/Mpc$, or $\Omega_{\rm b} h_{\rm 100}^2=0.012$, consistent with the theory of 
primordial nucleosynthesis):
\[ \nu I(\nu)|_{\rm opt} \simeq \]
\[20 \ 10^{-9} h_{\rm 50}^2 {\Omega_{\rm b} \over 0.05} 
{f_\ast \over 0.1} \left(2.5 \over 1+z_\ast\right) {\epsilon \over 0.001}  {\rm Watt/m^2/sr} .
\]
A local density in low-mass stars is generated in this way consistent with the observations 
(Ellis et al. 1996), and based on standard mass to light ratios:
\begin{equation} 
\rho_{\rm b}(stars) \simeq 7\ 10^{10} f_\ast \Omega_{\rm b} \simeq
3.4\ 10^8\ {\rm M_\odot/Mpc^3} .
\label{star}
\end{equation} 
For typical solar metallicities, this corresponds to a local density in metals of
\[  \rho_Z(stars) \simeq   \]
\begin{equation} 
1.67\ 10^{9} f_\ast {Z\over Z_\odot} \Omega_{\rm b}\ {\rm M_\odot/Mpc^3\simeq 8\ 10^6 
 M_\odot/Mpc^3} .
\label{Z}
\end{equation} 
%

\subsubsection{Explaining the CIRB background}

Following our previous assumption that luminous starbursting galaxies
emit negligible energy in the optical-UV and most of it in the far-IR,
we coherently assume that the energy resident in the CIRB background 
(eq.[\ref{CIRB}]) originates from 
star-forming galaxies at median $z_\ast \simeq 1.5$. The amount of baryons 
processed in this phase and the conversion efficiency $\epsilon$
have to account for the combined constraints set by eqs.(\ref{star}),
eqs.(\ref{Z}) and (\ref{CIRB}), 
that is to provide a huge amount of energy without contributing much 
stellar remnants to the locally observed amount. A viable solution
is then to change the assumptions about the stellar IMF characterizing
the starburst phase, for example to a Salpeter distribution cutoff below 
$M_{\rm min}=2$ M$_\odot$, with a correspondingly higher efficiency $\epsilon=0.002$
(see Sect. 7.2). This may explain the energy density in the CIRB
(eq.[\ref{CIRB}]):
\[\nu I(\nu)|_{\rm FIR} \simeq \]
\[ 40 \ 10^{-9}  h_{\rm 50}^2 {\Omega_{\rm b} \over 0.05} 
{f_\ast \over 0.1} \left(2.5 \over 1+z_\ast\right){\epsilon \over 0.002}\ {\rm Watt/m^2/sr} ,
\]
assumed that a similar amount of baryons, $f_\ast\simeq 10\%$, as processed with low 
efficiency during the ``inactive'' secular evolution, are processed with higher 
radiative efficiency during the starbursting phases, 
producing a two times larger amount of metals:
$\rho(metals)\sim 1.5\ 10^7\ M_\odot/Mpc^3$.
Note that by decreasing $M_{\rm min}$ during the SB phase would decrease the efficiency 
$\epsilon$ and increase
the amount of processed baryons $f_\ast$, and would bring to exceed the locally
observed mass in stellar remnants (eq.[\ref{star}]).

The above scheme is made intentionally extreme, to illustrate the point. The reality
is obviously more complex, e.g. by including a flattening at low mass values in the 
Salpeter law (e.g. Zoccali et al 1999) for the solar-neighborhood SF and, likewise, 
a more gentle convergence of the starburst IMF than a simple low-mass cutoff.

\subsection{Galactic winds and metal pollution of the inter-cluster medium}

A direct prediction of the above scheme is that most of the metals produced
during the starburst phase have to be removed by the galaxies to avoid  
exceeding the locally observed metals in galactic stars (eq.[\ref{Z}]). 
There is clear evidence in local starbursts, based on optical and 
X-ray observations, for large-scale super-winds out-gassing high-temperature
enriched plasmas from the galaxy. Our expectation would be that the 
amounts of metals originating from the SF processes producing the CIRB
are hidden in hot cosmic media.

Where all these metals are? Most likely the polluted plasmas are hidden
in the diffuse (mostly primordial and un-processed) inter-cluster medium 
with densities and temperatures preventing to detect them.
On the other hand, an interesting support to our scheme is provided by
observations of rich clusters of galaxies, considered as closed boxes 
from a chemical point of view, as well as representative samples of the universe.
%
%
The mass of metals in the ICP plasma can be evaluated from the total amount of 
ICP baryons ($\sim$5 times the mass in galactic stars)
and from their average metallicity, $\sim$40\% solar. The mass of ICP metals
is then $M_{\rm metals, ICP} \simeq 5\times 0.4\ (Z/Z_\odot)\ M_{\rm stars}$, 
i.e. two times larger than the mass of the metals present in galactic stars
and consistent with the mass in metals we inferred to be produced during the 
SB phase.

{\sl Hence, the same starbursts producing the ICP metals are also likely responsible
for the origin of the CIRB.} In a similar fashion,  Mushotzky \& Loewenstein (1997) 
used their metallicity measurements of clusters to estimate the contribution of 
spheroidal galaxies to the SFR density (see Fig. \ref{sfr}).

\subsection{A two-phase star-formation: origin of galactic disks and spheroids}

Our reference model implies that 
{\sl star formation in galaxies has proceeded along two phases: a quiescent one 
taking place during most of the Hubble time, slowly building stars with 
standard IMF 
from the regular flow of gas in rotational supported disks; and a transient actively
starbursting phase, recurrently triggered by galaxy mergers and interactions. }
During the merger,
{\it violent relaxation} redistributes old stars, producing de Vaucouleur profiles typical 
of galaxy spheroids, while young stars are generated following a top-heavy IMF.

Because of the geometric (thin disk) configuration of the diffuse ISM and the modest
incidence of dusty molecular clouds, the quiescent phase is only moderately
affected by dust extinction, and naturally produces most of the optical/NIR background 
(including NIR emission by early-type galaxies completely deprived of an ISM).

The merger-triggered active starburst phase is instead characterized by a large-scale 
redistribution of the dusty ISM, with bar-modes and shocks, compressing a large
fraction of
the gas into the inner galactic regions and triggering formation of molecular clouds.
As a consequence, this phase is expected to be heavily extinguished
and the bulk of the emission to happen at long wavelengths, naturally originating
the cosmic IR background.
Based on dynamical considerations, we expect that during this violent SB phase 
the elliptical and S0 galaxies are formed in the most luminous IR SBs 
(corresponding to the SCUBA source population), whereas galactic bulges in later-type 
galaxies likely originate in lower IR luminosity starbursts (the ISO mid-IR population). 

The presently available IR-selected galaxy samples, dominated as they are by 
K-correction and selection effects, cannot allow to establish the precise 
evolutionary timescales as a function of source luminosity.
In our best-fit model, both SCUBA-selected ULIRG and ISO-selected LIRG
galaxies have the same evolution history: if anything, SCUBA 
sources originating massive E/S0s might evolve on a faster cosmic timescale.
This could be still in line with the expectations of hierarchical clustering models 
if we consider that SCUBA sources likely trace the
very high-density environments (galaxy clusters) with an accelerated merging rate at
high-z, while ISO sources are likely related with lower-density environments 
(galaxy groups or the field) entering the non-linear collapse phase at later cosmic
epochs (e.g. Franceschini et al. 1999).

Finally, if indeed the IMF characteristic of the starburst phase is deprived 
of low-mass stars, as suggested in
the previous paragraphs, a consequence would be that the excess blue stars
formed during the starburst would quickly die and disappear, leaving the colors of the 
emerging remnant as typically observed for early-type galaxies and keeping 
consistent with the evidence that the stellar mass in spheroidal galaxies does 
not change much for $z<1$.

\section{SUMMARY AND CONCLUSIONS}

We have analyzed a large dataset derived from deep galaxy surveys at
long wavelengths, exploiting in particular new ISO and published SCUBA observations, 
but also including data from IRAS, COBE and IRAM.
This study of galaxy evolution at long wavelengths benefit also by the
unique situation to combine constraints coming from faint resolved sources
with data on the integrated source emission provided by the spectral intensity of the 
cosmic IR background. The main results of our analysis are hereby summarized.

\begin{itemize}

\item
(1) The deep surveys by ISOCAM LW3 at 15 $\mu$m provide the most precise quantification
of statistical properties of faint IR galaxies, in the form of deep source counts 
and redshift distributions for faint complete samples. 
The numerous ISO-selected 15 $\mu$m sources also allow to 
perform detailed physical investigations of the high-z IR population, since their
optical counterparts are rather straightforward to identify. 
Analyses of the mid-IR to far-IR flux correlations suggest that not only at the ISOCAM 
LW3 flux limit a major fraction of the mid-IR CIRB intensity is resolved into sources,  
but also that the same sources are likely to contribute a substantial fraction of the 
bolometric energy density in the CIRB (Elbaz et al. 2001). 
Our present comparative study of the multi-wavelength counts and
LF's entirely confirms this conclusions, by showing that the various statistics 
on IR sources can only be reconciled each other by assuming 
IR SEDs typical of starbursts, in which case $\sim 50\%$ of the bolometric CIRB
is contributed by ISOCAM 15 $\mu$m sources brighter than S$_{\rm 15}=40\ \mu$Jy.
The sources of this important cosmological component, including $\sim 70\%$ of the 
integrated bolometric emission by galaxies, can then be investigated in the ISOCAM 
population.  
By contrast, our model predictions suggest that the extreme luminosities and 
redshifts and low spatial density of the SCUBA-selected population are such that 
only a modest fraction of the bolometric CIRB energy can be produce by them,
although they can dominate it at the long-$\lambda$.
For a constant comoving volume emissivity as a function of $z$ (see Fig. \ref{sfr}),
high-redshift sources suffer a $(1+z)^{-3.5}$ penalty factor in their contribution
to the CIRB.

\item
(2) The most robust conclusion coming from the present analysis appears to be that of 
a very 
rapid increase of galaxy long-wavelength volume emissivity with redshift, paralleled 
by an
increased incidence in high-redshift sources of dust extinction and thermal dust
reprocessing, with respect to locally observed galaxies. This is the faster evolution
rate observed for galaxies at any wavelengths [$\rho(L,z)\propto (1+z)^{5}$ if averaged
over the whole galaxy population, see Fig. \ref{sfr}], 
and is higher than the rates inferred for quasars.
We have found that the shape of the LW3 counts (a roughly Euclidean behavior down to 
S$_{\rm 15}\simeq 5$ mJy followed by a very fast upturn) is better fit by assuming 
that for only a fraction of local galaxies the IR emissivity evolves back in time.
In this case strong evolution both in number density and in the average source 
luminosity is indicated.

\item
(3) The combined constraints set by the z-distributions
and by the spectral shape of the CIRB impose that this fast evolution rolls-over
around redshift 1, and keeps flat above. The evolution of the luminosity density
is so rapid up to $z\simeq 1$ that no much further increase is left for the higher 
redshifts.          Consequently, the history 
of galaxy formation traced by long-wavelength observations does not appear
fundamentally different from
the one inferred from optical observations, and shows a maximum around $z\sim 1$.
Scenarios in which a substantial fraction of stellar formation happens at very high-z 
(e.g. producing the bulk of stars in spheroidal galaxies at $z>2-3$) 
are not supported by our analysis, and appear to conflict in particular with the 
observed shape of the CIRB at $1000<\lambda<100\ \mu$m.
Obviously, for a definitive proof of these statements it will be necessary
to resolve a significant
fraction of the CIRB at its peak ($\lambda\sim 100$ to 300 $\mu$m). The next important 
step in this sense is expected from the Herschel Space Observatory.

\item
(4) Our present results, together with preliminary spectroscopic studies of the faint 
IR sources, suggest that only a minor fraction of their IR flux originates from AGN 
activity, the bulk of it being likely due to star formation. Our best-fits to the SEDs
indicate massive systems hosting violent starbursts ($SFR\sim 100$ M$_\odot$/yr).
We have suggested that the most natural interpretation for the strong observed
evolution is to assume that the evolving starbursting population consists of otherwise
normal galaxies observed during a dust-extinguished short-lived but 
luminous starburst event. The strong increases with redshift
of the {\sl probability} of interactions (as partly due to a plain geometrical effect
in the expanding universe) and of the {\sl effects} of interactions (due to the 
more abundant fuel avaliable in the past), likely explain the observed rapid
evolution. Galaxy interactions and mergers are emphasized by the present analysis
as a crucial cosmogonic driver for galaxy evolution.

\item
(5) Our suggested evolutionary scheme considers bi-modal star-formation in
galaxies: SF during a long-lived quiescent phase, and enhanced SF taking place in
transient starburst phases recurrently triggered by interactions and merging. 
The former would be responsible for building of galaxy disks, 
the latter for the assembly of spheroidal components in galaxies.
Very schematically, we have attributed the optical background to emission during
the quiescent mode, and explain the CIRB as mostly due to dust-extinguished emission
by young stars during the starbursting phase. 

\item
(6) The large energy content in
the CIRB is not easy to explain, if we consider the modest efficiency of the 
mass-energy transformations allowed by stellar evolution. 
The most obvious way to alleviate the
combined constraints set by the local observed amounts of low-mass stars and metals
would be to assume that a stellar IMF somewhat deprived in low-mass stars is
characteristic of the IR starburst phase. 
In any case, we expect that as much as 2 times the amount of
metals in stars should be present in the diffuse intergalactic medium
as a remnant of the ancient SB phase. A support to this concept is provided by
observations of the abundance of metals in hot Intra-Cluster plasmas.

\item
(7) Many of the phenomena revealed by long-wavelength observations were largely
unexpected based on UV-optical-NIR observation: among others, 
the rate of cosmic evolution of IR
galaxies, the CIRB energetics, the luminosities and rates of SF for LIRGs and ULIRGs.
The attempts so far to infer the IR properties based on UV-optical observations are
producing modest results: there does not seem to be an alternative to long-wavelength
observations if we aim at an exhaustive and reliable description of the history 
of SF in galaxies.

\end{itemize}

\begin{acknowledgements}
We have pleasure to acknowledge fruitful discussions and exchanges in particular
with S. Bressan, J.L. Puget, M. Harwit and H. Flores.
This research has been supported by the Italian Space Agency (ASI) and the
European Community RTN Network "POE", under contract HPRN-CT-2000-00138.
\end{acknowledgements}

\centerline{APPENDIX}

\centerline{PREDICTIONS FOR FUTURE TESTS}

We report in this Section predictions based on the reference
model, which could be useful for testing it, as well as
for planning of future observations.

The confusion noise, which is the fundamental limiting factor for space IR instrumentation,
is evaluated in this Section as well as in the paper from the criterion to
accept sources down to a flux limit corresponding to 1/27 independent beams. 
For Euclidean counts, this roughly corresponds to a $3 \sigma$ confidence limit.
Note however that for non-Euclidean counts, a case often encountered in the infrared,
this criterion may under- or over-predict the $3 \sigma$ limit according to
the count slope (Franceschini 2000).

\begin{figure}
\psfig{figure=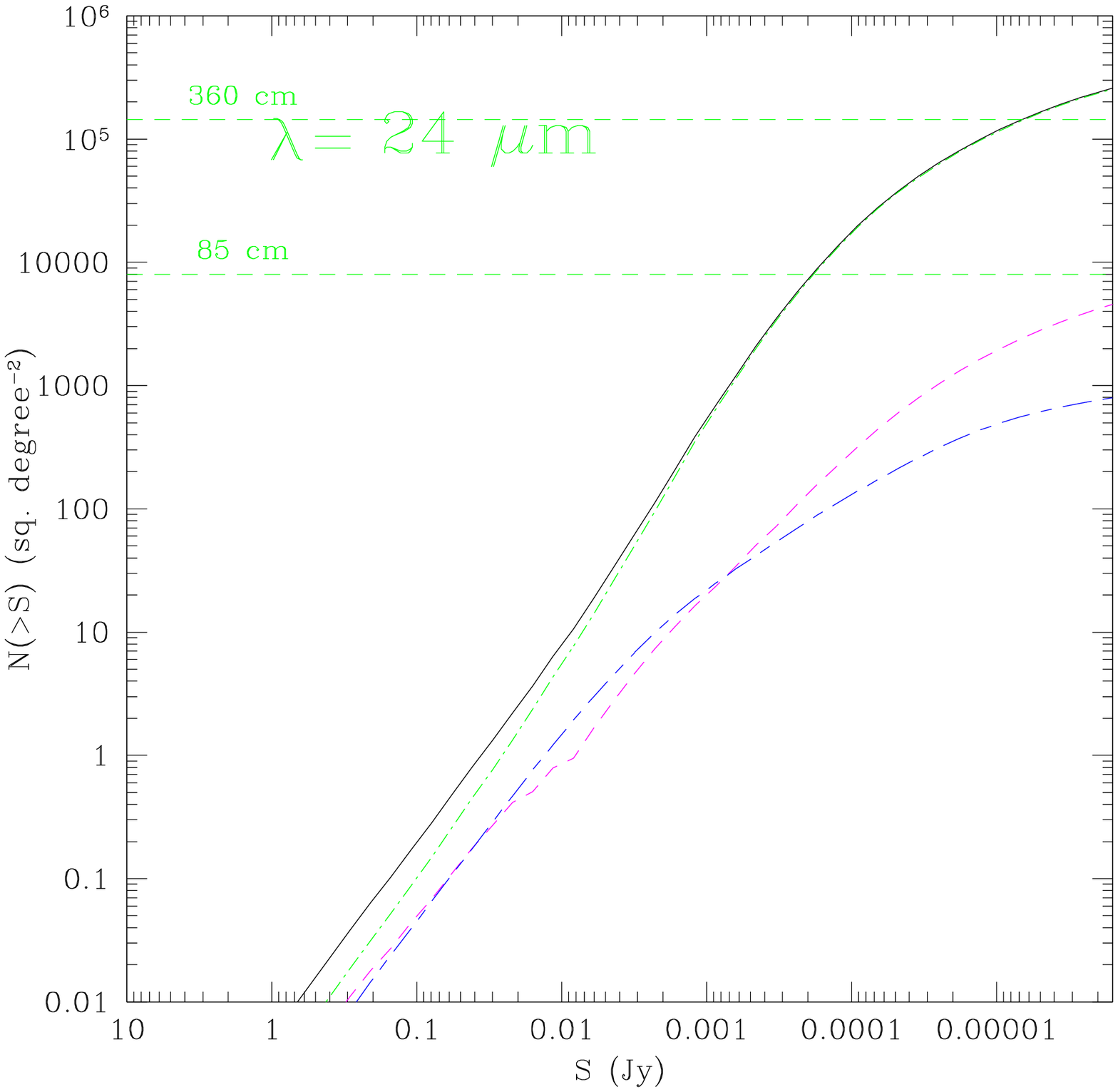,height=70mm,width=80mm}
\psfig{figure=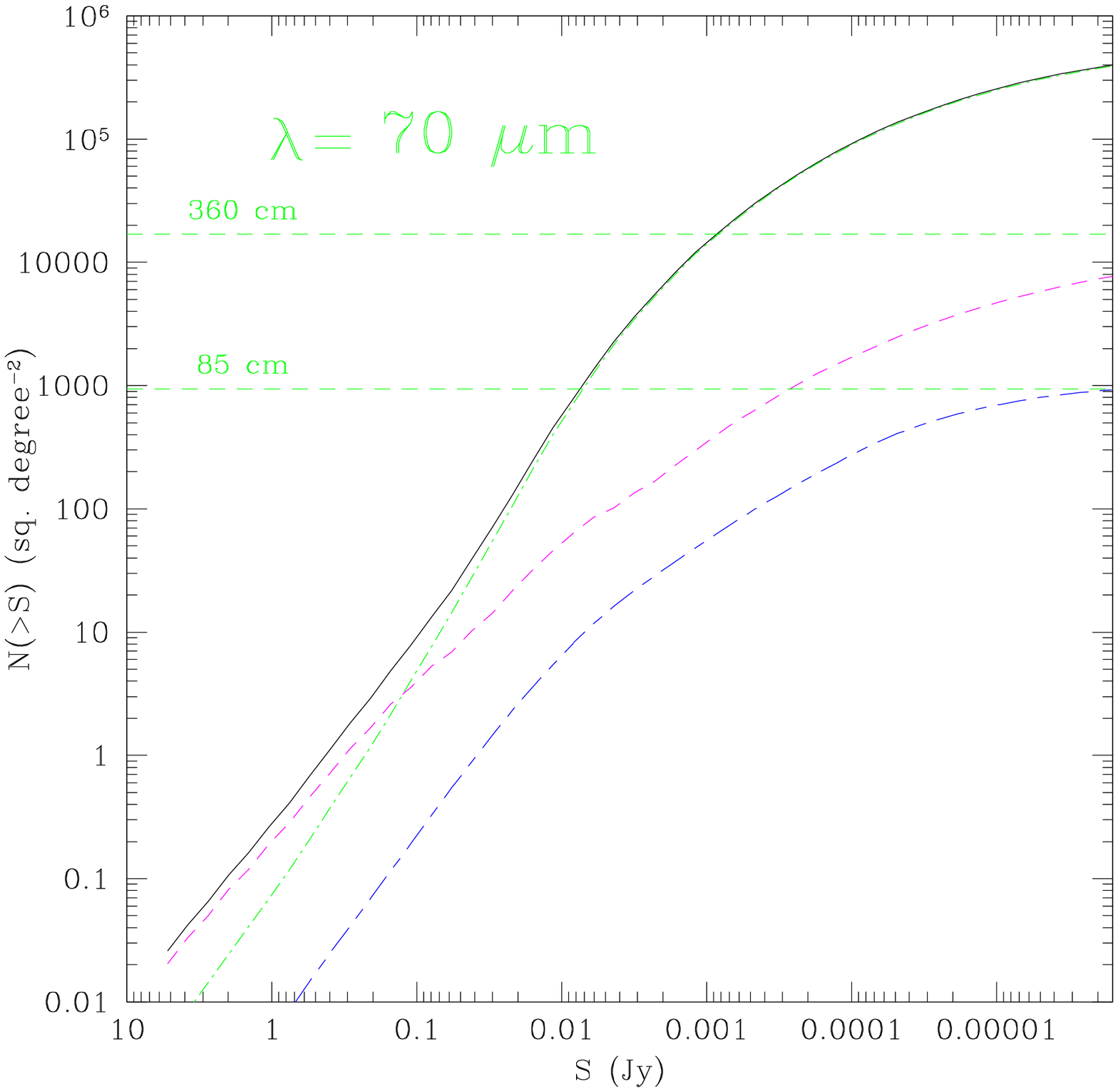,height=70mm,width=80mm}
\vspace*{1.cm}
\caption{Integral number counts at $\lambda_{\rm eff}=24$ and 70 $\mu$m. The SIRTF
confusion limits are referred to as the 85 cm lines.
For comparison, an 8m {\sl New Generation Space Telescope} observing at 
24 $\mu$m would be confusion limited at a source areal density of 
$10^6\ sources/sq.deg.$ (at the level of the top axis).
}
\label{A1}
\end{figure}

SIRTF will operate an imaging camera (MIPS) with broad-band filters centered
at 24, 70 and 170 $\mu$m. Fig. \ref{A1} (see also Fig. \ref{c175}) 
reports predicted counts at these wavelengths.

\begin{figure}
\psfig{figure=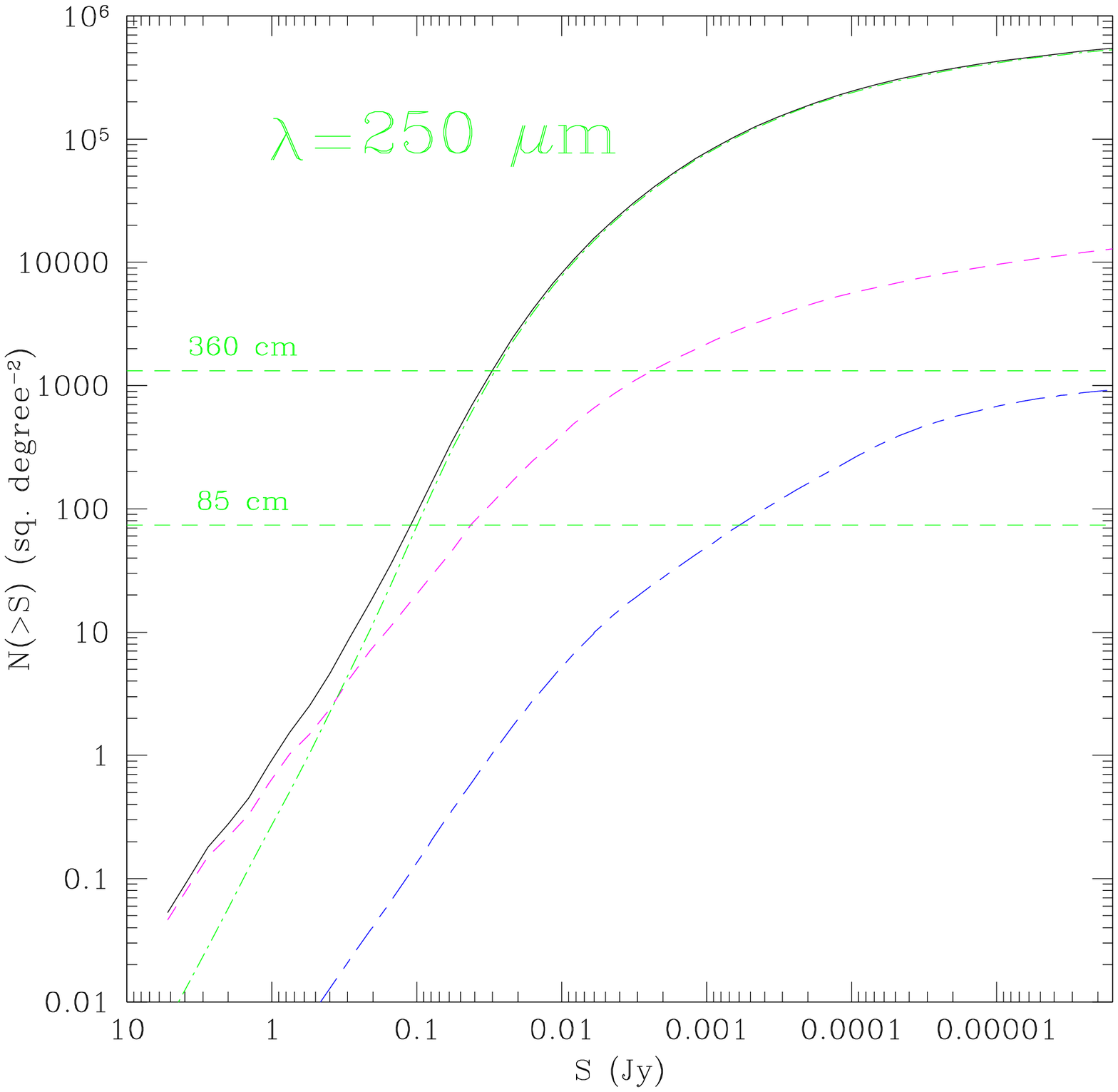,height=70mm,width=80mm}
\psfig{figure=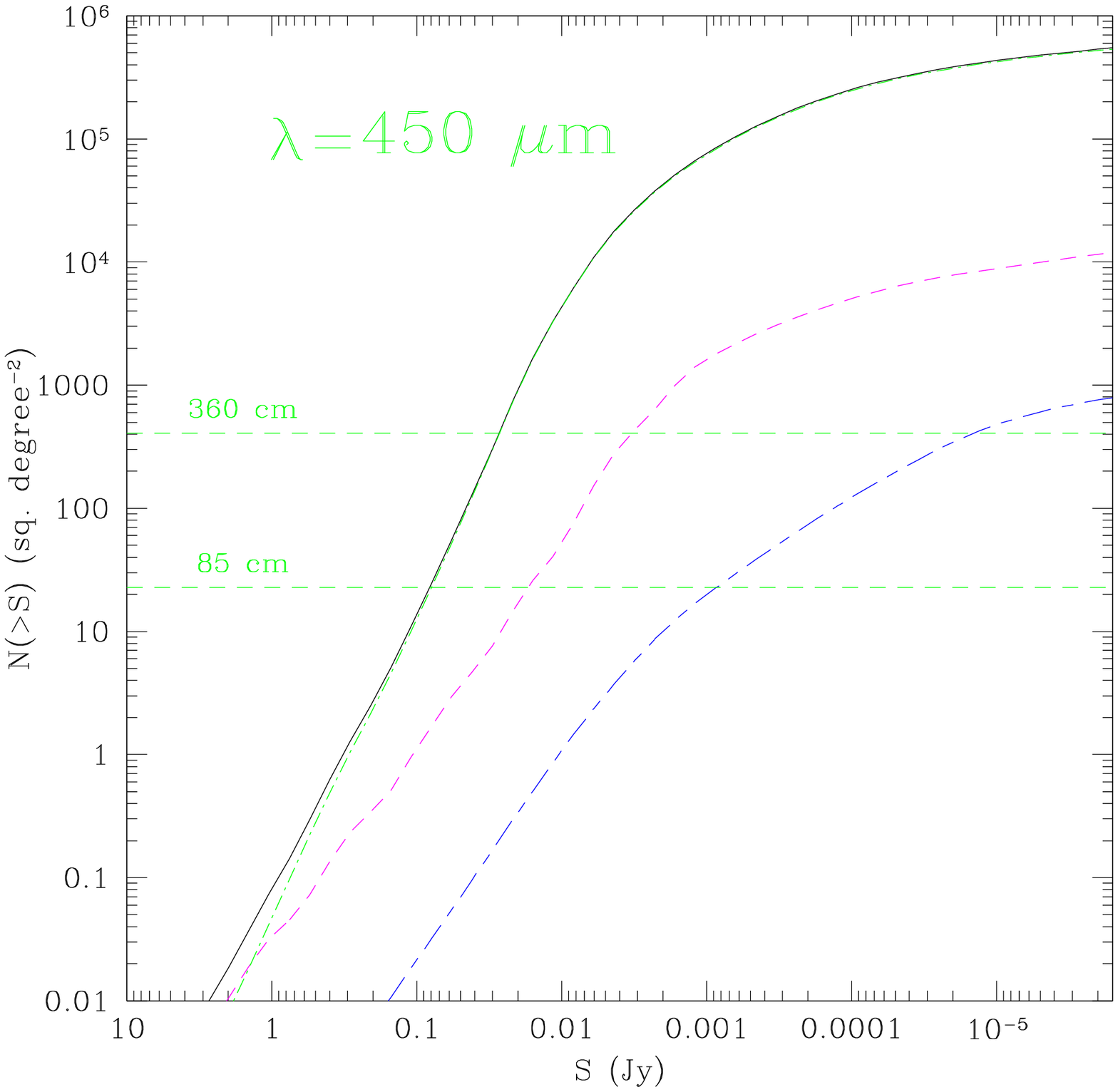,height=70mm,width=80mm}
\vspace*{1.cm} 
\caption{Integral number counts at $\lambda_{\rm eff}=250$ and 450 $\mu$m, relevant
for future surveys with the Hershel Observatory. See also caption to Fig. \ref{A1}.
}
\label{A2}
\end{figure}

The Herschel Observatory (former ESA Cornerstone FIRST) will
characterize the far-IR (70 to 500 $\mu$m) emission by galaxies at any redshifts,
by pushing down in flux the limit of confusion with its large 3.6m primary mirror.
Fig. \ref{A2} illustrates galaxy counts and the confusion threshold in two of the
Herschel long-$\lambda$ channel at 250 and 450 $\mu$m 
(other information can be retrived from Figs. \ref{c175} and \ref{A1}).

\begin{figure}
\psfig{figure=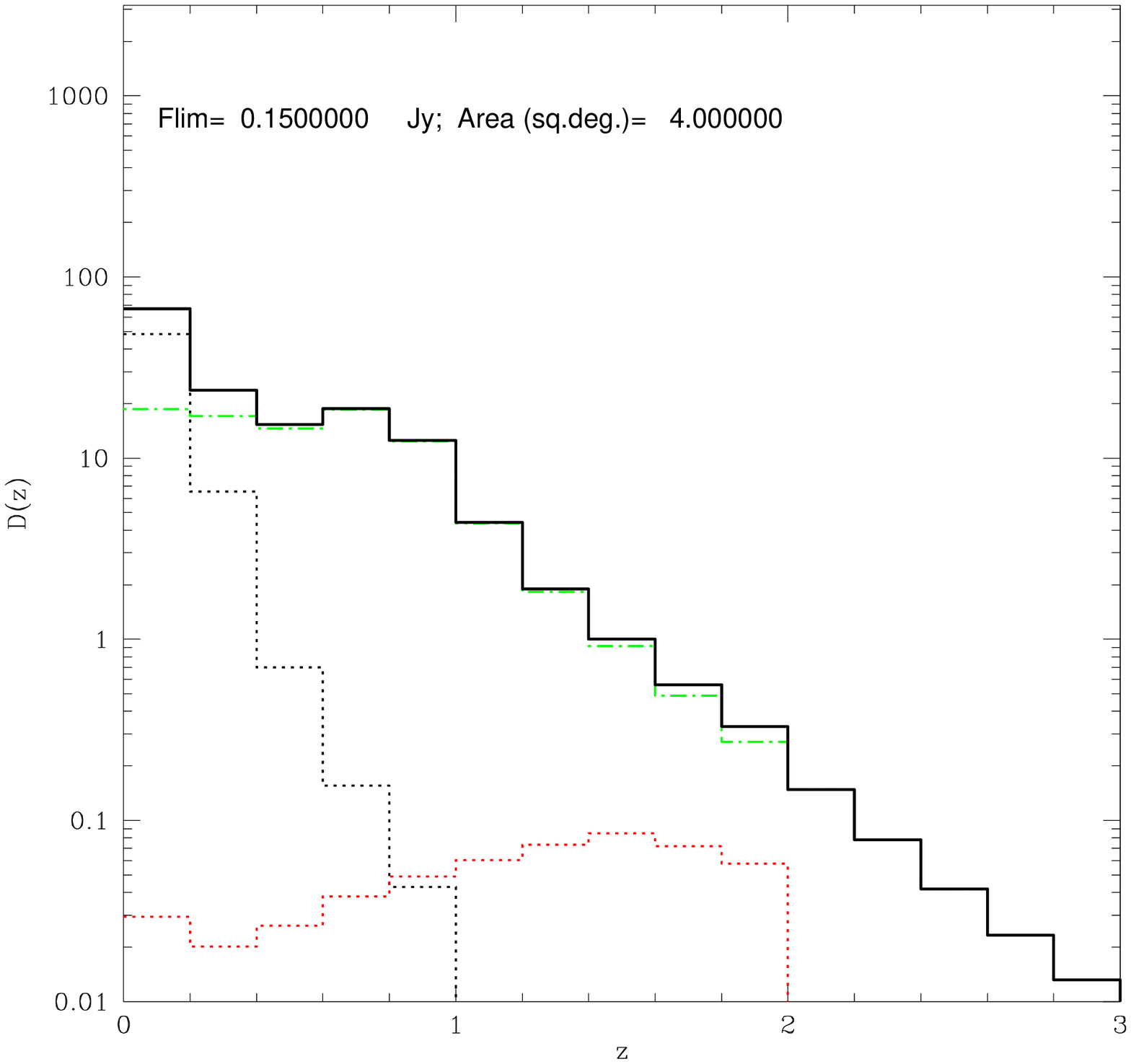,height=70mm,width=80mm}
\psfig{figure=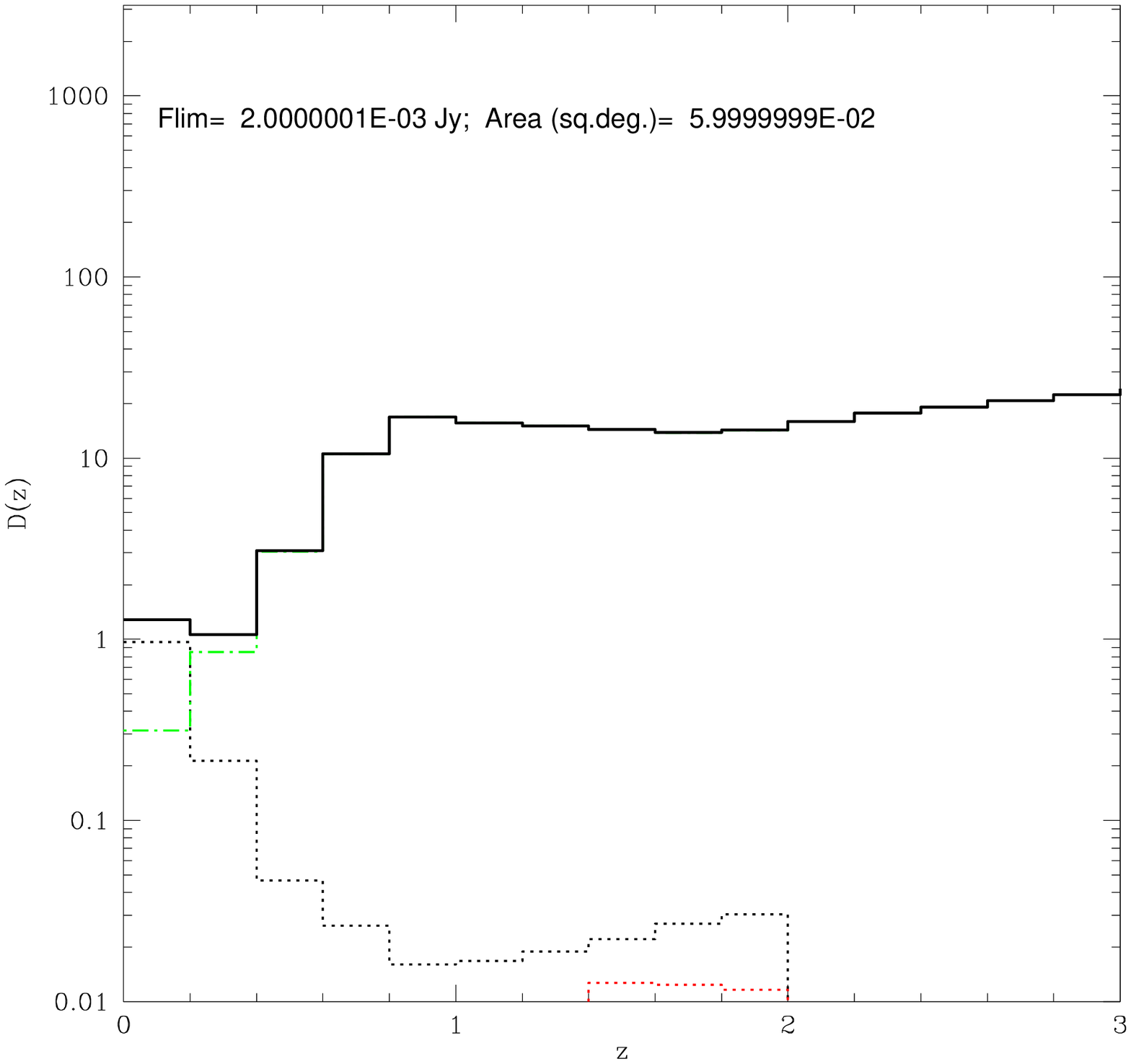,height=70mm,width=80mm}
\vspace*{1.cm} 
\caption{Distributions of redshifts for flux-limited samples at 170 (top panel) 
and 850 $\mu$m (bottom), at the limits of confusion for ISO and SCUBA.
}
\label{A3}
\end{figure}

The distributions of redshifts for flux-limited samples with complete identification
are the other fundamental statistical observable for faint distant sources.
Identifications of FIRBACK/ELAIS sources detected by ISO at 170 $\mu$m will
require an extensive effort of deep radio imaging to reduce the errorbox.
A similar effort will be required to identify SCUBA- or IRAM-detected high-z 
galaxies. Alternatively, the errorbox will be reduced by following-up
SCUBA sources with mm interferometers (Plateaux de Bure, ALMA).
We report in Fig. \ref{A3} predicted z-distributions for both kind of surveys 
at the typical limiting fluxes.

\begin{figure}
\psfig{figure=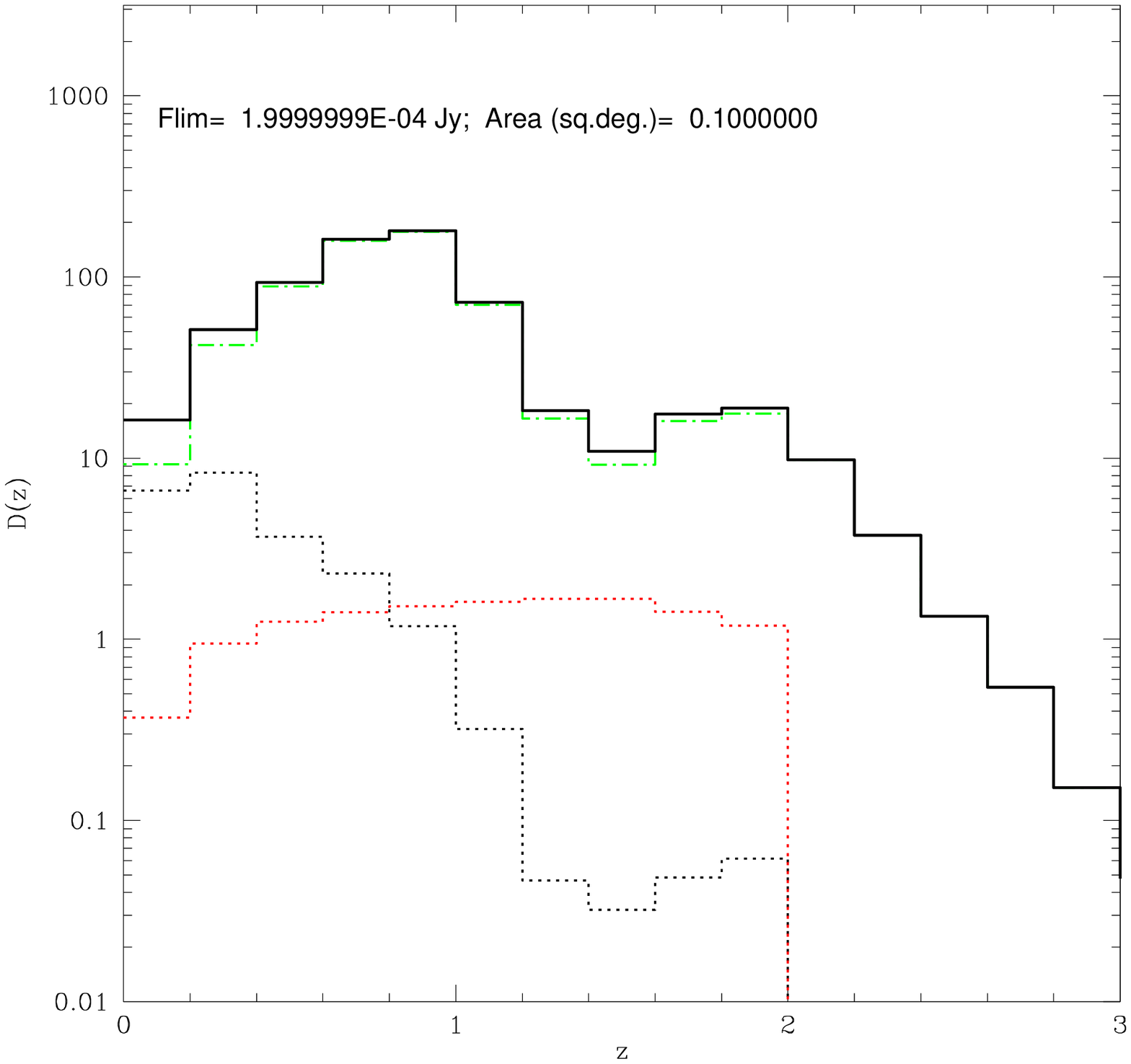,height=70mm,width=80mm}
\psfig{figure=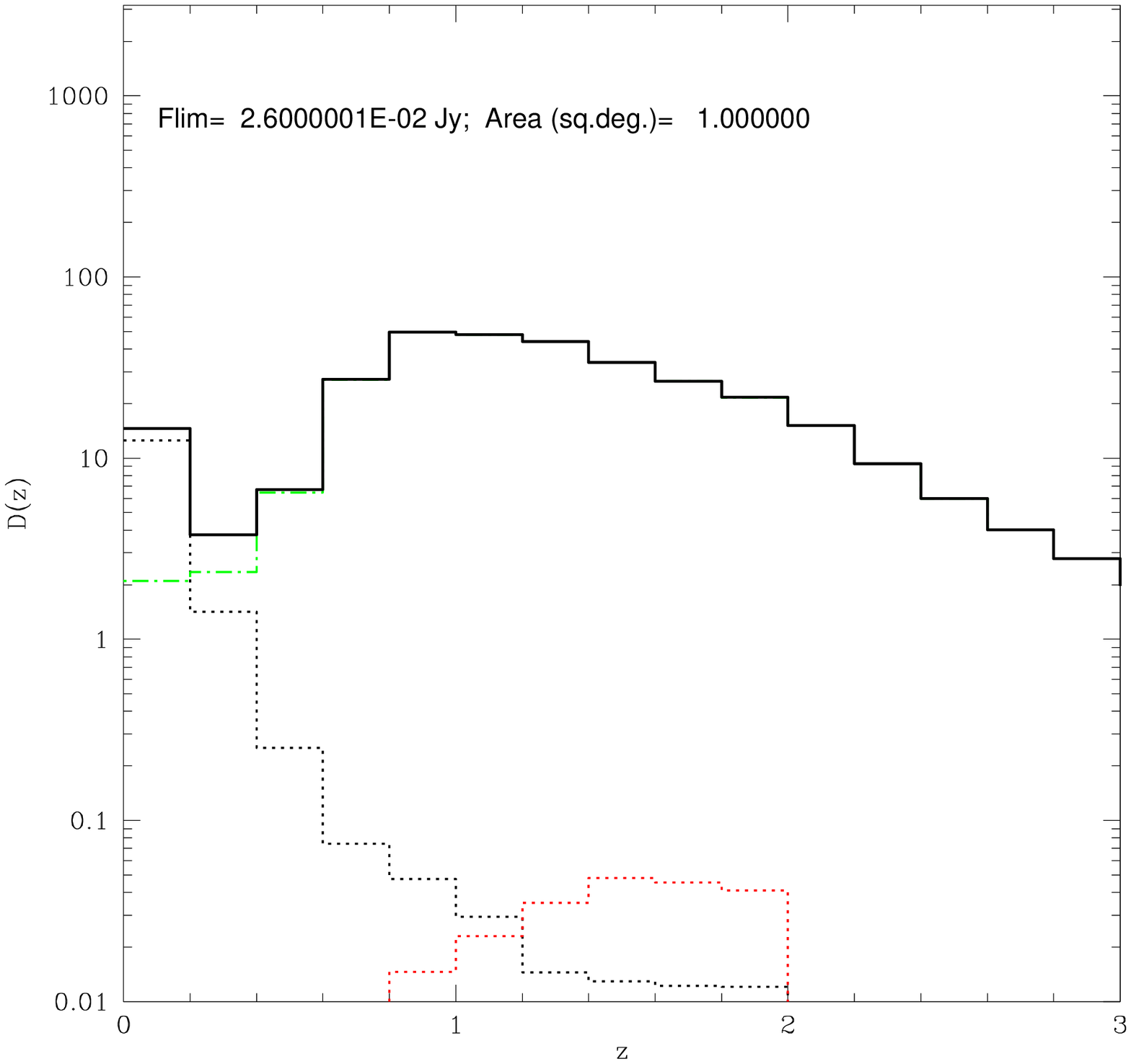,height=70mm,width=80mm}
\vspace*{1.cm} 
\caption{Distributions of redshifts for flux-limited samples at 24 (top panel) 
and 450 $\mu$m (bottom), at the limits of confusion for SIRTF and Herschel.
}
\label{A4}
\end{figure}

Finally, we report in Figures \ref{A4} z-distributions for confusion-limited 
surveys by SIRTF and Herschel at 24 and 450 $\mu$m, respectrively.
In both cases, the K-correction for typical starburst spectra plays in favour 
of the detection of galaxies well above $z=1$. 
Note, in particular, the secondary peak at $z\simeq 2$ in $D(z)$ for the
24 $\mu$m selection, due to the PAH emission bundle in the rest-frame 
$\sim 8$ $\mu$m entering the observable bandwidth at such redshift, provides
an attractive feature of the forthcoming SIRTF surveys (e.g. Lonsdale 2001;
Lonsdale et al. 2001; Dickinson 2001).

\begin{thebibliography}{} 

  \bibitem[]{}
Adelberger, K. L., Steidel, C. C., 2000, ApJ 544, 218.

  \bibitem[]{}
  Aharonian, F., Akhperjanian, A. G., Barrio, J. A., et al., 1999, A\&A 349, 11.

\bibitem[]{}
Alexander, D., \& Aussel, H.,  2000, in "ISO Surveys of a Dusty Universe,", in press 
(astroph/0002200).

       \bibitem[]{}
 Almaini, O., Lawrence, A., Boyle, B.J.,  1999, MNRAS 305, L59.

  \bibitem[]{}
Altieri, B., Metcalfe, L., Kneib, J.P., et al. , 1999, A\&A 343.

  \bibitem[]{}
Andreani, P., Franceschini, A., 1996, MNRAS 283, 85.


  \bibitem[]{}
Aussel, H., 1998, PhD thesis, CEA Saclay, Paris.

  \bibitem[]{}
Aussel H., Cesarsky C., Elbaz D., Starck, J.L. 1999, A\&A 342, 313.

  \bibitem[]{}
Aussel, H., et al., 2001, in preparation.

  \bibitem[]{}
Aussel, H., Coia, D., Mazzei, P., De Zotti, G., Franceschini, A., 
2000, A\&A Suppl. 141, 257

  \bibitem[]{}
 Barger, A. J., Cowie, L. L., Sanders, D. B., ulton, E., Taniguchi, Y.,
 Sato, Y., Kawara, K., Okuda, H.,  1998, Nature 394, 248.

  \bibitem[]{}
 Barger, A. J., Cowie, L. L., Sanders, D. B., 1999, ApJ 518, L5.

  \bibitem[]{}
 Barger, A. J., Cowie, L. L., Smail, I.,
 Ivison, R. J., Blain, A. W., Kneib, J.-P, 1999, AJ 117, 2656.
 
\bibitem[]{}
  Barger, A., Cowie, L., Mushotzky, R., Richards, E., 2001, 
 AJ 121, 662.
  


  \bibitem[]{}
Bassani, L., Franceschini, A., Malaguti, G., Cappi, M., Della Ceca,
R., 2001, A\&A submitted.
 
  \bibitem[]{}
Bautz, M.W., Malm, M. R., Baganoff, F. K., et al., 2000, ApJ 543, L119.
 
  \bibitem[]{}
Bernstein, R.A., 1998, PhD Thesis, California Institute of Technology, 
Source DAI-B 59/11. 


  \bibitem[]{}
  Bertoldi, F., Carilli, C. L., Menten, K. M., et al., 2000, A\&A 360, 92.

  \bibitem[]{}
 Bertoldi, F., Menten, K.M., Kreysa, E., Carilli, C.L., Owen, F., 2001,
 in Proc. JD9, IAU Manchester 2001, `Cold Gas and Dust at High Redshift',
 D.J. Wilner Ed., Highlights of Astronomy Vol. 12. PASP, in press (astroph/0010553).
  
%
%
%
%

  \bibitem[]{}
 Biviano, A., Metcalfe, L., Altieri, B., et al., 2001, ASP Conference
 Proceedings on "Clustering at high redshifts", A. Mazure et al. Eds., 
 in press (astro-ph/9910314).
 
   \bibitem[]{}
 Blain, A.W., Longair, M.S., 1993, MNRAS 264, 509.
 
   \bibitem[]{}            
 Blain, A. W., Kneib, J.-P., Ivison, R. J., Smail, I., 1999, ApJ 512, L87.
 
  \bibitem[]{}
 Carilli, C.L., Yun, M.S., 1999, ApJ Letters 513, L13.


  \bibitem[]{}
Chapman, S.C., Scott, D., Steidel, C., et al., 2000, MNRAS 319, 318.

  \bibitem[]{}
Chary, R., Elbaz, D., 2001, ApJ 556, 562.



  \bibitem[]{} 
Cohen, J., Hogg, D.W., Blandford, R., et al., 2000, ApJ 538, 29.

%

  \bibitem[]{}
  Connolly, A., Szlay, A., Dickinson, M., Subbarao, M., Brunner, R., 1997, ApJ 486, L11.
  
  \bibitem[]{}
Coppi, P., Aharonian, F., 1999, Astropart. Phys. 11, 35-39
               
  \bibitem[]{}
 Danese  L.,  De  Zotti L.,  Franceschini A.,  Toffolatti  L.,  1987,
 ApJL 318, L15.

%

  \bibitem[]{} 
Desert F.X., Puget, J.-L., Clements, D., et al. , 1999, A\&A 342, 363.

  \bibitem[]{}
 Devriendt, J. E. G., Guiderdoni, B., Sadat, R., 1999, A\&A 350, 381.

  \bibitem[]{}
  Dey, A., Ivison, R., Smail, I., Wright, J., Liu, M., 1999, ApJ 519, 610.


  \bibitem[]{} 
Dickinson, M., 2001, "The Great Observatories Origins Deep Survey (GOODS)",
AAS 198, 2501.

  \bibitem[]{}
Dole, H., Gispert, R., Lagache, G., et al. , 2000, in {\sl  ISO Beyond Point Sources, 
Studies of Extended Infrared Emission},  R. Laureijs, K. Leech and M. Kessler Eds., 
ESA-SP 455, 167. 
 
  \bibitem[]{}
Dole, H., Gispert, R., Lagache, G., et al., 2001, A\&A 372, 364.
 
%

  \bibitem[]{}
 Dunne, L., Eales, S. Edmunds, M., Ivison, R., Alexander, P., Clements, D., 
 2000, MNRAS 315, 115.


  \bibitem[]{}
Dwek, E., Arendt, R. G., 1998, ApJ Letters 508, L9.


  \bibitem[]{}
 Eales, S., Lilly, S., Webb, T., Dunne, L., Gear, W., Clements, D., Min, Y.,  
 2000, AJ 120, 2244.

\bibitem[]{}
Efstathiou, A., Oliver, S., Rowan-Robinson, M., et al., 2000, MNRAS 319, 1169.

  \bibitem[]{} 
Elbaz D.,  Aussel H., Cesarsky C.J., Desert F.X., Fadda D., Franceschini A., 
Puget J.L., Starck J.L, 1998, in 
Proc. of 34th Liege International Astrophysics Colloquium on the "Next Generation
Space Telescope", Belgium  (astroph/9807209).

  \bibitem[]{} 
Elbaz D., Cesarsky, C.J., Fadda, D., et al., 1999, A\&A Letters 351, L37.

  \bibitem[]{} 
Elbaz D., Cesarsky, C., Chanial, P., Aussel, H., Franceschini, A., Fadda, D.,
Chary, R., 2001, A\&A submitted.


  \bibitem[]{}
Ellis, R. S.,  1997, ARAA 35, 389.

  \bibitem[]{}
 Ellis, R. S., Colless, M., Broadhurst, T., Heyl, J., Glazebrook, K., 1996, 
 MNRAS 280, 235.


  
\bibitem[]{}
  Fabian, A., Smail, I., Iwasawa, K., et al.,  2000, MNRAS 315, 8.
  
\bibitem[]{}
Fadda, D., Flores, H., Hasinger, G., et al., 2001, A\&A submitted.
  
  \bibitem[]{}
 Fang F., Shupe, D.L., Xu, C., Hacking, P.B. 1998, ApJ 500, 693.

  \bibitem[]{}
Finkbeiner, D.P., Davies, M., Schlegel, D.J., 2000, ApJ 544, 81.

 \bibitem[]{} 
Fixsen D.J., Dwek, E., Mather, J.C., et al. 1998, ApJ 508, 123.
 
  \bibitem[]{} 
 Flores H., Hammer F., Thuan T., et al. 1999, ApJ 517, 148.


  \bibitem[]{}
Franceschini, A., Mazzei, P., De Zotti, G., Danese, L., 1994,  ApJ 427, 140.
 
  \bibitem[]{} 
 {Franceschini, A.}, La Franca, F., Cristiani, S., Martin-Mirones J.,
1994b, MNRAS 269, 683. 


\bibitem[]{} 
Franceschini, A., Gratton, R., 1997, MNRAS 286, 235. 


  \bibitem[]{}
Franceschini, A., Hasinger, G., Miyaji, T., Malquori, D., 1999,
{MNRAS} 310, L5.


  \bibitem[]{}
Franceschini, A., 2000, in "Galaxies at High Redshifts",
Proceedings of the XI CANARY ISLANDS WINTER SCHOOL OF ASTROPHYSICS,
F. Sanchez, I. Perez-Fournon, M. Balcells, F. Moreno-Insertis Eds.,
Cambridge University Press.

  \bibitem[]{}
Frayer, D.T., Ivison, R.J., Scoville, N.Z., 
et al., 1999, ApJ Letters 514, L13.

  \bibitem[]{}
Genzel, R., \& Cesarsky, C.J., 2000, ARAA 38, 761.

  \bibitem[]{}
Genzel, R., Lutz, D., Sturm, E., et al. , 1998, ApJ 498, 579.


  \bibitem[]{}
Giacconi, R., Rosati, P.,  Tozzi, P., et al., 2001, ApJ 551, 624.

  \bibitem[]{}
 Gispert, R., Lagache, G., Puget, J. L., 2000, AA 360, 1.



  \bibitem[]{}
Gorjian, V., Wright, E. L., Chary, R. R.,  2000, ApJ 536, 550.




  \bibitem[]{}
Granato, G.L., Silva, L., Monaco, P., et al., 2001, MNRAS 324, 757.




  \bibitem[]{}
Guiderdoni, B., Bouchet, F., Puget, J.-L., et al. , 1997, Nature 390, 257.

  \bibitem[]{}
 Haarsma, D. B., Partridge, R. B.,  1998, ApJ Letters 503, L5.

 


  \bibitem[]{}
 Harwit, M.,  1999, ApJ Letters 510, L83.
  

 \bibitem[]{}
 Harwit, M., Protheroe, R. J.,
 Biermann, P. L., 1999, ApJ Letters 524, 91.


   \bibitem[]{}
Hauser, M.G., Arendt, R.G., Kelsall, T., et al. 1998, ApJ 508, 25.


  
  
  \bibitem[]{}
 Hornschemeier,  A.E., Brandt, W.N., Garmire, G.P., 2000, ApJ 554, 742.
 

  \bibitem[]{}
Hughes, D., Serjeant, S., Dunlop, J., et al. , 1998, Nature 394, 241.

 \bibitem[]{}
Ivison R. J., Smail I., Barger A. J., Kneib J.-P., Blain A. W., Owen F. N.,
Kerr T., Cowie L. L., 2000a, MNRAS 315, 209.

  \bibitem[]{}
Ivison, R.J.,  Dunlop, J.S., Smail, I., Dey, A., Liu, M., Graham, J., 2000b, 
ApJ 542, 27.



  \bibitem[]{}
Juvela, M., Mattila, K., \& Lemke, D., 2000, A\& A 360, 813.

   \bibitem[]{}
Kauffmann, G., Charlot, S., 1998, MNRAS 297, L23. 


  \bibitem[]{}
Kormendy, J., Sanders, D., 1992, ApJ Letters 390, 53.

  \bibitem[]{}
Krawczynski, H., Coppi, P.S., Maccarone, T., Aharonian, F.A., 2000, A\&A   353,  97.
 
  \bibitem[]{}
Lagache, G., Puget, J. L., 2000, A\&A 355, 17L.

    \bibitem[]{}
Lagache G., Abergel, A., Boulanger, F., Desert, F.X., Puget J.L., 1999, 
A\&A 344, 322L.


   \bibitem[]{}
Lari, C., Pozzi, F., Gruppioni, C., et al. , 2001, MNRAS 325, 1173.


  \bibitem[]{}
 Le Fevre, O., Abraham, R., Lilly, S.J., et al. , 2000, MNRAS 311, 565.




  \bibitem[]{}
  Lilly, S., Le Fevre, O., Hammer, F., Crampton, D., 1996, ApJ 460, L1.

  \bibitem[]{}
 Lilly, S.J., Eales, S.A., Gear, W., et al. ,  1999, ApJ 518, 641.

  \bibitem[]{}                         
  Lonsdale, C. J.,  2001, AAS 198, 2502
   
  \bibitem[]{}                   
Lonsdale, C.J., Smith, H.E., Rowan-Robinson, M., et al., 2001, 
"The SIRTF Wide-Area Infrared Extragalactic
Survey (SWIRE)", http://www.ipac.caltech.edu/SWIRE.


   \bibitem[]{}
Madau, P., Pozzetti, L.,  2000, MNRAS Letters 312, 9.
 
  \bibitem[]{}
Madau, P., Ferguson, H.C., Dickinson, M.E., et al. , 1996, MNRAS 283, 1388.


  \bibitem[]{}
  Matsuhara, H., Kawara, K., Sato, Y., et al., 2000, A\& A 361, 407.
  

  \bibitem[]{}
 Mazzei, P., Aussel, H., Xu, C., Salvo, M., de Zotti, G., Franceschini, A., 2001, 
 New Astronomy 6, 265.
 
 




  \bibitem[]{}
 Mushotzky, R. F., Loewenstein, M.,  1997, ApJ Letters 481, 63.


  \bibitem[]{}
Oliver, S., Rowan-Robinson, M., Broadhurst, T.J., et al., 1996, MNRAS  280, 673. 

  \bibitem[]{}
Oliver, S., Goldschmidt, P., Franceschini, A., et al. , 1997, MNRAS  289, 471.

   \bibitem[]{}
Oliver, S., Rowan-Robinson, M., Alexander, D.M., et al., 
2000a, MNRAS 316, 749.

  \bibitem[]{}
Oliver, S., Mann, R.G., Carballo, R., et al. , 2000b, MNRAS submitted.

  \bibitem[]{}
Omont, A., McMahon, R. G., Cox, P., Kreysa, E., Bergeron, J., Pajot, F.,
Storrie-Lombardi, L. J., 1996, A\&A 315, 1



 \bibitem[]{}
Padovani, P., Matteucci, F.,  1993, ApJ 416, 26

  \bibitem[]{}
Pearson, C., Rowan-Robinson, M., 1996, MNRAS 283, 174.

%

 

  \bibitem[]{}
Poggianti, B.M., Bressan, A., \& Franceschini, A., 2001, ApJ 550, 195.

  \bibitem[]{}
 Poggianti, B.M., Wu, H., 2000, ApJ 529, 157.



\bibitem[]{}
 Puget J.-L., Abergel, A., Bernard, J.-P., et al. 1996, A\&A 308, L5.

\bibitem[]{}
Puget, G.L., Lagache, G., Clements, D., Reach W., Aussel, H., Bouchet, F.,
Cesarsky, C., Desert, F., Dole, H., Elbaz, D., Franceschini, A., Guiderdoni,
B., Moorwood, A., 1999, A\& A 345, 29.

\bibitem[]{}
Renault, C., Barrau, A., Lagache, G., G.L., 2001, A\& A in press

  \bibitem[]{}
Richards, E. A., Kellermann, K.I., Fomalont, E.B., Windhorst, R.A., Partridge, R.B.,
1998, AJ 116, 1039.



  \bibitem[]{}
Rigopoulou, D., Franceschini, A., Aussel, H., et al. , 2000,
 ApJ Letters, 537, L85.

\bibitem[]{}
 Risaliti, G., Gilli, R., Maiolino, R.,  Salvati, M., 2000, A\& A 357, 13
               
  \bibitem[]{}
Roche, N., Eales, S.A., 1999, MNRAS 307, 111.
                        
  \bibitem[]{}
Rodighiero, G., Franceschini, A., Fasano, G., 2001, MNRAS 324, 491.




  \bibitem[]{}
{Rowan-Robinson}, M., Mann, R. G., Oliver, S. J., et al. , 1997, MNRAS 289, 482.

  \bibitem[]{}
{Rowan-Robinson}, M., 2001, ApJ 549, 745.

  \bibitem[]{}
 Rush, B., Malkan, M.A., Spinoglio, L., 1993, ApJS 89, 1.

J. Wiley \& Sons.

  \bibitem[]{}
Sanders, D., Soifer, B.T., Elias, J.H., et al. , 1988, ApJ 325, 74.

  \bibitem[]{}
Sanders, D., \& Mirabel, I.F., 1996, ARAA 34, 749.

  \bibitem[]{}
Sanders, D.B., 2001, in "New results in Far-Infrared and Sub-millimeter
Astronomy", T. de Graauw and R. Szczerba, in press.

  \bibitem[]{}
Saunders, W., Rowan-Robinson, M., Lawrence, A., Efstathiou, G., Kaiser, N., Ellis, R. S.,
Frenk, C. S., 1990, MNRAS 242, 318.



  \bibitem[]{}
Scott, D., Lagache, G., Borys, C., et al., 2000, A\&A Letters 357, L5.

  



  \bibitem[]{}
Silva, L., Granato, G.L.,  Bressan, A., Danese, L., 1998, ApJ 509, 103.


   \bibitem[]{}
Smail, I., Ivison, R. J., Blain, A. W.,  1997, ApJ Letters 490, L5.

  \bibitem[]{}
Smail, I., Ivison, R.J., Kneib, J.-P., et al. , 1999, MNRAS 308, 1061.
                   
 \bibitem[]{}
 Smail, I., Ivison, R. J., Owen, F.N.,  Blain, A. W., Kneib, J.-P., 
 2000 ApJ 528, 612.



  \bibitem[]{}
Stanev, T., Franceschini, A., 1998, ApJ Letters 494, L159.

  \bibitem[]{}
 Starck, J. L., Aussel, H., Elbaz, D., Fadda, D.,  Cesarsky, C., 1999, A\& AS 138, 365.

  \bibitem[]{}
 Stecker, F., De Jager, O., Salamon, M., 1992, ApJL 390, L49. 

 \bibitem[]{}
Steidel, C., Adelberger, K., Giavalisco, M., Dickinson, M., Pettini, M., 1999, ApJ 519, 1.
 
  \bibitem[]{}
 Stickel, M., Bogun, S., Lemke, D., et al., 1998, A\&A 336, 116.
 
   \bibitem[]{}
 Takeuchi, T., Ishii, T., Hirashita, H., Yoshikawa, K., Matsuhara, H., Kawara, K., Okuda, H.,
 2001, PASJ 53, 37.

  \bibitem[]{}
Taniguchi, Y., Cowie, L.L., Sato, Y., Sanders, D.,
Kawara, K., 1997, A\&A 328, L9 







  \bibitem[]{}
Tozzi, P., Rosati, P., Nonino, M., et al., 2001, ApJ in press (astroph/0103014).

  \bibitem[]{}
  Tran, Q.D., Lutz, D., Genzel, R., et al., 2001, ApJ 552, 527.
  
   
\bibitem[]{}
van der Werf, P., Knudsen, K., Labb'e, I., Franx, M., 2001,
in "The far-infrared and submillimeter spectral energy distributions of active 
and starburst galaxies", I. van Bemmel, B. Wilkes and P. Barthel Eds., 
Elsevier New Astronomy Reviews, in press (astroph/0011217)

\bibitem[]{}
van der Werf, P., Moorwood, A.F.M., Yan, L., 2001, Proceedings of 2nd Hellenic Cosmology 
Workshop, eds. M. Plionis et al., Kluwer in press (astroph/0108161).

  \bibitem[]{}
 Vigroux, L., Charmandaris, V., Gallais, P., et al. , 1999, in {\sl The Universe as seen by ISO}, ESA-SP 427, 805.
 
 
  \bibitem[]{}
 Xu, C., Hacking, P., Fang, F., et al. , 1998, ApJ 508, 576.

  \bibitem[]{}
 Xu, C., Lonsdale, C.J., Shupe, D.L., O'Linger, J., Masci, F., 2001,
 ApJ in press (astro-ph/0009220).
  
  \bibitem[]{}
 Zoccali, M., Cassisi, S. Frogel, J.A., et al. ,  2000, ApJ 530, 418.
 
\end{thebibliography}
\end{document}